\documentclass[usenatbib]{mnras}

\usepackage{graphicx}	
\graphicspath{{figs/}}
\usepackage{amsmath}	
\usepackage{amssymb}	
\usepackage{multicol}   
\usepackage{array}      
\usepackage{bm}		
\usepackage{pdflscape}	
\usepackage{xcolor}     
\usepackage{orcidlink}  

\newcommand{\msun}{M_{\odot}}

\newcommand{\disperse}{$\mathrm{DisPerSE}$\ }
\newcommand{\rockstar}{\textsc{Rockstar}\ }
\newcommand{\mflip}{M_\mathrm{flip}}

\usepackage[T1]{fontenc}
\usepackage{ae,aecompl}
\usepackage{mathptmx}

\title[Channelled vs Isotropic Group Accretion]{
Channelled versus Isotropic Accretion of Galaxy Groups onto Clusters:
Kinematics, Spin Alignment, and Assembly Bias
}

\author[Singh et al.]{
Ankit Singh\orcidlink{0000-0001-5427-4515}$^{1}$\thanks{Contact e-mail: \href{mailto:ankit.singh@nottingham.ac.uk}{ankit.singh@nottingham.ac.uk}}, Frazer Pearce\orcidlink{0000-0002-2383-9250}$^{1}$, Meghan Gray\orcidlink{0000-0002-6301-5870}$^{1}$, Ulrike Kuchner\orcidlink{0000-0002-0035-5202}$^{1}$, Julian Onions\orcidlink{0000-0001-5192-6856}$^{1}$,
\newauthor
Gustavo Yepes\orcidlink{0000-0001-5031-7936}$^{2,3}$, Alexander Knebe\orcidlink{0000-0003-4066-8307}$^{2,3,4}$
\\
$^{1}$School of Physics and Astronomy, University of Nottingham, University Park, Nottingham NG7 2RD, United Kingdom\\
$^{2}$Departamento de F\'isica Te\'orica M-8, Universidad Aut\'onoma de Madrid, Cantoblanco, 28049, Madrid, Spain\\
$^{3}$Centro de Investigaci\'on Avanzada en F\'isica Fundamental (CIAFF), Facultad de Ciencias, Universidad Aut\'onoma de Madrid, Cantoblanco, 28049, Madrid, Spain\\
$^{4}$International center for Radio Astronomy Research, University of Western Australia, 35 Stirling Highway, Crawley, Western Australia 6009, Australia
}

\date{Accepted XXX. Received YYY; in original form ZZZ}

\pubyear{2026}

\begin{document}
\label{firstpage}
\pagerange{\pageref{firstpage}-\pageref{lastpage}}
\maketitle

\begin{abstract}
Galaxy clusters grow by accreting groups from the cosmic web, but whether the route of infall (along a cluster-connected filament or from the open field) imprints a group's kinematics, spin, and orbit remains unmeasured at scale. We classify $212{,}704$ group-cluster infall pairs across $27{,}029$ clusters in the $(1\,h^{-1}\,\mathrm{Gpc})^3$ MDPL2 simulation into primary (cluster-connected), secondary (tributary), primary-outskirt, and isotropic channels using \disperse filaments. (i) In both channelled classes, group spin axes lie mainly parallel to the filament below a `spin-flip' mass and perpendicular above it, $\log_{10} \mflip\;[h^{-1}\,\msun] = 13.07$ and $13.12$, a $0.05$-dex offset consistent with zero, lying $1.3$--$1.4$\,dex above the field, while isotropic groups show no flip; $\mflip$ stays radially flat out to $7\,R_{200}$. (ii) Within the primary class $\mflip$ rises with the linear mass density of the host filament, from $12.96$ (thin tendrils) to $13.44$ (dense bridges); matching group mass, cluster mass and distance keeps the rise positive but leaves its size uncertain ($0.07$--$0.58$\,dex). (iii) Isotropic groups fall below random alignment at $2\,R_{200}$ ($41$ per cent within $60^\circ$, a $4.3\sigma$ deficit). Primary groups show $85$ per cent velocity-filament alignment at $2\,R_{200}$, and isotropic groups infall $150$--$180\,\mathrm{km\,s^{-1}}$ faster, and more radially, out to $5\,R_{200}$. At matched distance isotropic groups are therefore predicted to reach orbital pericentres up to $10$ per cent smaller than primary groups, whose smaller pooled pericentres reflect only their proximity to the cluster. The topological route of accretion, and within it the host filament's linear density, shapes how groups enter clusters, testable with wide-area spectroscopic surveys.
\end{abstract}




\section{Introduction}
\label{sec:intro}

Galaxy clusters reside at the nodes of the cosmic web where multiple dark-matter filaments converge \citep{deLapperant1986, Bond1996}, and their growth is far from isotropic: cluster shapes and orientations align with the connecting large-scale filamentary structure \citep{West1995}, and $N$-body simulations confirm that matter is preferentially channelled along intercluster filaments \citep{Colberg2005}. This geometry is now directly measurable in observations. \citet{BaierSoto2025} find that 82\,per~cent of X-ray clusters are connected to their nearest optical filament, and the first measurement of the velocity field along a cluster-connecting filament \citep{splitinfall2025} reveals a coherent split-infall flow with a stagnation point at the filament midpoint and a $>5\sigma$ detection, demonstrating that filaments are dynamically active channels.

The number of filaments attached to a cluster, its connectivity $\kappa$, scales with cluster mass from first principles \citep{Codis2018} and is imprinted in the cluster's own dynamical state: at fixed mass, high-$\kappa$ clusters are more elongated and more unrelaxed than low-$\kappa$ clusters \citep{Gouin2021}. Connectivity is imprinted on the galaxies at the nodes as well: at fixed stellar mass, galaxies at more highly connected nodes are less star-forming and less rotationally supported, in both SDSS and hydrodynamical simulations \citep{Kraljic2020}. The route of matter infall leaves detectable traces in cluster properties.

What happens to the galaxies and galaxy groups being accreted onto clusters is less well understood. Galaxies and groups are environmentally processed before they reach the virial radius \citep{Wetzel2013}, and 30--45\,per~cent of galaxies reach clusters by travelling along filaments \citep{Kuchner2022}, consistent with the anisotropic satellite accretion found in cosmological simulations \citep{Deason2011}. Filament galaxies, backsplash objects, and field infalls occupy distinct regions of projected phase space \citep{Kuchner2022, DarraghFord2023}, the filamentary structure feeding cluster outskirts and the velocity field of the material collapsing onto it have been characterised in zoom simulations of 324 clusters \citep{Rost2021}, and roughly half of a group's members become gravitationally unbound by first pericentre passage \citep{Haggar2023}. These studies establish which objects travel via filaments and how they survive the cluster encounter. Does the route of infall (along a cluster-feeding filament or from the open field) leave a systematic imprint on the orbital kinematics and angular momentum of the infalling group itself? This has not yet been measured.

The theoretical framework predicts that it should, and makes quantitative predictions on two fronts. First, in the Zel'dovich approximation, the velocity field and the filament skeleton are both expressions of the same primordial tidal deformation tensor \citep{Zeldovich1970, Bond1996}: groups flowing along a cluster-connected filament should move along the filament axis, unlike groups approaching from inter-filament voids. Second, constrained tidal torque theory \citep[cTTT, building on tidal torque theory;][]{White1984, Porciani2002, Codis2015} predicts that the mass at which a group's spin transitions from parallel to perpendicular alignment with the filament (the spin flip mass $\mflip$) is set by the local tidal environment. The flip arises because low-mass haloes acquire their spin within the plane of the collapsing wall (parallel to the filament they later join), whereas high-mass haloes form later by accreting matter along the filament and are torqued about an axis perpendicular to it; $\mflip$ marks this crossover and, because it scales with the local mass of non-linearity, is expected to rise in denser environments \citep{Codis2015}, as measured for thicker filaments \citep{GaneshaiahVeena2018, GaneshaiahVeena2021}.

A saddle point is a critical point of the density field where two filaments meet (a local density minimum along the filament axis but a maximum in the two transverse directions) and marks the densest, most anisotropic point between two nodes. Such saddle-point environments produce the strongest angular momentum imprint. The spin flip has been established in $N$-body and hydrodynamic simulations \citep{Codis2012, Dubois2014} and confirmed observationally using the SAMI Galaxy Survey \citep{Welker2020}, the morphology-dependent spin alignments of \citet{Tempel2013}, and galaxy orientation catalogues \citep{Trujillo2006}. These results are established for individual haloes and galaxies; here we ask whether the same transition holds for galaxy groups, more massive and later-collapsing objects whose spin is nonetheless expected to follow the same tidal-torque mechanism, making them a natural extension of the prediction.

The group-scale spin--filament relation has itself now been measured. The orbital angular momentum of group member galaxies lies preferentially perpendicular to the filament spine, both in SDSS \citep{Rong2025} and in the IllustrisTNG300-1 simulation \citep{WangW2025}. That quantity traces the halo's own spin only loosely, however. \citet{WangW2025} find a mean misalignment of ${\approx}38^\circ$ between the spin inferred from member galaxies and that of all dark-matter particles in the same haloes. Neither study finds a reversal to parallel alignment in any group mass or richness bin it probes. \citet{AragonCalvo2014} found evidence for a hierarchical $\mflip$ dependence on filament level in the general field, and \citet{Wang2017, WangKang2018} showed that the transition mass is set by whether haloes form before or after migrating from wall to filament environments (part of the general flow of matter outward from voids, into two-dimensional walls or sheets, then along one-dimensional filaments, and finally into cluster nodes), with the strength of the spin--structure correlation depending on the anisotropy of the local tidal field. The mass at which the \emph{halo} spin flips has not, however, been measured in the cluster-outskirt environment, nor mapped onto the linear density of cluster-feeding filaments.

Assembly bias, the dependence of halo clustering and internal properties on formation history at fixed mass \citep{Hahn2007, Gao2005, Wechsler2006, Dalal2008}, further predicts \citep[ZOMG;][]{Borzyszkowski2017, Musso2018} that inter-filament groups are tidally stalled (their mass growth arrested because no filament feeds them fresh material), so that they form earlier and are more concentrated than groups actively accreting along filaments, a prediction tested in only seven zoom simulations.

These predictions connect the geometry of the cosmic web to measurable group properties. Testing these predictions requires a sample far larger than the zoom simulations typically available for cluster-outskirt studies: one must bin simultaneously by group mass, cluster mass, and accretion channel while retaining enough objects per bin for reliable statistics. We meet this requirement using the \disperse filament catalogue of the full $(1\,h^{-1}\,\mathrm{Gpc})^3$ MultiDark Planck 2 (MDPL2) simulation \citep{paperI}, which provides 27,029 clusters with $M_\mathrm{vir} > 10^{14.0}\,h^{-1}\,\msun$ and ${\sim}2\times10^5$ group-cluster infall pairs, a statistical reach orders of magnitude beyond previous cluster-outskirt samples. Each infalling group is classified by its topological position in the cluster's filament network as primary, secondary, or isotropic (the classification algorithm is detailed in Section~\ref{sec:method}), so that we probe the filament's topological relationship to the cluster rather than mere proximity, distinguishing the dense bridges of the cosmic web from the thin tendrils that may cross the cluster shell without carrying a comparable tidal imprint.

Throughout, the analysis is confined to the infall regions around clusters. The probed volume lies within each cluster's gravitational sphere of influence; the cluster node is far more massive than the secondary density maxima feeding it (this is not a cluster--saddle--cluster configuration), and the bulk flow is directed toward the cluster. This infall-region setting is what licenses treating both primary and secondary filaments as ultimately feeding a single dominant cluster (Section~\ref{sec:dominant_cluster}); we are not characterising the cosmic web in general, but specifically the filamentary approaches to massive clusters.

Our aim is to establish whether these topological accretion channels leave measurable imprints on the kinematics, spin, and orbits of infalling groups and, going beyond a confirmation of existing theory, to identify which physical property of the host filament, its topological order or its density, is more closely tied to those imprints. Section~\ref{sec:data} describes the data. Section~\ref{sec:method} presents the classification and measurement framework. Section~\ref{sec:results} reports the results. Sections~\ref{sec:discussion} and \ref{sec:conclusions} discuss the physical interpretation and summarise the conclusions.


\section{Data}
\label{sec:data}

\subsection{The MDPL2 simulation and halo catalogue}
\label{sec:mdpl2}

We use the MultiDark Planck 2 (MDPL2) $N$-body simulation \citep{Klypin2016}, publicly available through the CosmoSim database \citep{Riebe2013}. MDPL2 follows $3840^3$ dark matter particles in a periodic box of side length $L = 1\,h^{-1}\,\mathrm{Gpc}$ from $z = 120$ to $z = 0$, adopting a flat $\Lambda$CDM cosmology consistent with \citet{Planck2014}: $\Omega_\mathrm{m} = 0.307$, $h = 0.678$, $\sigma_8 = 0.823$. Dark matter haloes are identified with the \rockstar phase-space halo finder \citep{Behroozi2013}. The full $z = 0$ catalogue contains 127,388,160 haloes spanning $3 \times 10^9$ (the \rockstar detection floor, ${\approx}2$ particles at the MDPL2 particle mass of $1.51 \times 10^9\,h^{-1}\,\msun$) to $3.5 \times 10^{15}\,h^{-1}\,\msun$. The filament construction and all analysis below use only the ${\geq}20$-particle sample ($M \gtrsim 3 \times 10^{10}\,h^{-1}\,\msun$; Section~\ref{sec:filaments}).

\subsection{The filament catalogue}
\label{sec:filaments}

Filaments are extracted with \disperse \citep[Discrete Persistent Structures Extractor;][]{Sousbie2011}, a widely-used cosmic-web finder \citep[one of the many cosmic-web classification methods compared by][]{Libeskind2018} whose algorithm is described in full in its original presentation \citep{Sousbie2011}; here we summarise only the elements needed to follow the classification. \disperse identifies filaments as topological features of the density field. It first constructs a Delaunay tessellation of the input halo positions and estimates the local density at each vertex via the Delaunay Tessellation Field Estimator \citep[DTFE;][]{Schaap2000}. A discrete Morse--Smale complex is then computed on this simplicial complex: simplices are paired along the density gradient, and unpaired simplices define \emph{critical points}, stationary points of the density field classified by their Morse index $\lambda$ (the number of negative eigenvalues of the Hessian). In three dimensions there are four types (Table~\ref{tab:critical_points}): minima ($\lambda = 0$; void centres), 1-saddles ($\lambda = 1$; wall junctions), 2-saddles ($\lambda = 2$; the points from which filaments emanate, where the density is a local minimum along the filament axis but a maximum in the two transverse directions), and maxima ($\lambda = 3$; cosmic web nodes corresponding to clusters and groups). Filaments are the integral lines of the density gradient that connect maxima to 2-saddles, the one-dimensional ascending manifold of the Morse--Smale complex. Each topological feature is assigned a \emph{persistence}, defined as the density contrast between the critical point pair that creates and destroys it. Features with persistence below a user-specified threshold $n_\sigma$ (in units of the expected Poisson noise) are cancelled, providing a single-parameter noise filter that retains only structures whose density contrast exceeds $n_\sigma$ times the noise level.

\begin{table}
    \centering
    \caption{Critical points identified by \disperse in the $(1\,h^{-1}\,\mathrm{Gpc})^3$ MDPL2 filament catalogue at $n_\sigma = 6.5$. Types 0--3 are Morse-theoretic critical points from the persistence-simplified skeleton; bifurcation points are introduced during post-processing (arc breaking, orphan removal, smoothing).}
    \label{tab:critical_points}
    \begin{tabular}{clcll}
        \hline
        $\lambda$ & Type & Hessian signature & Structure & Count \\
        \hline
        0 & Minimum      & $(+,+,+)$   & Void centre        & 9{,}478 \\
        1 & 1-saddle     & $(-,+,+)$   & Wall junction      & 229{,}381 \\
        2 & 2-saddle     & $(-,-,+)$   & Filament branch    & 363{,}420 \\
        3 & Maximum      & $(-,-,-)$   & Cluster/group node & 143{,}500 \\
        -- & Bifurcation  & --          & Break/junction     & 466{,}969 \\
        \hline
    \end{tabular}
\end{table}

The filament catalogue used in this work is taken from Paper~I \citep{paperI}, where the full construction and validation are described. In that work, we applied \disperse to the MDPL2 halo catalogue after imposing a mass cut requiring at least 20 particles per halo, resulting in a sample of ${\sim}92$ million haloes. Constructing a filament catalogue from such a large volume is computationally demanding. \disperse builds a three-dimensional Delaunay tessellation of the halo distribution, and performing this tessellation on the full MDPL2 catalogue at once would require ${\sim}600$--$1{,}300$\,GB of memory, far beyond the capacity of standard compute nodes.

Paper~I overcomes this bottleneck with a \emph{frozen-core method}. The simulation volume is decomposed into 1{,}000 overlapping tiles, each a $100\,h^{-1}$\,Mpc core region surrounded by a $60\,h^{-1}$\,Mpc padding region. Each tile is tessellated independently within bounded memory, and a circumsphere-based filter then retains only those tetrahedra guaranteed to belong to the global Delaunay tessellation. The tiled \disperse outputs are merged by a post-processing pipeline (core filtering, spatial deduplication, boundary stitching; single parameter $\delta = 0.5\,h^{-1}$\,Mpc), followed by a five-step topological cleaning pipeline: (i) removal of zero-length arcs ($L < 0.01\,h^{-1}$\,Mpc); (ii) removal of degree-1 nodes (filament endpoints with only a single attached arc) and that arc; (iii) removal of filaments shorter than $1\,h^{-1}$\,Mpc; (iv) merging of nearly collinear arcs at shared saddle points (angle threshold $60^\circ$, endpoint tolerance $0.5\,h^{-1}$\,Mpc); and (v) iterative smoothing (10 iterations) of interior sampling points to suppress discretisation noise from the Delaunay tessellation grid. Validation against a monolithic reference on a $300\,h^{-1}$\,Mpc subvolume shows 99.6 per cent total length recovery, 100 per cent recovery of density maxima and minima, and 94.7 per cent individual filament matching.

We adopt the persistence threshold $n_\sigma = 6.5$, which retains high-persistence filamentary ridges while preserving the secondary tributary network needed for the BFS classification. The resulting catalogue contains 824,460 filaments with a median length of $10.0\,h^{-1}$\,Mpc. Each filament is stored as an ordered sequence of three-dimensional sampling points.

\subsection{Group and cluster catalogues}
\label{sec:catalogues}

Clusters. We select all host haloes (i.e.\ not subhaloes of a larger system) with $M_\mathrm{vir} > 10^{14.0}\,h^{-1}\,\msun$, yielding 27,029 clusters. For each cluster we adopt as the reference radius $R_{200}$ the spherical-overdensity radius enclosing $200$ times the critical density, computed from the \rockstar $M_{200c}$ (we write $R_{200}$ throughout for this $r_{200c}$-equivalent radius):
\begin{equation}
    R_{200} = \left(\frac{3\,M_{200c}}{4\pi \cdot 200\,\rho_\mathrm{crit}}\right)^{1/3},
    \label{eq:r200}
\end{equation}
where $\rho_\mathrm{crit} = 2.775 \times 10^{11}\,h^2\,\msun\,(\mathrm{Mpc\,h^{-1}})^{-3}$. Clusters are selected on the virial mass $M_\mathrm{vir}$, defined using the redshift-dependent virial overdensity, which for the MDPL2 cosmology at $z = 0$ is $\Delta_\mathrm{vir} \approx 337$ times the mean matter density of the Universe (equivalently ${\approx}102$ times the critical density) rather than the traditional fixed overdensity of $200$ times the critical density. The reference radius $R_{200}$ used to normalise all radial profiles and classification thresholds is the $M_{200c}$-based value above, matching the convention of \citet{DarraghFord2023} and \citet{Rost2024}. Cluster $R_{200}$ values range from $0.37$ to $2.27\,h^{-1}$\,Mpc. The $M_\mathrm{vir} > 10^{14.0}\,h^{-1}\,\msun$ threshold reaches down to the massive-group / poor-cluster regime. For brevity we refer to all haloes above this cut as ``clusters'' throughout, but the sample is more accurately described as massive nodes of the cosmic web spanning the rich-group to cluster mass range.

Groups. We select host haloes with $10^{12.5} < M_\mathrm{vir} < 10^{14}\,h^{-1}\,\msun$, giving 1,386,794 groups. Throughout, a group is an isolated host halo (not a subhalo of a more massive system), selected purely on mass in the range above with no galaxy-occupation or richness information used. Each hosts a distinct dark-matter potential well and is the simulation analogue of a galaxy group in that sense; the correspondence to an observer's richness- or luminosity-selected group is statistical rather than object-by-object. The upper mass bound distinguishes group from cluster by convention rather than as a hard selection: because each cluster's infall region is populated from the full host-halo catalogue, cluster-mass haloes lying within the infall zone of a more massive neighbour formally enter the sample. Self-associations (a cluster relative to itself) are removed from the kinematic sample by the dominant-cluster and backsplash cuts (Section~\ref{sec:dominant_cluster}). The spin-alignment analysis, which uses all classified pairs, retains them, but all have $\log_{10} M_\mathrm{vir} \geq 14.0$, above the range $\log_{10} M_\mathrm{vir} < 14.0$ over which the spin-flip statistics are measured (Section~\ref{sec:spin_flip}). They and the cluster-mass companions (which supply $827$ of the $860$ isotropic pairs in that bin) affect only the $14.0$--$14.5$ mass bin, which we retain only to show the near-total exclusion of channelled labels above the group-mass cut. Essentially every surviving object there is primary-outskirt or isotropic (Section~\ref{sec:mass_dependence}). In addition to position and velocity, we use the following \rockstar properties: virial mass $M_\mathrm{vir}$ and radius $r_\mathrm{vir}$; NFW concentration $c = r_\mathrm{vir}/r_s$; Bullock spin parameter $\lambda$; shape axis ratios $b/a$ and $c/a$; the position offset $X_\mathrm{off} = |\mathbf{r}_\mathrm{com} - \mathbf{r}_\mathrm{peak}|/r_\mathrm{vir}$; virial ratio $T/|U|$; and angular momentum vector $\mathbf{J} = (J_x, J_y, J_z)$.

Infall selection. We consider all groups within $7\,R_{200}$ of any cluster as potentially infalling. This aperture is set by the physical extent of a cluster's dynamical influence rather than by an arbitrary metric distance. The boundary of the region a cluster gravitationally controls is its turnaround, or zero-velocity, radius $r_\mathrm{ta}$: the surface where infalling matter decouples from the Hubble flow and inside which the cluster potential governs the dynamics. Because the turnaround radius encloses a fixed mean interior overdensity ($\delta \approx 11$ relative to the background; \citealt{Korkidis2020}), the ratio $r_\mathrm{ta}/R_{200}$ is nearly invariant with both cluster mass and redshift. For haloes in precisely our regime ($M_{200c} > 10^{14}\,\msun$, $z \lesssim 1$) it is measured from mean radial-velocity profiles to be $r_\mathrm{ta} \approx 4.8\,R_{200}$, constant to a few per cent across mass and out to $z \approx 1$ \citep{Pizzardo2024}. Our $7\,R_{200}$ selection therefore lies $\approx 1.5\times$ beyond the turnaround radius.

This deliberately conservative envelope encloses the entire infall region with margin; the peak-infall (``depletion'') scale of that region is only $\approx 2$--$3\,R_{200}$ \citep{Pizzardo2024}. The envelope is also chosen so that secondary (order-2) connected groups, whose host filaments by definition lie beyond $3\,R_{200}$, are captured on first approach before entering the primary filament network. It extends well beyond the splashback radius \citep[${\sim}1$--$1.5\,R_{200}$;][]{More2015}.

\begin{figure*}
    \centering
    \includegraphics[width=\textwidth]{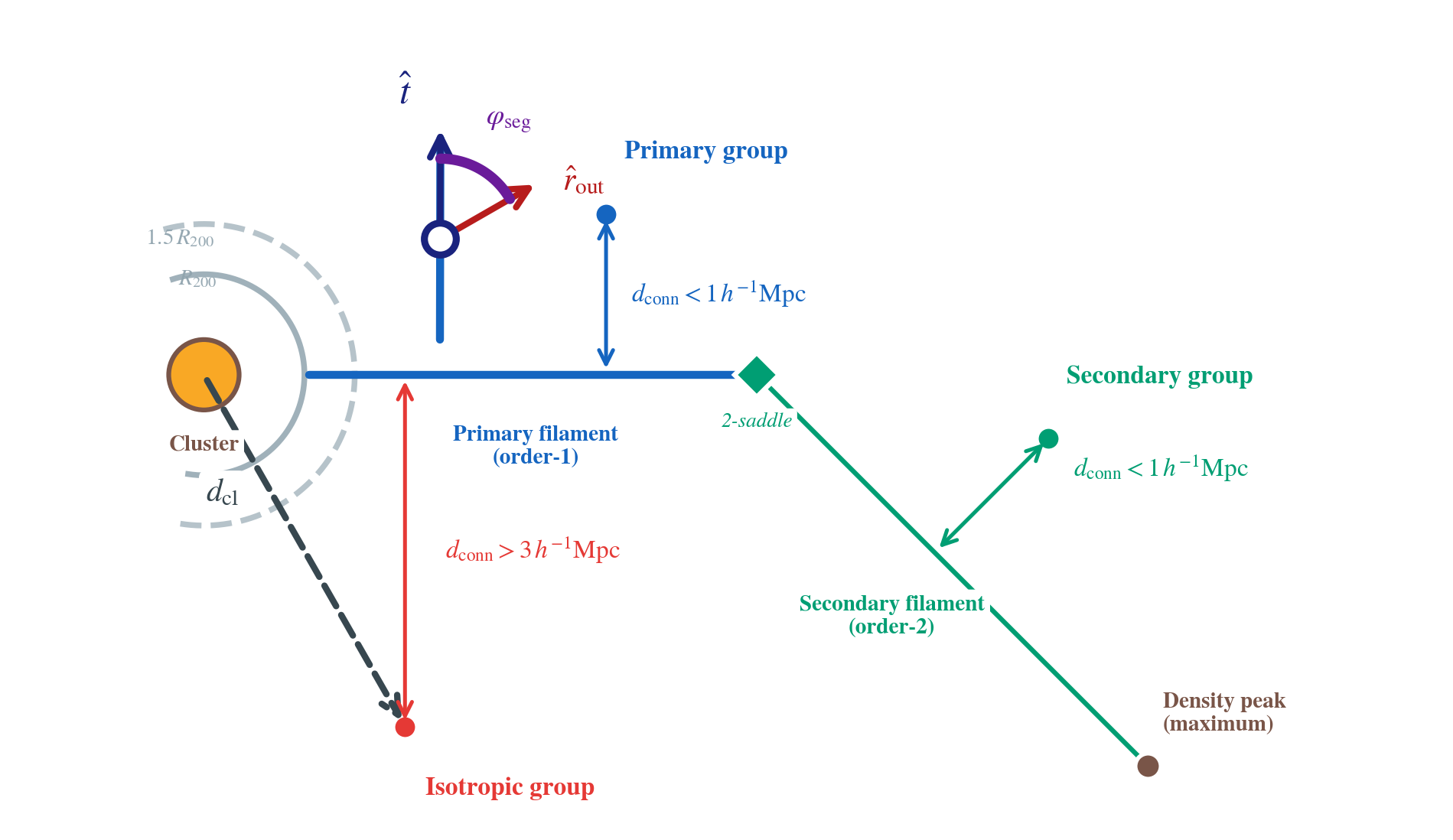}
    \caption{Geometric illustration of the distance metrics and Morse-theoretic topology used for group classification. The primary filament (order-1) connects the cluster maximum to a 2-saddle critical point; a secondary filament (order-2) connects a separate density peak (maximum) to the same 2-saddle. In Morse theory, exactly two filaments meet at each 2-saddle, each originating from a different maximum, and the material beyond the 2-saddle is conventionally assigned to the other maximum. We treat the secondary filament as connected to the cluster only because of the infall-region setting (Section~\ref{sec:intro}): the probed volume lies entirely within the cluster's gravitational sphere of influence, the cluster node is far more massive than the secondary maximum (this is not a cluster--saddle--cluster configuration), and the bulk flow is toward the cluster (Section~\ref{sec:dominant_cluster}). The `primary'/`secondary' labels therefore indicate each filament's topological relationship to the cluster (Section~\ref{sec:connectivity_network}), not an intrinsic hierarchy. Symbols are defined in Table~\ref{tab:symbols}.}
    \label{fig:schematic_topo}
\end{figure*}

\begin{figure*}
    \centering
    \includegraphics[width=\textwidth]{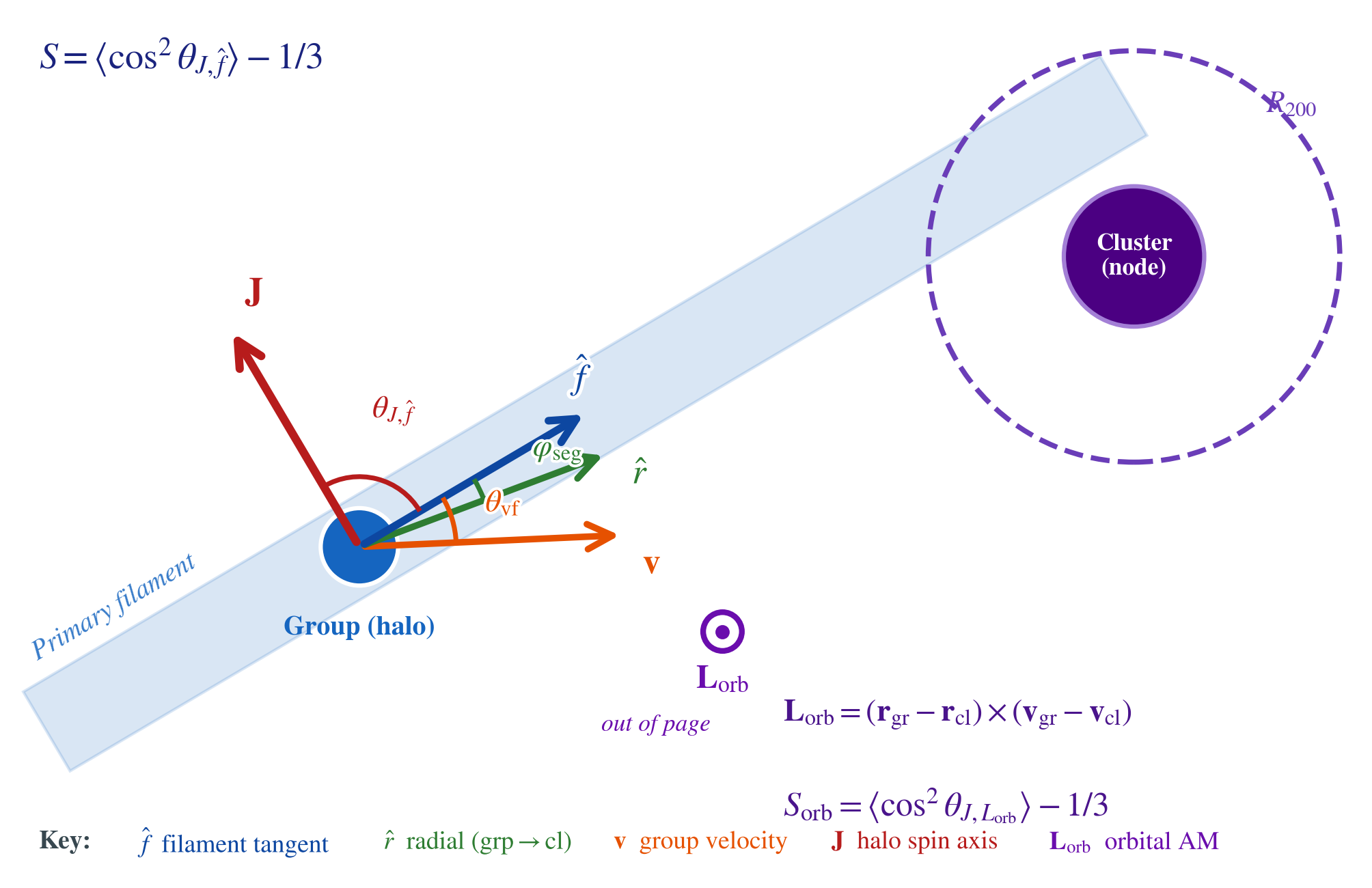}
    \caption{Schematic of the spin-filament alignment geometry for an infalling group (symbols defined in Table~\ref{tab:symbols}). Because $\hat{f}$ and $\hat{r}$ are approximately co-planar (primary filaments inclined at $\approx 15^\circ$), the orbital angular momentum $\mathbf{L}_\mathrm{orb} = (\mathbf{r}_\mathrm{gr} - \mathbf{r}_\mathrm{cl}) \times (\mathbf{v}_\mathrm{gr} - \mathbf{v}_\mathrm{cl}) \propto \hat{r} \times \hat{f}$ lies perpendicular to the filament axis, making the $\mathbf{J}$-vs-$\hat{f}$ and $\mathbf{J}$-vs-$\mathbf{L}_\mathrm{orb}$ signals geometric complements.}
    \label{fig:schematic_spin}
\end{figure*}

The dominant-cluster and kinematic steps (Section~\ref{sec:dominant_cluster}) then retain only groups falling toward the cluster; and because $r_\mathrm{ta}/R_{200}$ is nearly mass- and redshift-independent while the filament geometry and spin-flip signals are self-similar across $1.5$--$7\,R_{200}$ (Figs~\ref{fig:filament_inclination} and \ref{fig:radial_spatial}) and the classification is robust to the threshold choices (Appendix~\ref{sec:appendix_sensitivity}), no conclusion depends on the exact outer boundary. The distance metrics ($d_\mathrm{cl}$, $d_\mathrm{conn}$) and the spin-alignment geometry used in the classification and analysis that follow are illustrated in Figs~\ref{fig:schematic_topo} and~\ref{fig:schematic_spin}.

\section{Method}
\label{sec:method}

We first post-process the \disperse\ skeleton to remove spurious cluster-internal arcs (Section~\ref{sec:masking}) and build the cluster-connected filament network via a breadth-first search (Section~\ref{sec:connectivity_network}). We then measure the transverse filament density profile to place the distance thresholds on a physical footing (Section~\ref{sec:radial_profile}), and use these thresholds to classify every group-cluster pair into primary/secondary connected, primary-outskirt, or isotropic accretion channels (Section~\ref{sec:classification}). Finally, we define the kinematic (Section~\ref{sec:kinematics}) and spin-alignment (Section~\ref{sec:spin_method}) measurements applied to each class. Table~\ref{tab:symbols} collects the symbols and measured quantities used throughout.

\begin{table*}
  \centering
  \caption{Symbols and measured quantities used throughout the paper, with the section in which each is defined. Distances and radii are in $h^{-1}$\,Mpc and masses in $h^{-1}\,\msun$ unless stated otherwise.}
  \label{tab:symbols}
  \small
  \setlength{\tabcolsep}{4pt}
  \begin{minipage}[t]{0.48\textwidth}
  \vspace{0pt}
  \begin{tabular}{@{}l>{\raggedright\arraybackslash}p{0.36\textwidth}@{}}
    \hline
    Symbol & Definition \\
    \hline
    \multicolumn{2}{@{}l}{\emph{Group / halo properties (\textup{\rockstar}; Section~\ref{sec:mdpl2})}}\\
    $M_\mathrm{vir}$ & Group virial mass \\
    $r_\mathrm{vir}$ & Group virial radius \\
    $c$ & NFW concentration, $r_\mathrm{vir}/r_s$ \\
    $\lambda$ & Bullock spin parameter \\
    $b/a,\,c/a$ & Minor-, intermediate-to-major axis ratios \\
    $\mathbf{A}$ & Halo major (shape) axis \\
    $\mathbf{J}$ & Group angular-momentum vector \\
    $X_\mathrm{off}$ & CoM offset, $|\mathbf{r}_\mathrm{com}-\mathbf{r}_\mathrm{peak}|/r_\mathrm{vir}$ \\
    $T/|U|$ & Virial ratio \\
    $\dot{M}$ & Instantaneous mass accretion rate \\[2pt]
    \multicolumn{2}{@{}l}{\emph{Cluster \& filament network (Sections~\ref{sec:catalogues}--\ref{sec:radial_profile})}}\\
    $M_{200c}$ & Cluster mass within $r_{200c}$ \\
    $R_{200}\!\equiv\! r_{200c}$ & Cluster radius ($200\times$ critical, from $M_{200c}$) \\
    $\kappa$ & Connectivity: filaments crossing $1.5R_{200}$ \\
    $n_\sigma$ & \disperse\ persistence threshold \\
    $\epsilon$ & BFS endpoint proximity tolerance \\
    $L$ & Filament arc-length \\
    $s$ & Arc-length fraction ($0=$ cluster, $1=$ saddle) \\
    $\hat{f}$ & Unit tangent of nearest connected filament \\
    $\hat{t}$ & Filament segment unit tangent \\
    $\rho_\mathrm{lin}$ & Linear halo mass density, $\Sigma M_\mathrm{vir}/L$ (host-halo mass within $2\,h^{-1}$\,Mpc per unit $L$) \\
    $R_\mathrm{c},\,\alpha$ & $\beta$-model core radius, steepness exponent \\
    $R_\mathrm{drop}$ & Radius where $\beta$-model $=$ background \\
    $\rho_0/\rho_\mathrm{bg}$ & $\beta$-model central density contrast \\
    \hline
  \end{tabular}
  \end{minipage}\hfill
  \begin{minipage}[t]{0.48\textwidth}
  \vspace{0pt}
  \begin{tabular}{@{}l>{\raggedright\arraybackslash}p{0.36\textwidth}@{}}
    \hline
    Symbol & Definition \\
    \hline
    \multicolumn{2}{@{}l}{\emph{Geometry \& distances (Sections~\ref{sec:catalogues}--\ref{sec:classification})}}\\
    $d_\mathrm{cl}$ & Group--cluster separation \\
    $d_\mathrm{conn}$ & Perp.\ distance to nearest connected filament \\
    $\hat{r}$ & Radial unit vector, from cluster to group \\
    $\varphi_\mathrm{seg}$ & Filament inclination, $\arccos|\hat{t}\cdot\hat{r}|$ \\
    $r_\mathrm{ta}$ & Cluster turnaround (zero-velocity) radius \\
    $r_\mathrm{peri}$ & Predicted orbital pericentre \\[2pt]
    \multicolumn{2}{@{}l}{\emph{Kinematics (Section~\ref{sec:kinematics})}}\\
    $v_r$ & Radial velocity ($<0$: infall) \\
    $v_t$ & Tangential velocity \\
    $\beta$ & Velocity anisotropy, $1-\sigma_t^2/2\sigma_r^2$ \\
    $\theta_{vf}$ & Velocity--connected-filament angle \\
    $\theta_\mathrm{fil}$ & Velocity--nearest-filament angle (any type) \\
    $\theta_{vr}$ & Velocity--radial angle, from the infall direction $-\hat{r}$ \\
    $\mathbf{L}_\mathrm{orb}$ & Orbital angular momentum \\[2pt]
    \multicolumn{2}{@{}l}{\emph{Alignment statistics (Section~\ref{sec:spin_method})}}\\
    $\theta$ & Angle between $\mathbf{J}$ and $\hat{f}$ \\
    $S$ & Spin--filament, $\langle\cos^2\theta_{J,\hat f}\rangle-\tfrac{1}{3}$ \\
    $S_\mathrm{shape}$ & Shape--filament, $\langle\cos^2\theta_{A,\hat f}\rangle-\tfrac{1}{3}$ \\
    $S_\mathrm{orb}$ & Spin--orbit, $\langle\cos^2\theta_{J,L_\mathrm{orb}}\rangle-\tfrac{1}{3}$ \\
    $\mflip$ & Spin flip mass ($S$ zero-crossing) \\
    $\delta$ & Cliff's effect size \\
    \hline
  \end{tabular}
  \end{minipage}
\end{table*}

\subsection{Filament cleaning near cluster centres}
\label{sec:masking}

\disperse\ constructs the topological skeleton of the density field, and its output near massive clusters requires careful post-processing. Within the virial radius of a cluster, the halo distribution is dense and irregular, causing \disperse\ to generate a large number of short arcs connecting multiple saddle and maximum critical points arising from the cluster's internal substructure. These arcs are real topological features of the local density field, but they do not represent the large-scale filamentary channels feeding the cluster from the cosmic web. Classifying infalling groups as connected based on proximity to such cluster-internal arcs would contaminate the sample with groups orbiting within the cluster potential rather than approaching along a large-scale filament.

This is a recognised problem in the literature. \citet{Rost2024} address it before skeleton extraction by excising all dark matter particles within $2\,R_{200}$ of haloes above $10^{11}\,h^{-1}\,\msun$ from the input density field prior to running \disperse, so that the cluster density peaks are absent from the skeleton entirely. \citet{Bahe2025} apply a post-extraction mask: any filament sampling point lying within $3\,r_{200c}$ of a group- or cluster-mass halo ($M_{200c} > 10^{13}\,\msun$) is flagged as cluster-internal, and all filament segments connected to those points are removed from the skeleton. \citet{DarraghFord2023} adopt a complementary strategy at the analysis stage, restricting their group sample to $1.5 < r/R_{200} < 5.5$ to exclude the complex cluster-core region entirely without any modification to the skeleton.

We adopt a combination of the second and third approaches. First (analysis exclusion, following \citealt{DarraghFord2023}): our connectivity measurement uses filaments crossing the $1.5\,R_{200}$ shell, and our analysis is restricted to infalling groups at $r > 1.5\,R_{200}$. This inner boundary is physically motivated: $1.5\,R_{200}$ approximates the cluster splashback radius, the outermost apocentre of recently accreted material and hence the physical edge of the collapsed halo \citep[${\sim}1$--$1.5\,R_{200}$;][]{More2015, Pizzardo2024}. Groups inside it have already orbited through the cluster (members or backsplash), whereas those beyond it are on first infall; this scale is also the standard aperture for measuring cluster connectivity \citep{Codis2018} and the infall-sample boundary adopted by \citet{DarraghFord2023}.

Second (skeleton masking, following \citealt{Bahe2025} and \citealt{Rost2024}): we apply a post-extraction segment mask to the filament skeleton before computing group-filament distances. All filament sampling points within $2\,R_{200}$ of any cluster with $M_\mathrm{vir} > 10^{14.0}\,h^{-1}\,\msun$ are marked as cluster-internal; the filament is split at those points, and sub-filaments of length less than $1\,h^{-1}$\,Mpc are discarded. The outer portions are retained and used to form the \textit{masked filament catalogue}. We use this masked catalogue exclusively for computing group-filament distances (Section~\ref{sec:classification}). The original unmasked catalogue from Paper~I is retained for the connectivity measurement (Section~\ref{sec:connectivity_network}), where filaments crossing the cluster shell must be counted. The masking radius of $2\,R_{200}$ is chosen to be smaller than the $3\,r_{200c}$ of \citet{Bahe2025} to avoid removing the inner portion of large-scale filaments that are well resolved in our 1\,Gpc volume; sensitivity to this choice is tested in Appendix~\ref{sec:appendix_sensitivity}.

\begin{figure}
    \centering
    \includegraphics[width=\columnwidth]{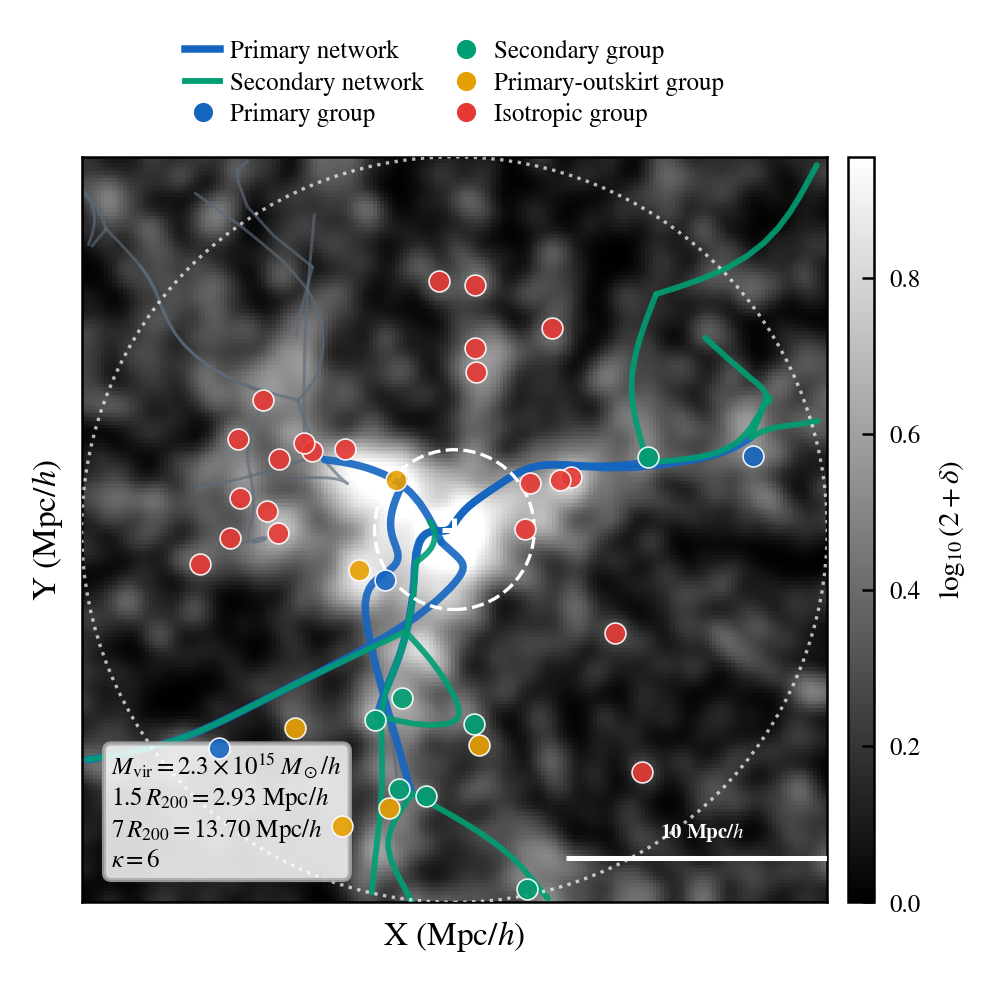}
    \caption{The $XY$ projection of a massive cluster ($M_\mathrm{vir} = 2.3 \times 10^{15}\,h^{-1}\,\msun$, $\kappa = 6$), showing the analysed infall sample (dominant-cluster, non-backsplash pairs) within a $\pm 10\,h^{-1}$\,Mpc slab in projected depth. Dark blue filaments and points: primary (order-1) filament network and primary connected groups. Green filaments and points: secondary (order-2) network and secondary connected groups. Orange points: primary-outskirt (intermediate) groups ($1 \leq d_\mathrm{conn} \leq 3\,h^{-1}$\,Mpc). Red points: isotropic groups. Faint blue-grey lines: the post-processed \disperse\ filament catalogue (after the five-step cleaning of Section~\ref{sec:filaments}), with this cluster's own order-1 and order-2 filaments drawn in colour instead. The dashed circle marks $1.5\,R_{200}$ (the connectivity shell); the dotted circle marks $7\,R_{200}$ (the outer infall/analysis boundary). Background: local halo number density.}
    \label{fig:illustration}
\end{figure}

Figure~\ref{fig:illustration} previews the classification developed in the rest of this section (formally defined in Section~\ref{sec:classification}) for a representative massive cluster ($M_\mathrm{vir} = 2.3 \times 10^{15}\,h^{-1}\,\msun$, connectivity $\kappa = 6$), shown here in the $XY$ projection. Groups on the primary filament spines that cross the $1.5\,R_{200}$ shell (dark blue points, on the dark blue primary network) and on the order-2 network (green) are visually distinct from isotropic field groups (red), which show no preferential alignment with the filament network.

\subsection{Cluster-connected filament network}
\label{sec:connectivity_network}

For each cluster we identify the set of filaments from the unmasked catalogue that physically cross the $1.5\,R_{200}$ shell, following the methodology of \citet{Santoni2024} and Paper~I. A filament is connected if at least one pair of consecutive sampling points straddles the shell. The number of such filaments defines the cluster connectivity $\kappa$.

We extend the connected set outward by tracing the filament network through shared saddle points using a breadth-first search (BFS), a standard graph-traversal algorithm that, starting from a set of seed nodes, visits all directly-linked neighbours, then their neighbours, and so on until no new node can be reached. Here the nodes are filaments and two filaments are linked if their endpoints lie close together. Starting from the outer endpoints of the shell-crossing filaments (endpoints lying outside the $1.5\,R_{200}$ shell and within $7\,R_{200}$ of the cluster centre), we iteratively add filaments whose endpoints lie within a proximity tolerance $\epsilon = 0.5\,h^{-1}$\,Mpc of any endpoint in the current set, stopping when no new filament within $7\,R_{200}$ can be added. Outer endpoints of shell-crossing filaments that lie beyond $7\,R_{200}$ are excluded from the initial BFS seed to prevent secondary filaments from being associated to the cluster solely through the far arms of long primary filaments. This breadth-first-search (BFS) expansion recovers the full tributary network feeding the cluster: filaments that join the primary cluster arms at saddle points in the inter-cluster density field.

The BFS-connected filaments are further classified by topological order:
\begin{itemize}
    \item Order-1 (primary) filaments: the filament physically crosses the $1.5\,R_{200}$ shell, i.e.\ it belongs to the shell-crossing set measured in Paper~I. These filaments have at least one endpoint inside $1.5\,R_{200}$ and are the direct cluster-feeding arms. In any 2-D projection they are seen to originate from within the cluster shell.
    \item Order-2 (secondary) filaments: the filament is BFS-reachable from the primary network via shared saddle points but does not itself cross the $1.5\,R_{200}$ shell. These filaments connect to the primary network at inter-cluster saddle points without directly entering the cluster environment; whether they dynamically feed the cluster is established later, at the dominant-cluster assignment (Section~\ref{sec:dominant_cluster}) and the kinematic analysis, rather than assumed here.
\end{itemize}

\subsection{Filament transverse density profile}
\label{sec:radial_profile}

\begin{figure*}
    \centering
    \includegraphics[width=\textwidth]{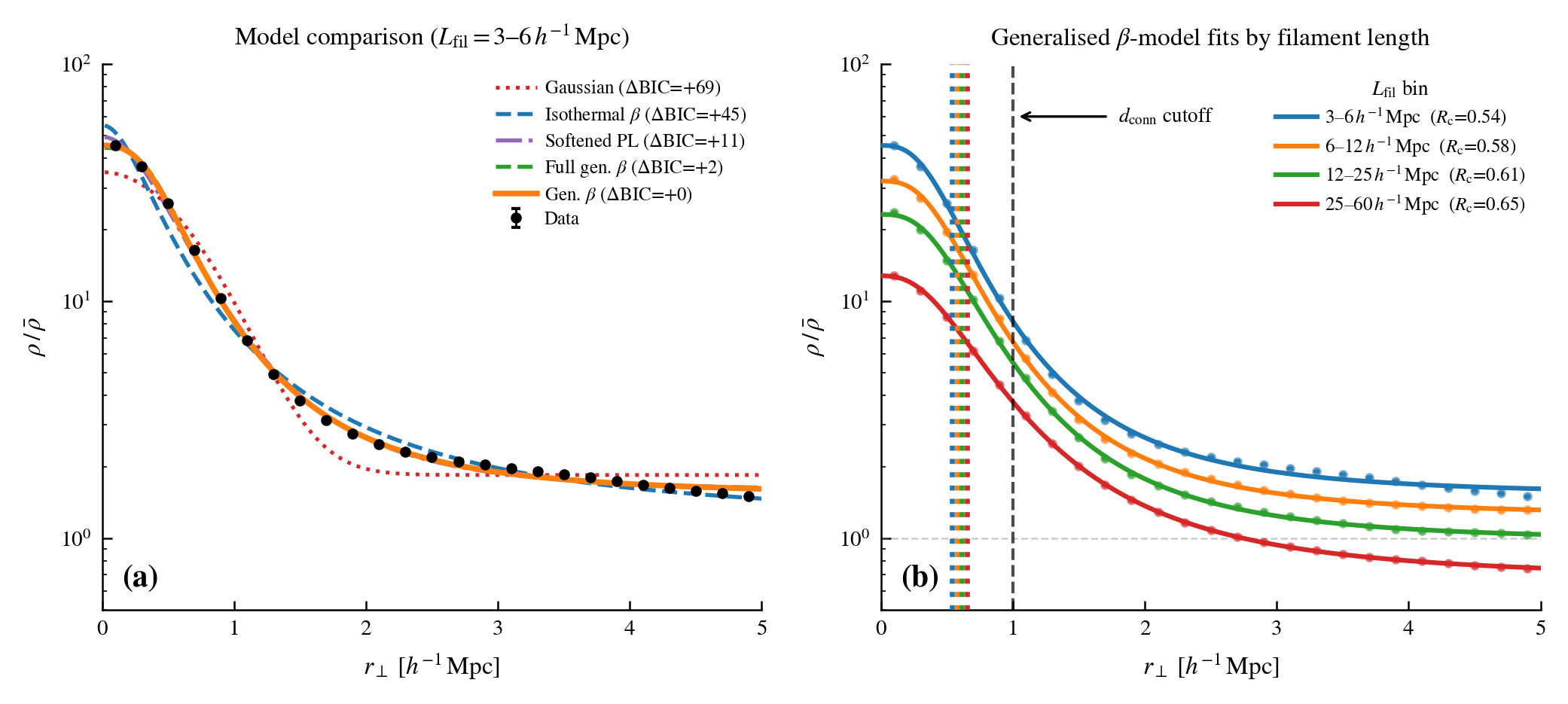}
    \caption{Stacked transverse dark matter density profiles of MDPL2 filaments ($n_\sigma = 6.5$, $2048^3$ CIC grid). \textit{(a)} Model comparison for short filaments ($L_\mathrm{fil} = 3$--$6\,h^{-1}$\,Mpc), where model disparity is largest. The generalised $\beta$-model (orange) is preferred by BIC over all alternatives. Data points with error bars show azimuthally averaged stacked profiles (2000 filaments). \textit{(b)} Generalised $\beta$-model fits for four filament-length bins. Dotted vertical lines mark $R_\mathrm{c}$ for each bin; the dashed black line at $r_\perp = 1\,h^{-1}$\,Mpc marks the $d_\mathrm{conn}$ classification threshold. $R_\mathrm{c}$ is stable at $0.54$--$0.65\,h^{-1}$\,Mpc across all lengths.}
    \label{fig:radial_profile}
\end{figure*}

To place the classification thresholds on a physical footing we measure the stacked transverse dark matter density profile of MDPL2 filaments. The full $3840^3$-particle distribution is deposited onto a $2048^3$ cloud-in-cell (CIC) grid ($\Delta x = 0.49\,h^{-1}$\,Mpc), and the overdensity $\rho/\bar{\rho}$ is sampled at 16 azimuthal angles around each filament segment at perpendicular distances $r_\perp = 0$--$5\,h^{-1}$\,Mpc. Profiles are stacked in four filament-length bins ($3$--$6$, $6$--$12$, $12$--$25$, and $25$--$60\,h^{-1}$\,Mpc; 2000 filaments per bin) and fitted with five models: an isothermal $\beta$-model ($\rho_0/(1 + (r/R_\mathrm{c})^2) + \rho_\mathrm{bg}$), a generalised $\beta$-model with free slope $\alpha$, a fully generalised $\beta$-model with an additional outer exponent, a Gaussian, and a softened power law. Bayesian Information Criterion (BIC) model selection (which ranks models by goodness of fit while penalising the number of free parameters, lower being preferred) (Fig.~\ref{fig:radial_profile}a) favours the generalised $\beta$-model,
\begin{equation}
    \frac{\rho(r_\perp)}{\bar{\rho}} = \frac{\rho_0}{1 + (r_\perp / R_\mathrm{c})^{\alpha}} + \rho_\mathrm{bg},
    \label{eq:genbeta}
\end{equation}
with $\Delta\mathrm{BIC} \geq +2$ for all alternatives ($+2.1$ for the fully generalised model, $+10.8$ for the softened power law, $+44.8$ for the standard $\beta$-model, $+68.6$ for the Gaussian; evaluated in the $3$--$6\,h^{-1}$\,Mpc bin where model disparity is largest).

Table~\ref{tab:radial_profile} lists the best-fit parameters. The core radius $R_\mathrm{c}$ is stable across lengths, rising only from $0.54 \pm 0.02$ to $0.65 \pm 0.004\,h^{-1}$\,Mpc ($0.80$--$0.96$\,Mpc in physical units for $h = 0.678$). The steepness parameter $\alpha \approx 2.5$--$2.8$ produces a sharper-than-Lorentzian roll-off: the profile drops to the background at the \emph{drop radius} $R_\mathrm{drop} \approx 1.8$--$2.0\,h^{-1}$\,Mpc, defined as the radius at which the fitted $\beta$-model density equals $\rho_\mathrm{bg}$ ($R_\mathrm{drop} = R_\mathrm{c}\,(\rho_0/\rho_\mathrm{bg} - 1)^{1/\alpha}$; Fig.~\ref{fig:radial_profile}b). The central contrast $\rho_0/\rho_\mathrm{bg}$ decreases from ${\sim}29$ for short filaments to ${\sim}18$ for long ones, consistent with shorter filaments preferentially tracing denser environments between closely spaced clusters.

These values are consistent with the literature: \citet{Yang2025} report a median filament width of ${\sim}1.0$--$1.5$\,Mpc at $z=0$ (defined as the radius where the dark matter density drops to the cosmic mean), \citet{Wang2024}, locating the filament boundary at the minimum of the logarithmic slope of the stacked galaxy density profile, find radii of $0.81$\,Mpc (SDSS) and $0.98$\,Mpc (MillenniumTNG) at $z = 0$, comparable in scale to our core radius though differently defined, and \citet{Cautun2014} find filament diameters of $2$--$6\,h^{-1}$\,Mpc. Our $R_\mathrm{drop} \approx 1.8$--$2.0\,h^{-1}$\,Mpc is a radius (diameter $3.6$--$4.0\,h^{-1}$\,Mpc), placing our cluster-connected filaments at the upper end of the \citet{Yang2025} radius range and within the \citet{Cautun2014} $2$--$6\,h^{-1}$\,Mpc diameter range, as expected for the preferentially thickest structures in the cosmic web.

Filament width is not constant along a filament. \citet{Wang2024} find that filaments thicken by ${\approx}30$ per cent between 3 and 1\,Mpc from the cluster they connect to, with a hint (within the errors) of a stronger increase for more massive clusters. We return to this point when measuring filament linear density (Section~\ref{sec:spinflip_quality}). The self-similar density profiles described by a generalised $\beta$-model are also in agreement with \citet{Zhu2021} and \citet{GalarragaEspinosa2020}, who find that the same functional family fits both dark matter and galaxy density profiles across a range of filament populations, as well as with the filament-width characterisation of our own MDPL2 catalogue in Paper~I \citep{paperI}.

These measurements motivate the classification thresholds adopted in Section~\ref{sec:classification}: $d_\mathrm{conn} = 1\,h^{-1}$\,Mpc selects groups within the dense filament core ($r_\perp \lesssim 2\,R_\mathrm{c}$), while $d_\mathrm{conn} = 3\,h^{-1}$\,Mpc lies beyond $R_\mathrm{drop}$ where the profile has converged to the cosmic mean. The operative filament thickness used throughout is the $1\,h^{-1}$\,Mpc dense core. The $3\,h^{-1}$\,Mpc value is not a filament radius (a $3\,h^{-1}$\,Mpc-radius filament would be implausibly thick relative to a cluster) but an outer exclusion boundary beyond which a group is treated as unaffiliated (isotropic). The vertical dashed line in Fig.~\ref{fig:radial_profile}(b) marks the $d_\mathrm{conn} = 1\,h^{-1}$\,Mpc threshold.

\begin{table}
    \centering
    \caption{Best-fit generalised $\beta$-model parameters (Eq.~\ref{eq:genbeta}) for the stacked transverse dark matter density profile, in four filament-length bins. $R_\mathrm{c}$: core radius. $\alpha$: steepness exponent. $\rho_0/\rho_\mathrm{bg}$: central contrast. $R_\mathrm{drop}$: radius at which the model equals the background, $R_\mathrm{drop} = R_\mathrm{c}\,(\rho_0/\rho_\mathrm{bg} - 1)^{1/\alpha}$. All lengths are in $h^{-1}$\,Mpc.}
    \label{tab:radial_profile}
    \begin{tabular}{lcccc}
        \hline
        $L_\mathrm{fil}$ & $R_\mathrm{c}$ & $\alpha$ & $\rho_0/\rho_\mathrm{bg}$ & $R_\mathrm{drop}$ \\
        \hline
        $3$--$6$   & $0.54 \pm 0.02$ & $2.76 \pm 0.07$ & $28.8$ & $1.79$ \\
        $6$--$12$  & $0.58 \pm 0.01$ & $2.81 \pm 0.03$ & $24.8$ & $1.80$ \\
        $12$--$25$ & $0.61 \pm 0.01$ & $2.77 \pm 0.03$ & $22.9$ & $1.87$ \\
        $25$--$60$ & $0.65 \pm 0.004$ & $2.49 \pm 0.01$ & $18.0$ & $2.03$ \\
        \hline
    \end{tabular}
\end{table}

\subsection{Accretion channel classification}
\label{sec:classification}

The classification pipeline proceeds in four steps: (1) assignment of each group to a dominant cluster when it falls in the infall zone of multiple clusters; (2) computation of the perpendicular distance to the nearest cluster-connected filament and assignment of a topological class; (3) identification of backsplash groups via the radial velocity--clustercentric distance phase space; and (4) application of a strict kinematic criterion that requires the group velocity to be directed toward the cluster and aligned with the filament axis. Here a \emph{group-cluster pair} denotes one group associated with one cluster (not a single object tracked across two snapshots); a group near several clusters generates several pairs until the dominant-cluster step resolves it.

\subsubsection{Step 1: dominant-cluster assignment}
\label{sec:dominant_cluster}

A group located within $7\,R_{200}$ of two or more clusters represents an ambiguous case: using the topological criterion of Section~\ref{sec:connectivity_network}, the same group could be classified independently with respect to each cluster, potentially appearing as connected for one cluster and isotropic for another. To ensure a unique and physically motivated assignment, we define the \emph{dominant cluster} for each group as the cluster toward which the group has the most negative radial velocity $v_r$ (Section~\ref{sec:kinematics}), i.e.\ the cluster it is falling toward most strongly. It is at this step that the infall-region framework is earned: because each group is assigned to the single cluster it is most strongly falling toward, and the probed volume lies within that cluster's gravitational sphere of influence, the bulk flow of both primary- and secondary-connected groups is toward the dominant cluster, justifying the description of these filaments as ultimately feeding the cluster.

Classification is performed for every group-cluster pair. The dominant assignment, together with the backsplash removal of Step~3 below, defines the \emph{clean sample} used for all the infall-kinematic measurements (radial velocity, velocity--filament angle and orbital pericentre), which require a single, unambiguous cluster per group. The spin-alignment analysis (Section~\ref{sec:spin_method}) instead retains every classified pair with a defined filament tangent, since a group's angular momentum is an intrinsic property that can be compared with each cluster-connected filament it approaches (the repeated pairs share the same $\mathbf{J}$ and are not statistically independent; Section~\ref{sec:spin_method}). Our catalogue contains 212,704 group-cluster pairs from 180,577 unique groups, of which 27,846 (15.4\,per cent) lie in the infall zones of multiple clusters; retaining only the dominant assignment removes 32,127 pairs, leaving 180,577 dominant group-cluster pairs for the kinematic analyses.

\subsubsection{Step 2: filament proximity and topological class}

For each group we compute the perpendicular distance $d_\mathrm{conn}$ from the group to the spine of its nearest cluster-connected filament in the masked catalogue (Section~\ref{sec:masking}). We measure the distance to the nearest connected filament only, not to the globally nearest filament of any type. If an unconnected background filament passes closer to the group than any cluster-connected filament, a naive nearest-filament criterion would misclassify the group as isotropic even though it lies within the cluster's filamentary accretion network. By restricting to the connected subset, we ensure that the distance criterion directly probes the group's relationship to the cluster's feeding channels. Distances are computed as minimum point-to-line-segment distances over all segments of all connected filaments in the masked catalogue, with periodic boundary conditions, after the BFS expansion of Section~\ref{sec:connectivity_network}. The identity of the nearest connected filament determines the topological order of the group's class.

Groups are classified as follows:
\begin{itemize}
    \item Primary connected ($d_\mathrm{conn} < 1\,h^{-1}$\,Mpc; nearest connected filament is order-1): the group lies within the spine of a filament that directly crosses the $1.5\,R_{200}$ shell.
    \item Secondary connected ($d_\mathrm{conn} < 1\,h^{-1}$\,Mpc; nearest connected filament is order-2): the group lies within a filament that joins the primary network at an inter-cluster saddle point.
    \item Primary-outskirt ($1 \leq d_\mathrm{conn} \leq 3\,h^{-1}$\,Mpc): the group lies in the outskirts of a connected filament, beyond the dense core but within the exclusion radius. This class is excluded from the primary kinematic comparison because its filament affiliation is not clear-cut.
    \item Isotropic ($d_\mathrm{conn} > 3\,h^{-1}$\,Mpc): the group has no close proximity to any cluster-connected filament and approaches from the general field.
\end{itemize}

The thresholds $d_\mathrm{conn} = 1$ and $3\,h^{-1}$\,Mpc bracket the typical filament width as measured from the stacked transverse dark matter density profile (Section~\ref{sec:radial_profile}; Fig.~\ref{fig:radial_profile}): $d_\mathrm{conn} = 1\,h^{-1}$\,Mpc $\approx 2\,R_\mathrm{c}$ selects groups within the dense filament core, while $d_\mathrm{conn} = 3\,h^{-1}$\,Mpc lies beyond $R_\mathrm{drop}$ where the profile has converged to the cosmic mean, consistent with the filament widths reported by \citet{Cautun2014}, \citet{Zhu2021}, and \citet{Yang2025}. Sensitivity to these choices is tested in Appendix~\ref{sec:appendix_sensitivity}. Applied to all 212,704 group-cluster pairs, the classification yields 51,444 primary connected (24.2\,per cent), 29,408 secondary connected (13.8\,per cent), 88,446 primary-outskirt (41.6\,per cent), and 43,406 isotropic (20.4\,per cent) group-cluster pairs. Primary- and secondary-connected groups are collectively the \emph{connected} class. The classification is purely topological (based only on filament connectivity and proximity), so we use the neutral term \emph{connected} throughout the Data and Method sections (Sections~\ref{sec:data} and~\ref{sec:method}); only once the kinematic analysis of Section~\ref{sec:kinematics_results} demonstrates that these groups preferentially stream toward the cluster do we adopt the flow-laden term \emph{channelled} for them.

\subsubsection{Step 3: backsplash removal}

The topological classification makes no requirement on the direction of motion. Groups with $v_r > 0$ (outward velocity) in the cluster rest frame include \emph{backsplash} groups that have previously passed through the cluster potential and are now on their outward trajectory, and \emph{field outflows} that were never gravitationally captured. Because backsplash groups entered the cluster on a prior orbit, their pre-infall accretion channel is not observable in the current snapshot and their kinematics reflect post-pericentre orbital properties rather than infall dynamics \citep{Haggar2023}. We identify and remove them by tracing the main progenitor branch of each group through the MultiDark consistent-trees merger catalogues \citep{Behroozi2013b}: a group is confirmed backsplash if its progenitor was ever within $R_{200}$ of the cluster's main progenitor at any past snapshot. Because every cluster is itself a host halo, each cluster also forms a pair with itself at zero separation. Of the 180,577 dominant pairs, 21,042 are such self-pairs, which this step removes trivially. A further 1,838 (1.0\,per cent of the dominant pairs) are confirmed backsplash groups, leaving a clean sample of 157,697 pairs.\footnote{During revision we found that the merger-tree walk used to derive the backsplash flags in the submitted version did not restrict the progenitor track to the main branch, and tested proximity against the progenitor's virial radius rather than its $R_{200}$. The flags used throughout this version are re-derived along the main branch and against $R_{200}$. Of the 1,838 confirmed backsplash groups, 1,082 were flagged before. The clean sample differs from the submitted one by 4 pairs (about 750 pairs leave it and as many join it). All results based on the clean sample were recomputed with the corrected flags.}

Confirmed backsplash groups lie close to the cluster, at a median $d_\mathrm{cl} = 1.53\,R_{200}$ (16th--84th percentiles $1.40$--$1.80\,R_{200}$), and are mostly primary-outskirt (66\,per cent) or primary (27\,per cent) groups, with secondary and isotropic groups contributing 3 and 4\,per cent. They make up 29\,per cent of the non-self dominant pairs inside $1.5\,R_{200}$, 6\,per cent of those within $3\,R_{200}$ and 0.08\,per cent of those beyond $2\,R_{200}$, and 93\,per cent (508 of 547) of the pairs inside $1.5\,R_{200}$ that are moving outward. No group other than a cluster's own self-pair lies inside $R_{200}$. This is expected, since a host halo that close to the cluster centre would be a subhalo of the cluster.

Field outflows ($v_r > 0$, not confirmed backsplash) are retained alongside infalling groups. Pre-selecting on $v_r < 0$ would truncate the slow tail of each class's velocity distribution and so bias its median infall velocity; in practice few clean pairs move outward in any class. Sensitivity tests confirm that re-applying a $v_r < 0$ filter changes the median infall velocity of isotropic groups by at most $11\,\mathrm{km\,s^{-1}}$, and that of any class by at most $14\,\mathrm{km\,s^{-1}}$, over $1.5$--$5\,R_{200}$, well below the observed $150$--$180\,\mathrm{km\,s^{-1}}$ primary-isotropic differences reported in Section~\ref{sec:results}.

\subsubsection{Step 4: Strict kinematic criterion}

As a sensitivity check on whether the topological (position-only) classification is robust to also demanding that groups be kinematically infalling and filament-aligned, we test a strict kinematic criterion requiring simultaneously $v_r < 0$ and $\theta_\mathrm{fil} < 60^\circ$ (where $\theta_\mathrm{fil}$ is the angle between the group velocity vector and the nearest filament tangent) in addition to spatial proximity. Results are reported in Appendix~\ref{sec:appendix_sensitivity}, and confirm that the class-averaged spin flip masses and velocity contrasts are unchanged within their uncertainties.

\subsection{Kinematic measurements}
\label{sec:kinematics}

For each group-cluster pair we compute the relative velocity in the cluster rest frame. The cluster peculiar velocity $\mathbf{v}_\mathrm{cluster}$ is taken directly from the MDPL2 halo catalogue as the bulk velocity of the cluster's host halo. The \emph{radial velocity} is the component directed toward the cluster centre:
\begin{equation}
    v_r = (\mathbf{v}_\mathrm{group} - \mathbf{v}_\mathrm{cluster}) \cdot \hat{r},
    \label{eq:vr}
\end{equation}
where $\hat{r}$ is the unit vector pointing from the cluster to the group. Negative $v_r$ indicates infall. The \emph{tangential velocity} is the residual:
\begin{equation}
    v_t = \left|(\mathbf{v}_\mathrm{group} - \mathbf{v}_\mathrm{cluster}) - v_r\,\hat{r}\right|.
    \label{eq:vt}
\end{equation}
The \emph{velocity anisotropy} parameter is computed in bins of $r/R_{200}$ by stacking all group-cluster pairs:
\begin{equation}
    \beta(r) = 1 - \frac{\sigma_t^2(r)}{2\,\sigma_r^2(r)},
    \label{eq:beta}
\end{equation}
where $\sigma_r$ and $\sigma_t$ are the dispersions of $v_r$ and $v_t$ across all pairs in the bin. $\beta = 0$ is isotropic, $\beta = 1$ is purely radial, and $\beta < 0$ is tangentially biased.

\subsection{Spin-filament alignment statistic}
\label{sec:spin_method}

For each classified group we compute the angle $\theta$ between its angular momentum vector $\mathbf{J} = (J_x, J_y, J_z)$ from \rockstar and the local tangent $\hat{f}$ of its nearest cluster-connected filament spine. The tangent is computed as a central difference at the nearest spine point; forward/backward differences are used at the endpoints. The alignment statistic is:
\begin{equation}
    S(\log_{10} M) = \left\langle \cos^2\theta(\mathbf{J},\,\hat{f})\right\rangle - \frac{1}{3},
    \label{eq:S}
\end{equation}
where $S > 0$ indicates spin parallel to the filament ($\mathbf{J} \parallel \hat{f}$), $S = 0$ indicates random orientation, and $S < 0$ indicates spin perpendicular to the filament ($\mathbf{J} \perp \hat{f}$). The absolute value $|\cos\theta|$ is used because the filament tangent direction is sign-ambiguous. Uncertainties on $S$ in each mass bin are estimated by 2000-sample bootstrap resampling (16th and 84th percentiles defining the $\pm 1\sigma$ band). The spin flip mass $\mflip$ is determined by linear interpolation between the two adjacent mass bins that straddle the $S = 0$ crossing.

For isotropic groups, the nearest cluster-connected filament lies at $d_\mathrm{conn} > 3\,h^{-1}$\,Mpc: the filament tangent is from a non-interacting reference structure. The isotropic class thus serves as a control sample; the expectation for random spin alignment is $S = 0$, and any departure reflects residual large-scale tidal correlations.

The spin statistic $S$ is computed over all classified group-cluster pairs with a defined filament tangent, rather than over the smaller dominant-cluster sample used for the kinematic measurements: a group's angular momentum is an intrinsic property, so its alignment against each cluster-connected filament it approaches is a distinct, valid test of tidal torque. Groups that approach more than one cluster are a non-negligible fraction of each channelled sample (${\approx}19$ per cent of primary and ${\approx}27$ per cent of secondary pairs; ${\approx}8$ and ${\approx}12$ per cent when the repeated approaches fall within the same class). Each approach tests the group's spin against a distinct cluster-connected filament, but the approaches are not statistically independent, since the same $\mathbf{J}$ enters each of them. We therefore checked that the uncertainties are not underestimated by repeating the bootstrap with unique groups, and with whole clusters, as the resampled unit (Appendix~\ref{sec:appendix_independence}). Resampling unique groups inflates the uncertainty on $S$ by a median of $4$--$6$ per cent per class and widens the primary $\mflip$ interval by $0.01$\,dex, while resampling clusters changes it by a median of less than 1 per cent. We retain the pair-level bootstrap throughout.

Figure~\ref{fig:schematic_spin} illustrates the geometric relationship between $\mathbf{J}$, $\hat{f}$, and the orbital angular momentum $\mathbf{L}_\mathrm{orb}$ for an infalling group.

\begin{figure*}
    \centering
    \includegraphics[width=\textwidth]{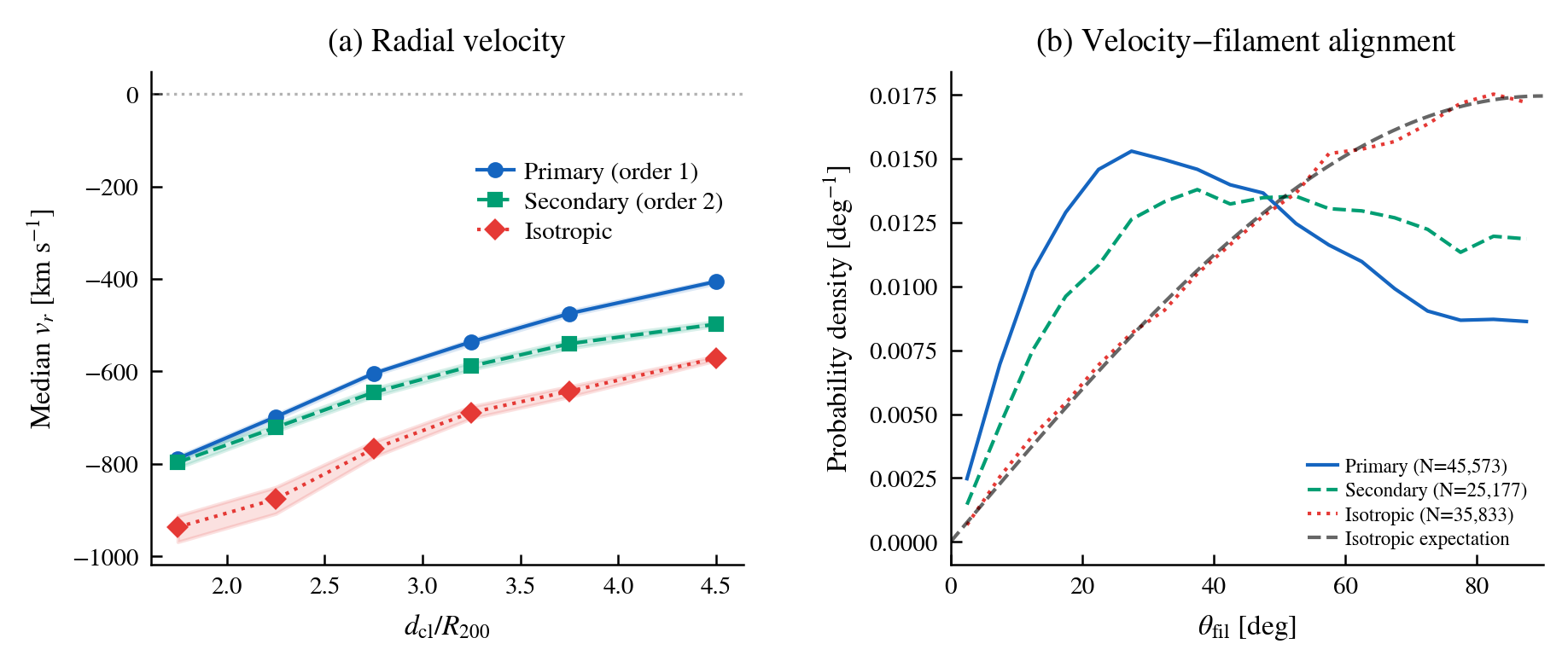}
    \caption{Kinematic properties of infalling groups by accretion class (primary channelled: dark blue; secondary channelled: green; isotropic: red). Shaded bands: $\pm 1\sigma$ bootstrap uncertainties. \textit{(a)} Median radial velocity $v_r$ vs cluster-centric distance $d_\mathrm{cl}/R_{200}$; negative values indicate infall. \textit{(b)} Probability density of the velocity-filament angle $\theta_\mathrm{fil}$ (measured to the nearest filament of any type); dashed grey curve: isotropic expectation $p \propto \sin\theta$. }
    \label{fig:velocity_profiles}
\end{figure*}


\section{Results}
\label{sec:results}

\subsection{Accretion channel fractions}
\label{sec:classification_illustration}

Of the 212,704 group-cluster pairs, 51,444 (24.2\,per cent) are primary connected, 29,408 (13.8\,per cent) are secondary connected, and 43,406 (20.4\,per cent) are isotropic (88,446 primary-outskirt are excluded from kinematic comparisons). The mass distributions of the three classes differ: the primary and secondary fractions rise steeply with group mass while the isotropic fraction falls (Section~\ref{sec:mass_dependence}), reflecting assembly bias: massive groups preferentially form in the dense environments of cluster-connected filaments. These are the raw classification totals over the full set of group-cluster pairs. The infall-kinematic analyses that follow are measured on the clean sample (each group assigned to its single dominant cluster, confirmed backsplash interlopers removed following Section~\ref{sec:classification}, and groups restricted to those with a defined nearest-filament angle), so the per-class counts in those figure legends are smaller than the totals above. For example, $45{,}573$ primary, $25{,}177$ secondary and $35{,}833$ isotropic groups enter the velocity--filament angle distribution of Fig.~\ref{fig:velocity_profiles}b.

\subsection{Infall kinematics}
\label{sec:kinematics_results}

We next examine the kinematics of the classified groups. The velocity--filament alignment results below show that primary- and secondary-connected groups stream preferentially toward the cluster along their filaments. This licenses the flow interpretation, and we therefore refer to them as primary- and secondary-\emph{channelled} from this point onward.

Infall velocities depend strongly on the large-scale environment: isotropic (field) groups fall into clusters significantly faster than filament-channelled groups across all radii (Fig.~\ref{fig:velocity_profiles}). The higher infall speeds of isotropic groups arise from environmental dynamics that increase their absolute velocities, rather than from differences in infall geometry alone. The median radial velocity profiles (Fig.~\ref{fig:velocity_profiles}a) show that isotropic groups infall $150$--$180\,\mathrm{km\,s^{-1}}$ faster (more negative $v_r$) than primary channelled groups at all radii $1.5$-$5\,R_{200}$, with secondary groups lying between the two. This ordering (field objects infalling faster than filament objects) is consistent with observational and simulation results in the literature: \citet{Ceccarelli2011} find that infall velocity amplitudes are systematically larger along the void (minor-axis) direction of the surrounding large-scale structure than along the filament (major-axis) direction, and \citet{Santucho2020} show that galaxies streaming onto groups from underdense directions arrive with significantly higher velocities than those arriving from overdense directions, attributing the difference to tidal braking by neighbouring structures. Our topological classification quantifies this environmental dependence with a factor-of-three larger sample and a physically motivated hierarchy.

The faster infall of isotropic groups is primarily a speed effect: isotropic groups have higher total speeds in the cluster rest frame, plausibly because they experience less tidal braking from neighbouring overdensities \citep{Santucho2020} and a coherent outward push from void expansion \citep{Peebles1980}, although we do not test these mechanisms (Section~\ref{sec:discussion_kinematics}). Isotropic groups are also slightly more radially directed at the same distance (mean $|\cos\theta_{vr}|$ higher by $0.02$--$0.09$ between $1.75$ and $7\,R_{200}$), so direction adds to the difference. When the difference in mean $|v_r|$ is split into a speed term and a direction term, the speed term contributes $69$--$87$ per cent and the direction term $9$--$29$ per cent. The contrast persists across the full group-mass range probed by the classification, with no sign reversal (Section~\ref{sec:mass_dependence}).

\begin{figure*}
    \centering
    \includegraphics[width=\textwidth]{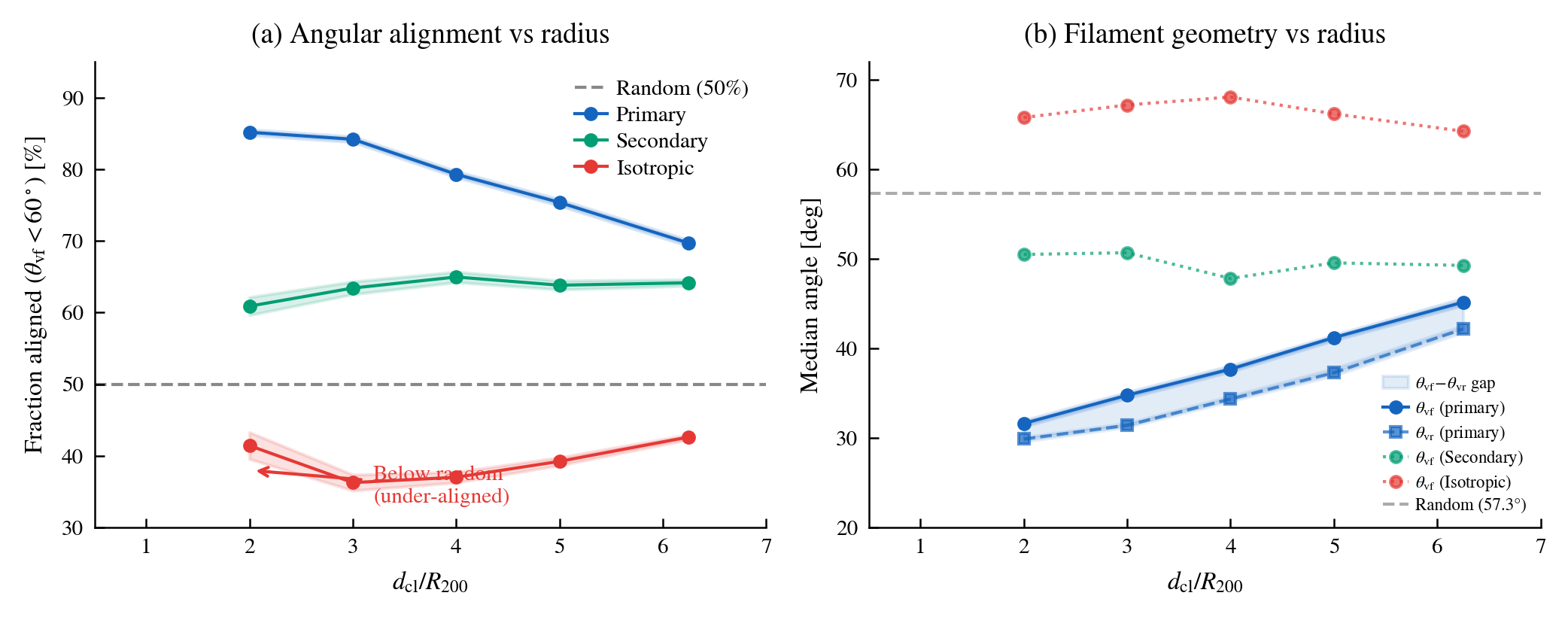}
    \caption{Radial profiles of alignment properties for infalling groups (colour scheme as in Fig.~\ref{fig:velocity_profiles}). \textit{(a)} Fraction of groups with $\theta_{vf} < 60^\circ$ (aligned within $60^\circ$ of the connected filament axis) vs $d_\mathrm{cl}/R_{200}$; dashed line: isotropic expectation (50 per cent). \textit{(b)} Median velocity-filament angle $\theta_{vf}$ (solid) and velocity-radial angle $\theta_{vr}$ (dashed) for primary channelled groups; dotted lines show $\theta_{vf}$ for secondary and isotropic. The shaded band shows the near-constant $\theta_{vf} - \theta_{vr} \approx 2$--$4^\circ$ gap for primary groups.}
    \label{fig:alignment_radial}
\end{figure*}

The distribution of the velocity-filament angle $\theta_\mathrm{fil}$ (Fig.~\ref{fig:velocity_profiles}b) is the strongest discriminant between accretion channels. Here $\theta_\mathrm{fil}$ is the angle to the \emph{nearest} filament of any type, distinct from the nearest-connected-filament angle $\theta_{vf}$ of Fig.~\ref{fig:alignment_radial}. Primary channelled groups cluster strongly near $\theta_\mathrm{fil} \sim 0^\circ$-$30^\circ$, while secondary channelled groups show an intermediate excess, and isotropic groups follow the isotropic expectation $p(\theta) \propto \sin\theta$ with no preferred direction (their nearest filament is typically an unconnected field filament; their anti-alignment with the cluster-connected network appears in Fig.~\ref{fig:alignment_radial}a).

\begin{figure*}
    \centering
    \includegraphics[width=\textwidth]{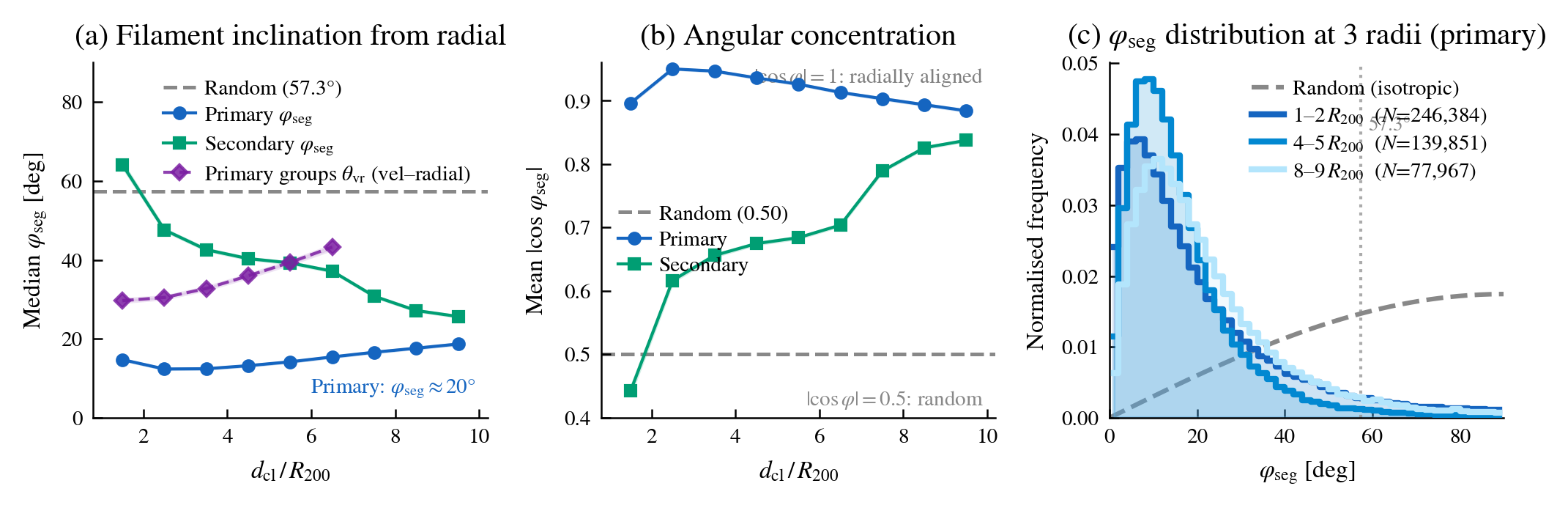}
    \caption{Direct measurement of filament segment inclination angles, stacked over all $26{,}616$ clusters with connected filaments. Primary filaments: dark blue circles. Secondary filaments: green squares. Dashed grey lines: isotropic expectation. \textit{(a)} Median inclination angle $\varphi_\mathrm{seg}$ vs $d_\mathrm{cl}/R_{200}$; purple diamonds show the median velocity-radial angle $\theta_{vr}$ of primary groups from Fig.~\ref{fig:alignment_radial}b. \textit{(b)} Mean angular concentration $|\cos\varphi_\mathrm{seg}|$ vs radius; random expectation is 0.50. \textit{(c)} Normalised distributions of $\varphi_\mathrm{seg}$ for primary filaments in three radial zones ($1$-$2$, $4$-$5$, $8$-$9\,R_{200}$) compared to the isotropic distribution $p(\varphi) \propto \sin\varphi$ (dashed).}
    \label{fig:filament_inclination}
\end{figure*}

\subsection{Angular alignment and filament geometry}
\label{sec:filament_geometry}

The coherent, filament-aligned infall of channelled groups raises the question of whether this kinematic alignment is inherited directly from the geometry of the filaments themselves, whether the groups simply trace filaments that are already pointed at the cluster. Figure~\ref{fig:alignment_radial}a shows the fraction of groups with velocity aligned within $60^\circ$ of the connected filament axis as a function of cluster-centric distance. Primary channelled groups have an alignment fraction of $85$ per cent at $2\,R_{200}$, declining monotonically to $70$ per cent at $6\,R_{200}$; both values substantially exceed the isotropic expectation of 50\,per cent. Secondary channelled groups maintain a flat alignment fraction of $\sim$63\,per cent at all radii. Isotropic groups, by contrast, fall below the random expectation at $2\,R_{200}$: only $41.4 \pm 2.0$ per cent are within $60^\circ$ of the nearest connected filament axis ($N \approx 640$ isotropic groups in this radial bin), a $4.3\sigma$ deficit below the random $50$ per cent. Because a within-$60^\circ$ deficit necessarily implies an excess at larger angles, these groups are mildly anti-aligned (preferentially perpendicular to the filament), arriving from the angular sectors not covered by the primary filament network.

The velocity-filament angle $\theta_{vf}$ (velocity versus the nearest cluster-connected filament tangent) and the velocity-radial angle $\theta_{vr}$ (velocity versus the infall direction $-\hat{r}$) for primary groups both rise from $\approx 30$--$32^\circ$ at $2\,R_{200}$ to $\approx 42$--$45^\circ$ at $6\,R_{200}$, while maintaining a nearly constant offset $\theta_{vf} - \theta_{vr} \approx 2$--$4^\circ$ at all radii (Fig.~\ref{fig:alignment_radial}b). This constant offset indicates that groups stream coherently along their connected filaments throughout the infall zone. The concurrent rise in both angles suggests that primary filaments become progressively more radially aligned toward the cluster centre, a geometric property of the filaments themselves rather than a kinematic property of the groups.

To test this directly, we measure the inclination angle $\varphi_\mathrm{seg}$ of each filament segment relative to the cluster radial direction (Fig.~\ref{fig:schematic_topo}). For $1{,}785{,}898$ segments of primary filaments and $1{,}142{,}892$ segments of secondary filaments stacked over all $26{,}616$ clusters with connected filaments, we compute $\varphi_\mathrm{seg} = \arccos|\hat{t} \cdot \hat{r}|$, where $\hat{t}$ is the filament segment tangent and $\hat{r}$ points from the segment midpoint toward the cluster centre. Primary filament segments point within $\varphi_\mathrm{seg} \approx 12$--$14^\circ$ of the cluster radial direction across the inner infall zone ($1.5$--$5\,R_{200}$), rising only mildly to $\approx 19^\circ$ by $10\,R_{200}$ (Fig.~\ref{fig:filament_inclination}), a factor of ${\approx}4$ below the isotropic expectation of $57.3^\circ$. This strong radial alignment is approximately flat across the inner infall zone and varies only gently with radius. The cluster potential holds its feeding filaments within $\approx 12$--$19^\circ$ of the radial direction throughout the entire $1.5$--$10\,R_{200}$ zone probed here (the cluster's gravitational influence zone, extending well beyond the virial radius). The angular concentration $|\cos\varphi_\mathrm{seg}| \approx 0.88$-$0.95$ at $1.5$-$10\,R_{200}$ ($1.8$-$1.9\times$ the random expectation of 0.50), peaking near $2$--$3\,R_{200}$.

Secondary filaments behave qualitatively differently: at $1.5\,R_{200}$, secondary segments have $\varphi_\mathrm{seg} \approx 64^\circ$ (above the random expectation), branching tangentially from primary arms at saddle points. Only at larger radii ($r \gtrsim 5\,R_{200}$) do secondary filaments gradually turn toward the cluster, a consequence of saddle-point geometry that we interpret in Section~\ref{sec:discussion_kinematics}.

The nearly flat orientation profile, $\varphi_\mathrm{seg}(r) \approx 12$--$19^\circ$, appears at odds with the increasing $\theta_{vr}(r)$ of primary groups. If groups streamed perfectly along radially oriented filaments, $\theta_{vr}$ would track this low $\varphi_\mathrm{seg}$ (modulo velocity scatter) at all radii. Instead, the increase in $\theta_{vr}$ is driven by a selection effect: groups at larger cluster-centric distances are preferentially found on the more inclined outer portions of long primary filaments, which are drawn from the high-$\varphi$ tail of the segment orientation distribution (Fig.~\ref{fig:filament_inclination}c). As a result, these groups exhibit systematically larger $\theta_{vr}(r)$ values.

\subsection{Mass dependence and filament quality}
\label{sec:mass_dependence}

\begin{figure}
    \centering
    \includegraphics[width=\columnwidth]{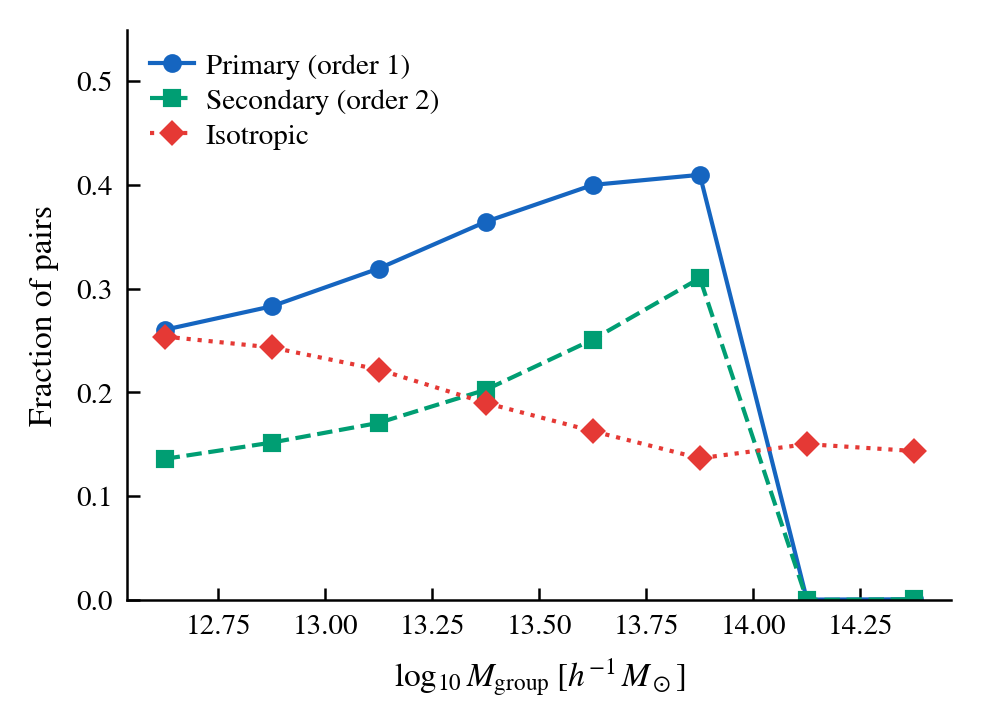}
    \caption{Fraction of group-cluster pairs classified as primary (dark blue), secondary (green), and isotropic (red) as a function of group virial mass. The rising channelled fractions and falling isotropic fraction reflect assembly bias: massive groups preferentially form along cluster-feeding filaments.}
    \label{fig:mass_dependence}
\end{figure}

More massive groups are increasingly likely to be associated with cluster-feeding filaments, while the dynamical difference between channelled and isotropic infall remains largely independent of group mass (Fig.~\ref{fig:mass_dependence}). The primary fraction rises from $\approx 26$ per cent at low group mass ($\log_{10} M \sim 12.6$) to $\approx 41$ per cent at $\log_{10} M \sim 13.9$; the secondary fraction rises more steeply, from $\approx 14$ per cent to $\approx 31$ per cent; and the isotropic fraction falls from $\approx 25$ per cent to $\approx 14$ per cent. This reflects assembly bias: massive groups preferentially form along cluster-feeding filaments. Above the $\log_{10} M = 14.0$ group-mass cut the sample retains only the few cluster-mass haloes lying within a more massive neighbour's infall zone, essentially all of which are labelled primary-outskirt or isotropic (the primary and secondary fractions fall to $<0.1$ per cent). The $14.0$--$14.5$ bins therefore reflect the group selection rather than a breakdown of the classification. The velocity contrast $\Delta v_r$ between channelled and isotropic groups is stable across the full mass range ($12.5 \le \log_{10} M < 14.0$), showing that the faster isotropic infall is not driven by a particular mass subpopulation.

To determine whether all primary filaments channel infall equally, we divided primary groups by the linear halo mass density of their host filament ($\rho_\mathrm{lin} = \Sigma M_\mathrm{vir}/L$; Table~\ref{tab:symbols}). In their bulk kinematics, thin-tendril (low-$\rho_\mathrm{lin}$) and dense-bridge (high-$\rho_\mathrm{lin}$) primary groups are alike: both stream along their host filament ($\gtrsim 75$ per cent of each have velocities within $60^\circ$ of the filament axis over $1.5$--$5\,R_{200}$, against ${\approx}40$ per cent for isotropic groups) and both infall with a comparably radial velocity anisotropy ($\beta \approx 0.6$--$0.8$, similar to that of the isotropic class). Bulk velocity therefore does not separate a thin tendril from a dense bridge. The two are distinguished instead by the tidal environment their host filament represents, which is imprinted not on the infall velocity but on the group spin (Section~\ref{sec:spinflip_quality}). Being connected to a cluster by a primary (order-1) filament, or even streaming along it, is thus not by itself sufficient to place a group in a strong tidal-torque environment; the density of the filament also matters.

\subsection{Spin-filament alignment}
\label{sec:spin_flip}

\subsubsection{The spin flip statistic and mass hierarchy}

The kinematic contrasts established above reflect the velocity field. We now test whether the tidal environment is also imprinted on halo angular momentum. Constrained tidal torque theory \citep[cTTT;][]{Codis2015} predicts that the mass at which halo spin vectors transition from parallel to perpendicular alignment with the host filament (the spin flip mass $\mflip$) is set by the local tidal environment, with higher $\mflip$ in denser filament regions. Our accretion classes sample distinct filament environments (cluster-connected filaments versus the inter-filament field), and within the primary class the filament linear density resolves a range of tidal environments, providing a test of this prediction.

\begin{figure*}
    \centering
    \includegraphics[width=\textwidth]{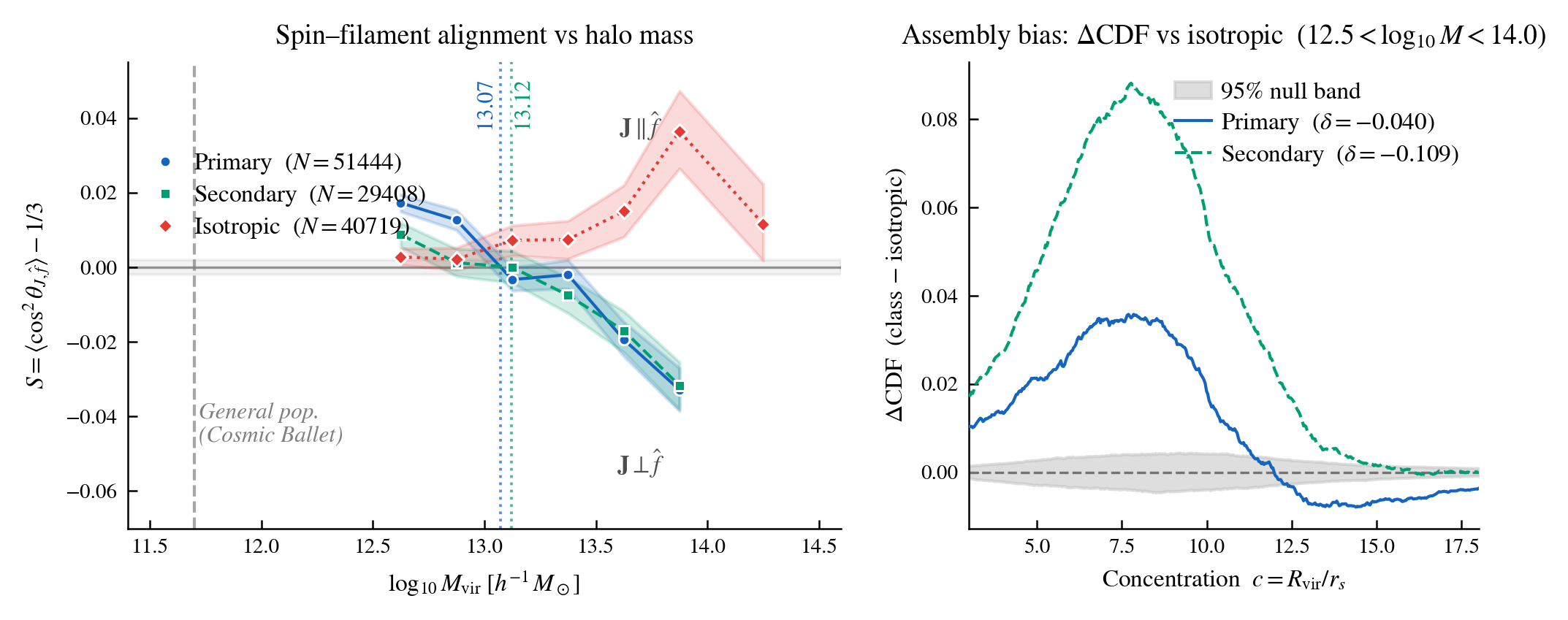}
    \caption{\textit{Left}: spin-filament alignment statistic $S(\log_{10} M) = \langle\cos^2\theta(\mathbf{J},\hat{f})\rangle - 1/3$ vs group virial mass for primary channelled (dark blue), secondary channelled (green), and isotropic (red) groups. Shaded bands: $\pm 1\sigma$ bootstrap uncertainties. Vertical dotted lines: interpolated spin flip masses $\mflip$ for primary and secondary classes (isotropic shows no flip). Dashed grey line: general field population flip mass from \citet{GaneshaiahVeena2018}. $S > 0$: spin parallel to filament; $S < 0$: spin perpendicular to filament. \textit{Right}: $\Delta\mathrm{CDF}(c) = \mathrm{CDF}_\mathrm{class}(c) - \mathrm{CDF}_\mathrm{isotropic}(c)$ for NFW concentration $c = r_\mathrm{vir}/r_s$ at matched mass, comparing primary (dark blue) and secondary (green) to isotropic. Grey band: 95\,per cent null band from bootstrap resampling. $\delta$: Cliff's effect size.}
    \label{fig:spin_flip}
\end{figure*}

Figure~\ref{fig:spin_flip} (left panel) shows the spin-filament alignment statistic $S(\log_{10} M) = \langle\cos^2\theta(\mathbf{J},\hat{f})\rangle - 1/3$ for primary, secondary, and isotropic groups (schematic geometry in Fig.~\ref{fig:schematic_spin}). Both channelled classes flip at closely comparable transition mass, well above the field, while the isotropic class shows no flip:
\begin{equation}
    \log_{10}\!\left(\frac{\mflip}{h^{-1}\,\msun}\right) = \begin{cases} 13.07 & \text{primary (node environment)} \\ 13.12 & \text{secondary (saddle environment)} \\ \text{no flip} & \text{isotropic (inter-filament)} \end{cases}
    \label{eq:mflip}
\end{equation}

The primary-secondary offset is $0.05$\,dex, consistent with zero, though the secondary flip mass is considerably less well determined than the primary: $\mflip^\mathrm{sec} = 13.12$ has a bootstrap $\pm1\sigma$ range $[12.82, 13.19]$, against $13.07\,[13.04, 13.13]$ for the primary, and bootstrapping the difference gives a 95 per cent interval $[-0.48, +0.27]$ that comfortably brackets zero. The BFS topological order alone (primary versus secondary) does not set a resolvable sub-hierarchy in $\mflip$. What does track $\mflip$ is the filament linear mass density: within the primary class alone the flip mass rises by $0.49$\,dex from the thinnest to the densest third of filaments, although after covariate matching the size of this rise is uncertain (Section~\ref{sec:spinflip_quality}). Both channelled classes lie $1.3$-$1.4$\,dex above the general field population ($\log_{10} \mflip \approx 11.7$ for intermediate-thickness filaments from the Cosmic Ballet measurements of \citealt{GaneshaiahVeena2018}, who report a range $\log_{10}\mflip \approx 11.0$-$12.3$ depending on filament diameter), reflecting the elevated tidal environment near clusters with $M > 10^{14.0}\,h^{-1}\,\msun$. Isotropic groups show $S > 0$ at all masses, with no zero-crossing, and roughly half ($47$--$53$ per cent, across bootstrap schemes and random seeds) of bootstrap resamples produce a crossing anywhere in the mass range from noise alone (Appendix~\ref{sec:appendix_independence}). The filament tangent at $d_\mathrm{conn} > 3\,h^{-1}$\,Mpc leaves no detectable flip in their spin orientation, so this class serves as a null control sample.

At low mass ($\log_{10} M \lesssim 12.9$), primary and secondary classes show $S > 0$: spin vectors are preferentially parallel to the filament, consistent with the tidal torque regime for group-mass haloes. At high mass ($\log_{10} M \gtrsim 13.5$), primary and secondary groups both show $S < 0$ at $\geq 3\sigma$ significance, while isotropic groups remain at $S > 0$ throughout, consistent with the null expectation.

In the high-mass bin $\log_{10} M \in [13.5, 14.0)$ both channelled classes show comparable perpendicular alignment: $S_\mathrm{pri} = -0.025$ (95 per cent CI $[-0.032, -0.018]$; $N = 6{,}920$) and $S_\mathrm{sec} = -0.023$ (95 per cent CI $[-0.032, -0.016]$; $N = 4{,}878$). The two confidence intervals overlap, so we detect no significant difference between the node-adjacent (primary) and saddle-point (secondary) environments. If anything, the point estimates run marginally counter to a saddle-stronger expectation ($|S_\mathrm{pri}| > |S_\mathrm{sec}|$). We therefore do not resolve the node-versus-saddle contrast in alignment strength that cTTT would predict, and treat the two channelled environments as producing equivalent high-mass perpendicular alignment within our uncertainties.

These measurements are drawn from the $210{,}017$ classified pairs with a defined filament tangent ($51{,}444$ primary, $29{,}408$ secondary, $88{,}446$ primary-outskirt, $40{,}719$ isotropic; the isotropic count is $2{,}687$ below the $43{,}406$ classified because those groups lie near clusters with no BFS-connected filament, so no tangent $\hat{f}$ is defined), with all mass bins at $\log_{10} M < 13.75$ containing $N \geq 1{,}700$ per class, ensuring reliable bootstrap uncertainties throughout.

\subsubsection{Environmental dependence and assembly bias}
\label{sec:environment_assembly}

If $\mflip$ is set by the tidal environment encoded in the filament structure, it should depend on halo formation history but not on the global connectivity of the host cluster. We test this by splitting the sample along two axes.

Splitting by cluster connectivity $\kappa$ into low ($\kappa \leq 2$), intermediate ($3 \leq \kappa \leq 4$), and high ($\kappa \geq 5$) bins (Fig.~\ref{fig:environment}a, Appendix~\ref{sec:appendix_environment}) reveals no significant dependence. The $S(\log_{10} M)$ curves overlap within the bootstrap confidence interval across all $\kappa$ bins. Splitting instead by NFW concentration tertiles (Fig.~\ref{fig:environment}b) shows the trend predicted by ZOMG/TTT: higher-concentration (earlier-forming) primaries display stronger perpendicular spin alignment at high mass, in the direction expected if $\mflip$ tracks formation epoch. The flip-mass shift between the concentration tertiles is only marginal, within the bootstrap uncertainties, because concentration anticorrelates with halo mass and the highest-concentration tertile is sparsely populated at high mass. The variable that better separates tidal environments is the filament linear density (Section~\ref{sec:spinflip_quality}).

The concentration $\Delta$CDF (right panel of Fig.~\ref{fig:spin_flip}) extends this result to the full population: both primary and secondary groups have lower NFW concentrations than isotropic groups at matched mass ($\Delta\mathrm{CDF} > 0$). The difference is well outside the 95\,per cent null band at all concentrations. The Cliff's effect size (a non-parametric measure equal to the probability that a random group of one class exceeds a random isotropic group minus the reverse probability, ranging from $-1$ to $+1$ with $0$ meaning no difference) is weak ($\delta \approx -0.04$ to $-0.11$, below the $|\delta| = 0.147$ threshold conventionally regarded as a ``small'' effect) but coherent in direction across the full concentration range.

The resulting concentration ordering (isotropic~$>$~primary~$>$~secondary) is in the direction expected from assembly bias, and admits a natural, though not unique, physical interpretation: isotropic groups in inter-filament environments may be tidally stalled, so that without a feeding filament to supply fresh material their mass growth is suppressed at earlier times, giving earlier formation epochs and higher concentrations at $z = 0$. This is the ZOMG mechanism identified by \citet{Borzyszkowski2017} in seven zoom-in haloes. Our ordering, measured at $N \approx 3$--$5\times10^4$ per class, is consistent with that mechanism, but the effect sizes are weak ($|\delta| \leq 0.11$) and the measurement establishes the ordering, not the stalling itself.

\subsection{Radial self-similarity and spin-orbital co-rotation}
\label{sec:radial_corotation}

The spin flip of Section~\ref{sec:spin_flip} was measured over the full infall zone. If $\mflip$ is set by short-range cluster tidal forces, it should decrease with cluster-centric distance as the tidal field weakens. If it is instead encoded in the large-scale filament structure, it should be approximately constant across the infall zone. We test this by repeating the $S(\log_{10} M)$ measurement in four cluster-centric distance bins ($1.5$--$2.5$, $2.5$--$4.0$, $4.0$--$5.5$, $5.5$--$7.0\,R_{200}$; $N = 9{,}069$--$15{,}343$ primary groups per bin).

\begin{figure*}
    \centering
    \includegraphics[width=\textwidth]{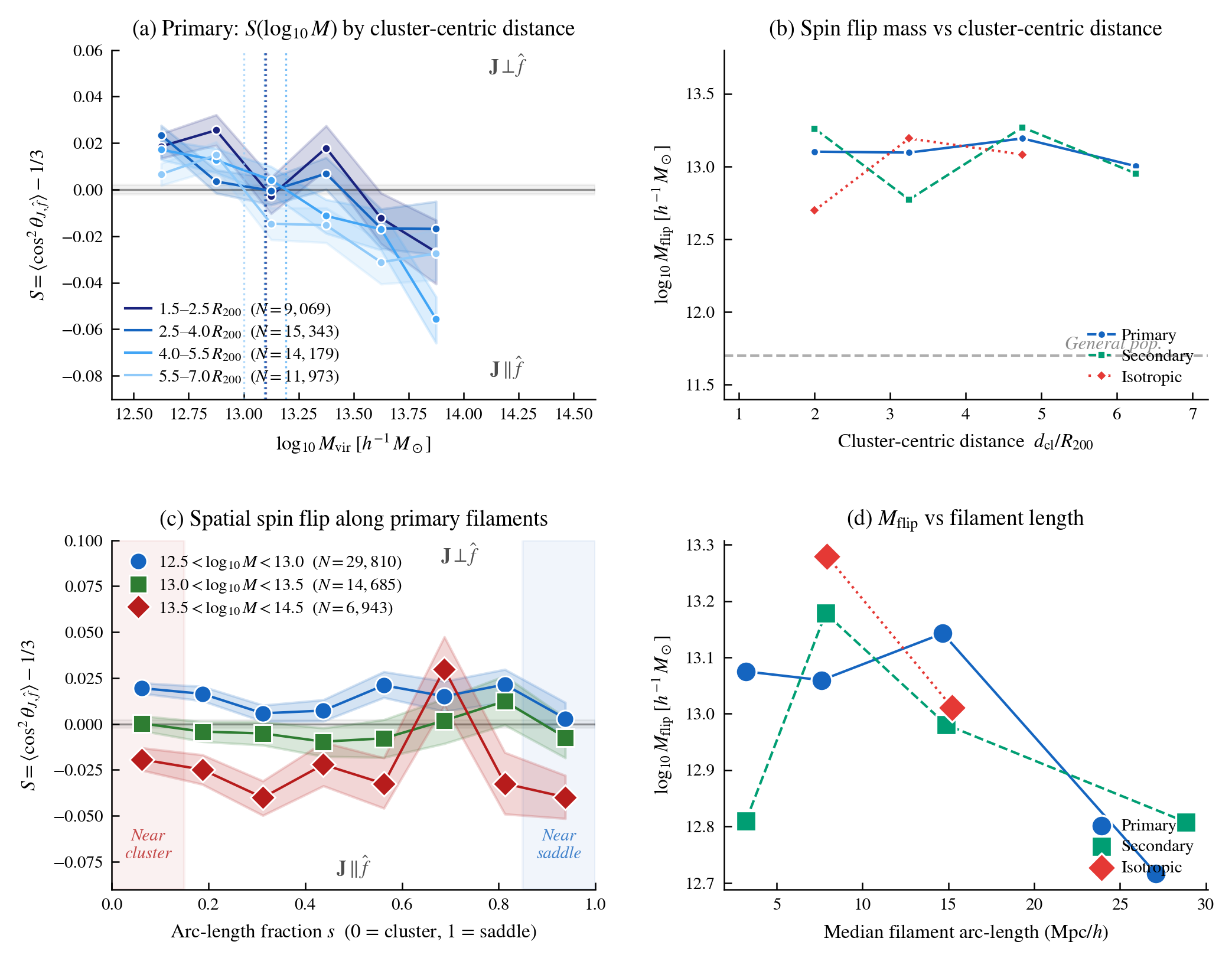}
    \caption{Radial self-similarity and spatial dependence of the spin flip. \textit{(a)} $S(\log_{10} M)$ for primary groups in four cluster-centric distance bins ($1.5$--$2.5$, $2.5$--$4.0$, $4.0$--$5.5$, $5.5$--$7.0\,R_{200}$); $\mflip$ is approximately constant with radius. \textit{(b)} Interpolated $\mflip$ vs $d_\mathrm{cl}/R_{200}$ for primary (dark blue), secondary (green), and isotropic (red); both channelled classes flip at similar mass at all radii. \textit{(c)} $S$ along the filament spine (arc-length fraction $s$; $0 =$ cluster, $1 =$ saddle) for primary groups in two mass bins, showing the expected cTTT tidal gradient within bootstrap uncertainties. \textit{(d)} $\mflip$ vs median filament arc-length per class. }
    \label{fig:radial_spatial}
\end{figure*}

The spin flip mass is approximately flat with cluster-centric distance for both channelled classes across $1.5$--$7\,R_{200}$ (Fig.~\ref{fig:radial_spatial}a,b): the primary $\mflip$ varies by only ${\approx}0.2$\,dex ($\log_{10}\mflip \approx 13.0$--$13.2$) with no systematic radial trend, and the two channelled classes flip at similar mass at all radii. This nearly flat $\mflip(r)$ profile is the spin-space counterpart of the strong, only-mildly-varying filament inclination $\varphi_\mathrm{seg}(r) \approx 12$--$19^\circ$ measured in Section~\ref{sec:filament_geometry}: in both cases, the cluster's tidal imprint on the surrounding structure is approximately self-similar across the infall zone. The parallel between these two profiles is consistent with the large-scale filament environment, rather than proximity to the cluster, setting both the spin flip mass and the kinematic alignment.

\begin{figure*}
    \centering
    \includegraphics[width=\textwidth]{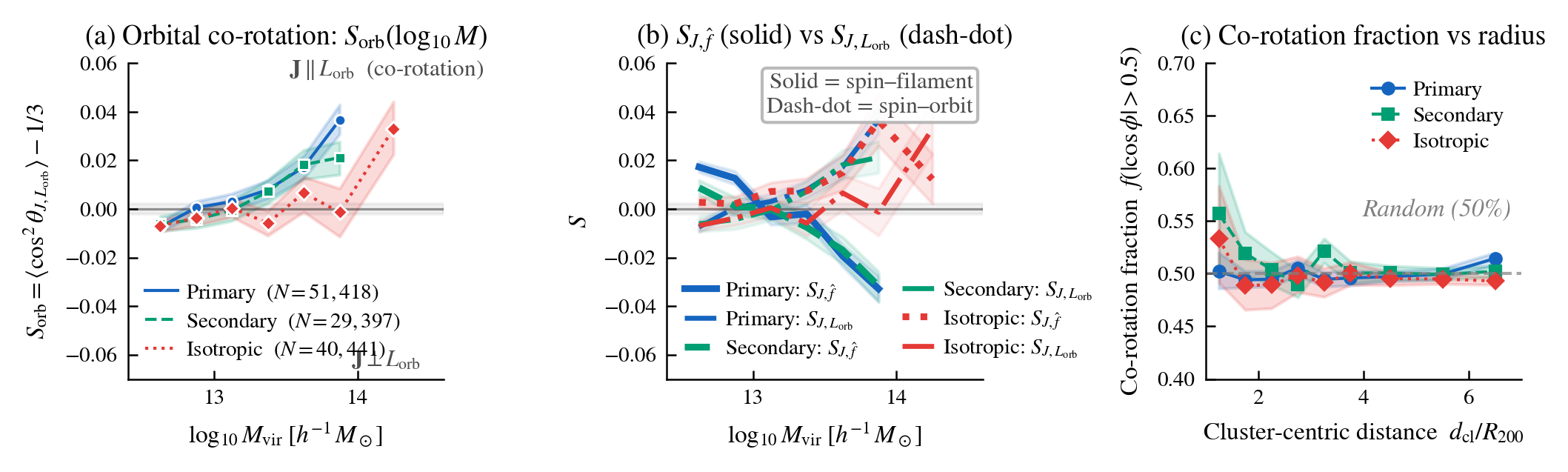}
    \caption{Spin-orbital co-rotation. \textit{(a)} Co-rotation statistic $S_\mathrm{orb}(\log_{10} M) = \langle\cos^2\theta(\mathbf{J},\hat{L}_\mathrm{orb})\rangle - 1/3$ for primary (dark blue), secondary (green), and isotropic (red) groups. $S_\mathrm{orb} > 0$: spin parallel to orbital angular momentum; $S_\mathrm{orb} < 0$: spin perpendicular. \textit{(b)} Overlay of $S_{J,\hat{f}}$ (solid) and $S_{J,L_\mathrm{orb}}$ (dash-dot), showing the mirror relationship. \textit{(c)} Co-rotation fraction ($|\cos\theta(\mathbf{J},\hat{L}_\mathrm{orb})| > 0.5$) vs cluster-centric distance; dashed line: random expectation (50 per cent).}
    \label{fig:corotation}
\end{figure*}

The radial self-similarity of $\mflip$ suggests that the spin flip is a physical feature of the tidal environment rather than an artefact of the alignment statistic. A further test of this is provided by the spin-orbital co-rotation. For groups streaming along filaments, the orbital angular momentum is
\begin{equation}
    \mathbf{L}_\mathrm{orb} \approx rv\,(\hat{r} \times \hat{f}),
    \label{eq:Lorb}
\end{equation}
which is perpendicular to both $\hat{r}$ and $\hat{f}$. Replacing $\hat{f}$ with $\hat{L}_\mathrm{orb}$ in equation~\ref{eq:S} therefore yields a geometric complement: any spin flip in $S(\log_{10} M)$ should produce a mirror-image transition in $S_\mathrm{orb}(\log_{10} M)$ at the same mass. We compute $S_\mathrm{orb}$ for the full sample with a defined spin and orbital angular momentum ($N = 51{,}418$ primary, $29{,}397$ secondary, $40{,}441$ isotropic group-cluster pairs). The result (Fig.~\ref{fig:corotation}a,b) shows the expected mirror symmetry: $S_\mathrm{orb}$ is negative at low mass ($\mathbf{J} \perp \mathbf{L}_\mathrm{orb}$) and positive at high mass ($\mathbf{J} \parallel \mathbf{L}_\mathrm{orb}$), crossing zero at the same $\mflip$ as the filament alignment. Below $\mflip$, group spin reflects its tidal birth environment (aligned with the filament wall); above $\mflip$, spin is co-planar with the orbital plane of cluster infall. The coincidence of both zero-crossing masses indicates that $\mflip$ is a characteristic of the tidal environment rather than a projection effect.

The co-rotation fraction ($|\cos\theta(\mathbf{J},\hat{L}_\mathrm{orb})| > 0.5$; Fig.~\ref{fig:corotation}c) is consistent with the $50$ per cent random expectation for primary channelled groups ($0.49$--$0.50$ at $r \lesssim 2\,R_{200}$) and approximately flat with radius for all three classes. This coarse fraction averages over mass and is therefore not itself elevated. The physically meaningful co-rotation signal is instead the mirror symmetry between $S_\mathrm{orb}(\log_{10} M)$ and the filament-alignment statistic $S(\log_{10} M)$ (Fig.~\ref{fig:corotation}a,b), which share a common zero-crossing at $\mflip$.

\subsection{Spin flip mass and filament quality}
\label{sec:spinflip_quality}

Having found that thin-tendril and dense-bridge primary groups are indistinguishable in their bulk infall kinematics (Section~\ref{sec:mass_dependence}), we now test whether they differ in the tidal-torque environment imprinted on the group spin flip mass.

\begin{figure}
    \centering
    \includegraphics[width=\columnwidth]{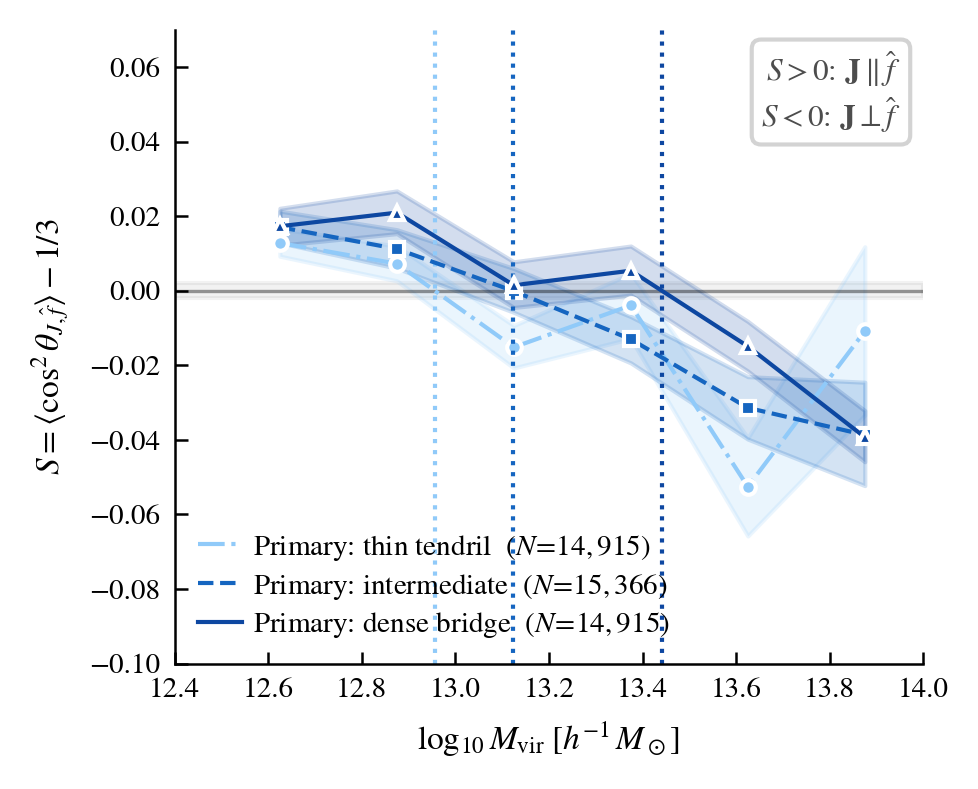}
    \caption{Spin-filament alignment $S(\log_{10} M)$ for primary groups stratified by filament linear halo mass density $\rho_\mathrm{lin} = \Sigma M_\mathrm{vir}/L$ (Table~\ref{tab:symbols}), split into three density tertiles (thin tendrils: light blue; intermediate: mid blue; dense bridges: dark blue). Vertical dotted lines mark the interpolated $\mflip$ per tertile. The $0.49$-dex spread in $\mflip$ within the primary class ($12.96$ to $13.44$) shows that the topological order-1 class contains groups on filaments of very different density. Tertile boundaries are the 33rd and 67th percentiles of the primary-filament $\rho_\mathrm{lin}$ distribution (Section~\ref{sec:spinflip_quality}).}
    \label{fig:filament_quality_spinflip}
\end{figure}

Figure~\ref{fig:filament_quality_spinflip} stratifies $S(\log_{10} M)$ for primary groups by the linear halo mass density of their connected filament ($\rho_\mathrm{lin}=\Sigma M_\mathrm{vir}/L$, measured on filament segments outside a fixed $2\,h^{-1}$\,Mpc cluster-core exclusion so that it is independent of $R_{200}$; tertiles: $\rho_\mathrm{lin} < 5.9\times10^{12}$, $5.9\times10^{12}$-$1.2\times10^{13}$, and $>1.2\times10^{13}\,h^{-1}M_\odot\,(h^{-1}\mathrm{Mpc})^{-1}$; boundaries are the 33rd and 67th percentiles of the $\rho_\mathrm{lin}$ distribution over primary group--cluster pairs and carry no independent physical significance). This split uses the clean, dominant-cluster sample, so each group enters it at most once. Groups on a common filament share its density (the tertiles contain $11{,}821$, $10{,}617$ and $9{,}490$ filaments). However, using whole filaments as the resampled unit leaves the uncertainties below unchanged (Appendix~\ref{sec:appendix_independence}). The three density tertiles trace a monotonic sequence in $\mflip$. Thin-tendril primary groups (lowest density tertile, $N=14{,}915$) have a spin flip mass $\log_{10}\mflip = 12.96$, the lowest of any channelled sub-population. Dense-bridge primary groups (highest density tertile, $N=14{,}915$) have $\log_{10}\mflip = 13.44$, $0.49\,\mathrm{dex}$ above the thin-tendril value. The intermediate tertile ($N=15{,}366$) falls between the two at $\log_{10}\mflip = 13.12$.

The primary $\mflip = 13.07$ reported in Section~\ref{sec:spin_flip} is the zero crossing of $S$ for all primary groups together, so it combines groups on filaments of very different density: dense bridges, which are cluster-feeding arms and plausibly sample a strong node-type tidal environment, and thin tendrils, which are topologically order-1 (crossing the $1.5\,R_{200}$ shell) but flip at the lowest mass of any channelled sub-population. The $\mflip$ of dense-bridge primaries is $0.32\,\mathrm{dex}$ above the secondary value ($\mflip^\mathrm{sec} = 13.12$), consistent with the environment of a massive bridge being more strongly tidal than that of a saddle-point tributary.

More broadly, $\rho_\mathrm{lin}$ provides a refined classification variable: ordering groups by the mass density of their host filament, rather than simply by topological order, produces a monotonic $\mflip$ sequence that spans $0.49\,\mathrm{dex}$ within the primary class alone. We adopt the mass-weighted $\rho_\mathrm{lin} = \Sigma M_\mathrm{vir}/L$ as the primary variable because the tidal field that drives the spin flip is sourced by mass rather than by halo number, and because a mass sum is far less sensitive than a count to the halo-mass resolution limit. The weighting matters: a simple halo count per unit arc length (a linear halo number density), although rank-correlated with the mass density (Spearman $0.79$), gives a weaker and non-monotonic $\mflip$ trend (thin/intermediate/dense $= 12.98/13.15/13.09$), with only $0.11$\,dex between the densest and thinnest thirds. The modulation is therefore a property of the mass density specifically, not of the number of haloes along the filament.

Because $\rho_\mathrm{lin}$ correlates with other properties of the group and its surroundings, we tested whether the trend is a proxy for one of them (Appendix~\ref{sec:appendix_rholin}, Fig.~\ref{fig:robust_rholin}). Among the $45{,}196$ primary pairs with a measured density, $\log\rho_\mathrm{lin}$ correlates with filament length (Spearman $\rho_s = -0.54$) and group mass ($+0.27$), but hardly with cluster-centric distance ($-0.07$) or cluster mass ($+0.03$). The weak dependence on cluster-centric distance matters because filaments thicken toward the clusters they connect \citep{Wang2024}. A fixed $2\,h^{-1}$\,Mpc aperture could therefore in principle make $\rho_\mathrm{lin}$ track proximity to the cluster. Matching the thin-tendril and dense-bridge tertiles cell by cell in group mass, cluster mass and cluster-centric distance (${\approx}2 \times 10{,}200$ pairs) leaves a flip-mass difference of $\Delta\log_{10}\mflip = 0.47$\,dex, most of the unmatched $0.49$\,dex. This difference is the median over 200 random matching draws, and the 16th--84th percentiles of the bootstrap pooled over the draws are $0.07$--$0.58$. The difference is positive in $95$ per cent of the pooled bootstrap samples. Additionally matching filament length raises the difference to $0.63$\,dex, as expected since the length enters $\rho_\mathrm{lin}$ as its denominator. These covariates therefore do not explain the sign of the modulation, but its size is poorly constrained. The dense-bridge $S$ is close to zero at $\log_{10} M = 13.125$ ($+0.0015$; Fig.~\ref{fig:filament_quality_spinflip}), so small changes to the sample move its first zero crossing (Appendix~\ref{sec:appendix_rholin}).

We next ask whether the accretion channel also affects the orbits on which groups enter the cluster.

\begin{figure}
    \centering
    \includegraphics[width=\columnwidth]{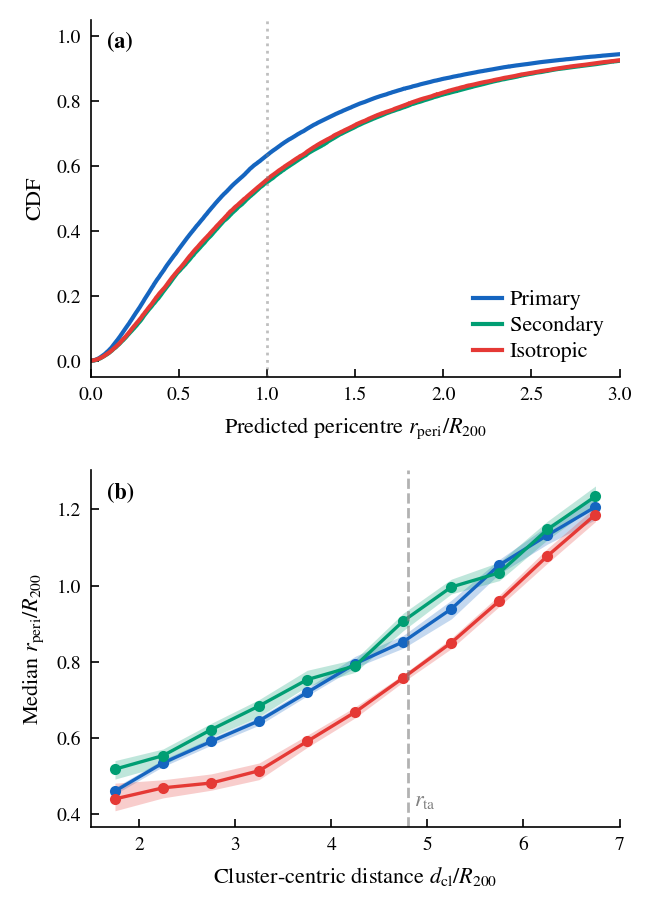}
    \caption{Predicted orbital pericentre $r_\mathrm{peri}/R_{200}$ for primary (dark blue), secondary (green), and isotropic (red) groups in the clean sample, for a static NFW cluster potential of mass $M_{200c}$. \textit{(a)} Cumulative distributions over all distances: primary groups have the smallest pooled median ($0.733\,R_{200}$, against $0.892$ for secondary and $0.876$ for isotropic groups); the dotted line marks $R_{200}$. \textit{(b)} Median $r_\mathrm{peri}/R_{200}$ in $0.5\,R_{200}$ bins of cluster-centric distance, with 16th--84th percentile bootstrap bands; the dashed line marks the turnaround radius $r_\mathrm{ta} = 4.8\,R_{200}$. At fixed distance isotropic groups have the smallest predicted pericentres. The pooled primary advantage in (a) reflects the smaller distances of primary groups (median $4.0\,R_{200}$, against $5.1$ and $5.7\,R_{200}$).}
    \label{fig:pericentre}
\end{figure}

\subsection{Predicted orbital pericentre}
\label{sec:pericentre}

To quantify whether filament channelling affects how close groups come to the cluster centre, we compute a predicted orbital pericentre for each group-cluster pair. Modelling each cluster as an NFW potential with mass $M_{200c}$ within $R_{200}$ and concentration from the \citet{Duffy2008} relation ($c_{200} = 5.71\,(M_{200}/2\times10^{12}\,h^{-1}\,\msun)^{-0.084}$, their Table~1, full sample, $z = 0$--$2$), we compute the orbital energy $E = \frac{1}{2}v^2 + \Phi_\mathrm{NFW}(r)$ and specific angular momentum $J_\mathrm{orb} = r\,v_t$ in the cluster rest frame, then find the inner turning point of the effective potential $\Phi_\mathrm{eff}(r) = \Phi_\mathrm{NFW}(r) + J_\mathrm{orb}^2/(2r^2)$. Of the $157{,}697$ clean group-cluster pairs, $148{,}233$ ($94.0$ per cent) are gravitationally bound, and the $138{,}757$ of these whose energies lie within our circular-orbit grid ($r \leq 20$\,Mpc) enter the analysis. Including the other $9{,}476$ bound pairs raises the medians below by $3$--$6$ per cent and changes none of the comparisons (Appendix~\ref{sec:appendix_pericentre}).

This is a predicted turning point in a static, spherical potential calibrated at $z = 0$, not a pericentre measured along the orbit. It neglects the growth of the cluster while the group falls in, the asphericity and substructure of the cluster, dynamical friction, and tidal mass loss. It also uses peculiar velocities only and an untruncated NFW profile. Adding the Hubble flow to $v_r$, or truncating the profile at $R_{200}$, changes the absolute medians (primary $0.71$--$0.80\,R_{200}$) but none of the comparisons below (Appendix~\ref{sec:appendix_pericentre}).

Because the subsequent pericentres of infalling groups lie in the future of the $z = 0$ snapshot, we validate the estimator on groups that have already passed pericentre (Appendix~\ref{sec:appendix_pericentre}, Fig.~\ref{fig:robust_pericentre}). We use $1{,}926$ groups whose main progenitor has already passed within the virial radius of the cluster's main progenitor. For these groups, the pericentre predicted by the same estimator at $z = 0$ and the closest approach to the cluster's main progenitor recorded in the merger trees have a Spearman rank correlation of $+0.70$. The prediction is smaller than the recorded value by a median factor of $0.89$ ($16$th--$84$th percentiles $0.88$--$0.91$), with $78$ per cent of systems within a factor of two. In physical units the median ratio is $0.97$, so most of this offset reflects the growth of $R_{200}$ since the passage. These groups lie closer to the cluster than the infalling sample analysed here and have mostly passed pericentre. The test therefore checks the estimator in a different regime; it does not calibrate it for first-infall groups. The biases of primary and isotropic groups are consistent within wide errors (the validation set contains only 74 isotropic groups). The pericentres recorded in the trees are the same for the two classes (median $0.45\,R_{200}$ for both, at the same median current distance of $1.6\,R_{200}$).

Pooled over all distances, primary channelled groups are predicted to reach the smallest pericentres (Fig.~\ref{fig:pericentre}a), with median $r_\mathrm{peri}/R_{200} = 0.733\,[0.728,\,0.737]$ (brackets: 68 per cent bootstrap intervals), against $0.892\,[0.885,\,0.898]$ for secondary and $0.876\,[0.868,\,0.882]$ for isotropic groups. This pooled contrast reflects where the classes lie rather than how they move. Predicted pericentres grow steeply with cluster-centric distance in every class (Fig.~\ref{fig:pericentre}b). Primary groups lie closer to the cluster (median distance $4.0\,R_{200}$, against $5.1$ and $5.7\,R_{200}$ for secondary and isotropic groups). Their proximity partly reflects the classification itself. Filaments converge on the cluster, so a group near it is more likely to lie close to one.

At fixed distance the primary--isotropic ordering reverses. In every distance bin between $2$ and $6\,R_{200}$ isotropic groups have the smallest predicted pericentres (Fig.~\ref{fig:pericentre}b). Reweighted to the distance distribution of the primary groups, their median is $0.660$, $10$ per cent ($95$ per cent interval $8$--$12$ per cent) below the primary $0.733$. No control for the covariates of distance (cluster and group mass, physical rather than scaled distance, groups near a single cluster, or the same $0.5\,R_{200}$ shell of the same cluster) gives the primary groups the smaller median. Inside the turnaround radius the pooled medians agree even without matching ($0.626$ and $0.613\,R_{200}$; Appendix~\ref{sec:appendix_pericentre}). The reason is the kinematic contrast of Section~\ref{sec:kinematics_results}. At a given distance isotropic groups fall in faster and more radially (for the pairs with a predicted pericentre at $3$--$4\,R_{200}$, median $v_r = -626$ against $-494\,\mathrm{km\,s^{-1}}$ and $\theta_{vr} = 26^\circ$ against $33^\circ$). As a result, in a static potential they are predicted to reach smaller pericentres. Secondary groups have the largest predicted pericentres at fixed distance ($0.761\,R_{200}$, reweighted to the primary distances). Filament channelling therefore does not deliver groups closer to the cluster centre at a given distance. We now test whether the tidal field leaves an imprint on halo morphology.

\subsection{Shape-filament alignment}
\label{sec:shape_alignment}

The tidal field that torques halo spins also stretches halo shapes along the filament. To test whether this imprint differs by accretion channel, we extract the major-axis eigenvector $\mathbf{A}$ from the \rockstar shape tensor for each group and compute $S_\mathrm{shape}(\log_{10} M) = \langle\cos^2\theta(\mathbf{A},\hat{f})\rangle - 1/3$, analogous to the spin flip statistic of equation~\ref{eq:S}.

\begin{figure*}
    \centering
    \includegraphics[width=\textwidth]{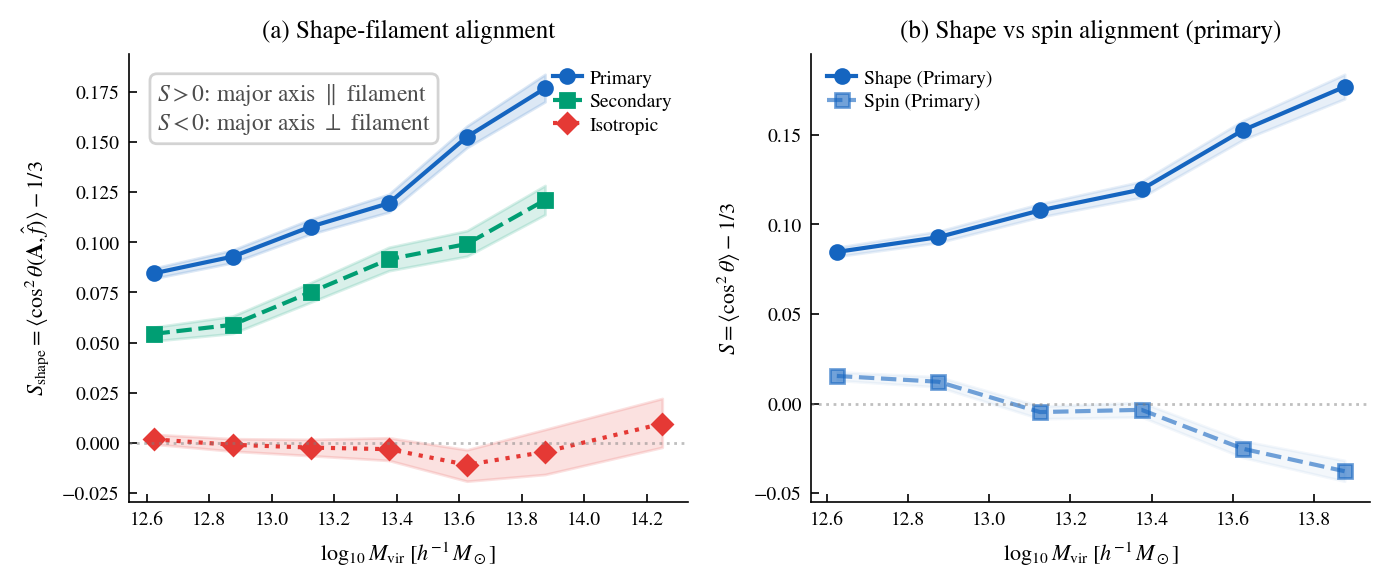}
    \caption{Shape-filament alignment by accretion channel. \textit{(a)} $S_\mathrm{shape}(\log_{10} M) = \langle\cos^2\theta(\mathbf{A},\hat{f})\rangle - 1/3$ for primary (dark blue circles), secondary (green squares), and isotropic (red diamonds) groups. Shape alignment is strongly positive for channelled groups and increases with mass; isotropic groups are consistent with random. \textit{(b)} Comparison of shape (solid circles) and spin (dashed squares) alignment for primary groups: the two signals are anti-correlated, with shape becoming more parallel and spin more perpendicular to the filament at high mass.}
    \label{fig:shape_alignment}
\end{figure*}

The shape alignment (Fig.~\ref{fig:shape_alignment}a) differs from the spin alignment. For channelled groups, $S_\mathrm{shape}$ is strongly positive at all masses ($+0.08$ to $+0.18$ for primary, $+0.05$ to $+0.12$ for secondary), indicating that halo major axes are preferentially parallel to the filament, and the alignment increases with mass. This is the opposite trend to the spin alignment, which transitions from parallel to perpendicular at $\mflip$. Isotropic groups show $S_\mathrm{shape} \approx 0$ (random alignment) at all masses, indicating that the signal is specific to filament-channelled accretion.

Panel~(b) of Fig.~\ref{fig:shape_alignment} directly compares $S_\mathrm{shape}$ and $S_\mathrm{spin}$ for primary groups. Shape and spin alignments are anti-correlated: as mass increases, the major axis becomes more parallel to the filament while the spin vector rotates perpendicular to it. This anti-correlation follows from tidal torque theory. The tidal field stretches haloes along the filament (shape $\parallel$ filament) while generating angular momentum perpendicular to the stretching axis (spin $\perp$ filament at high mass). No shape-flip mass is detected: shape alignment remains positive and monotonically increasing across the full mass range, unlike the spin which crosses zero. The velocity, spin, orbital, and shape results presented in Sections~\ref{sec:mass_dependence}--\ref{sec:shape_alignment} form a consistent picture in which the BFS topological classification separates three populations with distinct kinematics and shape alignment, with a spin flip in the two channelled classes only. We discuss the implications in Section~\ref{sec:discussion}.

\section{Discussion}
\label{sec:discussion}

\subsection{Physical interpretation of the kinematic contrasts}
\label{sec:discussion_kinematics}

The faster infall of field (isotropic) groups compared to filament-channelled groups (Section~\ref{sec:kinematics_results}, Fig.~\ref{fig:velocity_profiles}) has a direct analogue in the observational literature. \citet{Ceccarelli2011} showed that the infall velocity amplitude is largest along the void direction of the surrounding large-scale structure and smallest along the filament direction. \citet{Santucho2020} showed that galaxies streaming onto groups from underdense directions have significantly higher infall velocities ($>5\sigma$) than those arriving from overdense directions, attributing the difference to tidal braking by neighbouring masses that decelerates infall along high-density sight-lines. Our result quantifies the same effect in 3D and attributes a median ${\approx}72$\,per cent of the difference in mean $|v_r|$ to the speed contrast between populations and ${\approx}28$\,per cent to the more radially directed velocities of the isotropic groups.

Two mechanisms could produce the speed contrast, although we test neither directly. First, inter-filament groups may experience less tidal braking from neighbouring overdensities, allowing them to accumulate more infall speed by the time they reach a given cluster-centric distance \citep{Santucho2020}. Second, void expansion generates a coherent peculiar velocity component directed away from the void and toward bordering overdensities \citep[linear theory;][]{Peebles1980}, boosting the total speed of groups approaching the cluster from void-adjacent regions. The direction of motion does not compensate for the speed contrast. Although primary groups move along filaments that point almost radially, their velocities are less radially directed than those of isotropic groups at the same distance (median $\theta_{vr} = 31^\circ$ against $26^\circ$ at $2$--$3\,R_{200}$). The direction of motion therefore accounts for $9$--$29$ per cent of the difference in mean $|v_r|$ (Section~\ref{sec:kinematics_results}).

With $\varphi_\mathrm{seg} \approx 12$--$19^\circ$, the feeding filaments point, on average, to within ${\approx}15^\circ$ of the direct line to the cluster centre, nearly four times more radial than the $57.3^\circ$ expected for randomly oriented segments, so that the channels themselves approach the cluster nearly head-on rather than at a grazing angle (the velocities of the groups moving along them scatter more widely, $\theta_{vr} \approx 30$--$45^\circ$; Section~\ref{sec:filament_geometry}). That this strong radial pointing varies only mildly with radius (flat to within a few degrees across the inner infall zone) carries a further physical implication. The cluster does not substantially reorient its feeding filaments through short-range tidal forces as matter approaches the virial radius. Rather, the radial alignment is set by the large-scale gravitational collapse that initially formed both the cluster and its feeding arms; it is a fossil record of the Lagrangian tidal field during structure formation. This interpretation is supported by the approximately self-similar $\mflip(r)$ profile: the tidal environment imprinting spin on infalling groups does not change strongly as they approach the cluster, because the filament geometry itself is nearly radially self-similar throughout the infall zone.

The tangential branching of secondary filaments at saddle points ($\varphi_\mathrm{seg} > 57.3^\circ$ at $1.5\,R_{200}$) reflects the geometry of filament intersections: secondary filaments approach the junction from directions oblique to the cluster-to-cluster line, converging toward the cluster direction only at larger radii.

The 85\,per cent velocity-filament alignment fraction of primary groups at $2\,R_{200}$ follows from this tidal picture. As \citet{Bond1996} showed, both the filament orientation and the velocity field are expressions of the same primordial tidal deformation. The tidal field drives collapse along two axes (building the filament cross-section) while matter streams freely along the third, the filament spine. Groups are not threading their way toward the cluster by sensing the filament; they are simply following the tidal flow field that is already aligned with the filament spine \citep[formalised via the velocity shear tensor by][]{Libeskind2012,Tempel2014}. The nearly constant $\theta_{vf} - \theta_{vr} \approx 2$--$4^\circ$ gap at all radii (Fig.~\ref{fig:alignment_radial}b) is geometrically set by the $\varphi_\mathrm{seg} \approx 12$--$19^\circ$ filament inclination. The isotropic below-random alignment at $2\,R_{200}$ ($41.4 \pm 2.0$ per cent, $4.3\sigma$ below random) plausibly has the same tidal origin in reverse: these groups would occupy inter-filament zones where the local velocity shear directs motion away from the cluster, toward the surrounding void \citep[consistent with the split-infall velocity field measured by][]{splitinfall2025}. The three-class alignment pattern (primary far above random, secondary marginally above, isotropic below) is therefore consistent with three distinct tidal environments, the cluster node, the filament saddle and the inter-filament field, although we do not resolve a node-versus-saddle difference in spin (Section~\ref{sec:discussion_spin}).

\subsection{The spin flip in the cluster outskirts: connections to constrained tidal torque theory}
\label{sec:discussion_spin}

Our measurement of $\log_{10}\mflip = 13.07$ and $13.12$ for primary and secondary groups (with isotropic groups showing no spin flip) is, to our knowledge, the first measurement of the halo spin-flip mass relative to the filament spine within $7\,R_{200}$ of clusters with $M_\mathrm{vir} > 10^{14.0}\,h^{-1}\,\msun$. \citet{Wang2017} found a transition for haloes in Hessian-classified nodes, but measured against the axis of slowest collapse and without reference to cluster proximity. The two channelled classes flip at closely comparable mass ($0.05$-dex offset, consistent with zero); as we show below, it is the filament's linear density, not its BFS topological order, that tracks $\mflip$.

\textit{cTTT framework.} The constrained tidal torque theory \citep[cTTT;][]{Codis2015} predicts that the spin flip transition mass is set by the local tidal environment. Near filament-type saddle points, the tidal tensor is maximally anisotropic, driving $\mathbf{J} \perp \hat{f}$ at high mass and $\mathbf{J} \parallel \hat{f}$ at low mass. In the peak-background split, the transition mass scales with the local mass of non-linearity, and the overdensity near cluster node endpoints of filaments exceeds the saddle-point overdensity by a factor of $\sim$2 in the linear regime \citep{Codis2015}, enabling more massive haloes to form and elevating $\mflip$. The greater isotropy of the tidal field near nodes acts the other way. \citet{Wang2017} and \citet{WangKang2018} find halo spins more perpendicular to the large-scale structure at fixed mass, i.e.\ a lower transition mass, in Hessian nodes and low-anisotropy regions.

Our BFS topological classification maps onto this cTTT spatial gradient, with the caveat that the mapping is approximate. Primary groups span the full length of the cluster-feeding filament, from the node-adjacent inner region to the saddle-point midpoint, and therefore sample a range of tidal environments rather than a single node-type environment. Conversely, secondary groups reside on tributary filaments that join the primary network at saddle points, placing them in a predominantly saddle-type tidal environment. The flat $\mflip(r)$ profile measured in Section~\ref{sec:radial_corotation} indicates that the tidal imprint on spin alignment does not vary strongly with distance along the filament spine, implying that the primary class is not grossly biased by its node-end versus saddle-end subpopulations. A direct split of primary groups by arc-length fraction along the filament spine (Fig.~\ref{fig:radial_spatial}c) supports this: medium-mass groups ($13.0 < \log_{10} M < 13.5$) show $S > 0$ near the cluster end and $S < 0$ near the saddle end, consistent with the expected cTTT tidal gradient, but the variation is within the bootstrap uncertainties and does not significantly bias the class-averaged $\mflip$.

\textit{Filament density tracks the flip mass.} The cTTT density dependence is expressed not as a node-versus-saddle (primary-versus-secondary) offset, which we do not resolve (the difference is only $0.05$\,dex), but within the primary class itself: stratifying primary groups by filament linear mass density yields a monotonic $0.49$-dex spread in $\mflip$ ($12.96$ to $13.44$; Section~\ref{sec:spinflip_quality}), although after covariate matching its size is uncertain ($0.07$--$0.58$\,dex; Appendix~\ref{sec:appendix_rholin}).

In the peak-background split, $\mflip \propto \exp(\delta_\mathrm{LSS}/\delta_c)$, so $\Delta\log_{10}\mflip = \log_{10}(f)$ for a factor-of-$f$ density contrast. The $0.49$-dex spread corresponds to $f \approx 3$ (matched range $1.2$--$3.8$) between thin-tendril and dense-bridge environments, spanning the wide range of overdensities found among cluster-feeding filaments. This echoes the multi-scale result of \citet{AragonCalvo2014}, who found $\mflip$ to differ between filaments and sub-filaments (MMF-2 formalism), with higher $\mflip$ at the primary level. Within our cluster-connected network it is the density $\rho_\mathrm{lin}$, not the BFS order, that tracks $\mflip$. Topological order and density are correlated but not equivalent (a topologically order-1 filament can be a thin tendril whose groups flip at a markedly lower mass than those in dense bridges), which is why the primary-versus-secondary label alone leaves $\mflip$ unresolved. The absence of a spin flip in the isotropic class is the expected null: the filament tangent used for these groups lies at $d_\mathrm{conn} > 3\,h^{-1}$\,Mpc and carries no tidal information about their formation environment. While \citet{Codis2012} demonstrated the spin flip in $N$-body simulations, ours is, to our knowledge, the first mapping of $\mflip$ onto filament linear density in the cluster-outskirt environment, and the sign of the modulation survives matching in group mass, cluster mass and cluster-centric distance (Section~\ref{sec:spinflip_quality}).

\textit{Relation to measurements of angular momentum across scales.} Filaments themselves may carry angular momentum. \citet{Wang2021} report possible observational evidence for rotation of SDSS filaments. Their stacked rotation signal peaks near $100\,\mathrm{km\,s^{-1}}$ at ${\approx}1$\,Mpc from the spine, falls to zero beyond ${\approx}2$\,Mpc, and strengthens with the mass of the haloes at the filament end points. At group scales, the orbital angular momentum of member galaxies lies preferentially perpendicular to the filament spine in SDSS \citep{Rong2025} and in IllustrisTNG300-1 \citep{WangW2025}. The alignment is strongest for the richest groups in SDSS and for the most massive groups and those closest to the spine in TNG300-1. That quantity is not the halo spin measured here. \citet{WangW2025} find a mean ${\approx}38^\circ$ offset between the spin inferred from member galaxies and that of all dark-matter particles in the same haloes. Neither study finds a reversal to parallel alignment in any mass or richness bin probed. One plausible reason is that, over the mass range these studies probe (${\gtrsim}10^{12}\,h^{-1}\,\msun$), infall along a filament produces orbital angular momentum perpendicular to it. Within ${\approx}0.3\,h^{-1}$\,Mpc of the spine, satellite orbital planes are observed to contain the filament axis, although that preference disappears below $\log_{10} M_{200} \approx 11.8$ and reverses at $0.5$--$1\,h^{-1}$\,Mpc \citep{WangP2026}. The transition we locate is therefore complementary to those measurements. It is a property of the halo's own spin, and the primary-channel $\mflip = 10^{13.07}\,h^{-1}\,\msun$ lies within the mass range over which \citet{WangW2025} find the member-galaxy tracer perpendicular throughout.

Observations extend this chain to smaller scales. The spin of galaxy clusters, measured from the kinematics of their member galaxies, tends to lie perpendicular to the nearest filament and parallel to the spin of the central galaxy \citep{TangX2025}. Galaxy groups are spin-aligned with their central galaxies \citep{WangP2025}, and satellite orbits carry the imprint of the filament along which they were accreted \citep{WangP2026}. Together, these measurements trace angular momentum from filaments, through groups and clusters, down to their central galaxies. Our measurement concerns the spin of the dark-matter haloes within that chain.

\citet{Wang2017} and \citet{WangKang2018} showed that $\mflip$ depends on environment. In their analysis, a halo's spin--filament correlation is built up by anisotropic accretion and depends strongly on the anisotropy of the local tidal field. The sense of their dependence differs from ours. In their Hessian nodes and low-anisotropy regions spins are more perpendicular at fixed mass, i.e.\ the transition occurs at lower mass. We find $\mflip$ rising with filament linear density, as expected from its scaling with the local mass of non-linearity \citep{Codis2015} and as measured for thicker filaments \citep{GaneshaiahVeena2018, GaneshaiahVeena2021}. Our result measures this environmental dependence of $\mflip$ in the cluster outskirts, at fixed topological order. \citet{Tobar2026}, tracking halo spin in zoom-in simulations of ten filaments, report that the filament environment does not significantly modify halo spin alignment at fixed halo mass. Their filaments connect group-mass haloes ($10^{12.8}$--$10^{13.7}\,h^{-1}\,\msun$), and their environmental variable is distance to the filament spine. Ours connect clusters, and our variable is linear density. Their highest mass bin, which brackets our $\mflip$, is not statistically significant. The two results are therefore compatible, but a measurement of the density modulation in an independent simulation, and with an independent filament finder, would be valuable.

\textit{General field population.} The Cosmic Ballet series \citep{GaneshaiahVeena2018, GaneshaiahVeena2021} measures $\log_{10}\mflip \approx 11.0$--$12.3$ depending on filament diameter, with $\log_{10}\mflip \approx 11.7$ ($5 \times 10^{11}\,h^{-1}\,\msun$) for intermediate-thickness filaments. Our channelled values are $0.6$--$1.1$\,dex above even the thickest field filament bin ($\log_{10}\mflip \approx 12.3$). This elevation is the same peak-background effect at the cluster scale. The extreme overdensity near clusters of $M > 10^{14.0}\,h^{-1}\,\msun$ shifts $\mflip$ upward. Our measurement quantifies this elevation in the cluster-outskirt environment.

\textit{Saddle-versus-node alignment strength.} cTTT predicts that saddle environments, with their maximally anisotropic tidal fields, should imprint stronger perpendicular alignment than the more three-dimensionally isotropic node environments. We do not resolve this contrast: in the high-mass bin $\log_{10} M \in [13.5, 14.0)$ the primary and secondary alignment strengths are equal within their uncertainties ($S_\mathrm{pri} = -0.025$, $S_\mathrm{sec} = -0.023$; overlapping 95 per cent CIs), with the point-estimate ordering if anything reversed. Consistent with the negligible primary-secondary $\mflip$ offset (Section~\ref{sec:spin_flip}), our data show no significant node-versus-saddle difference in alignment strength, and we cannot claim a quantitative confirmation of this specific cTTT prediction. The reading of primary and secondary groups as sampling node-type and saddle-type environments is therefore a motivation for the classification rather than a result of this paper. The data do resolve the dependence on filament linear density.

\textit{ZOMG assembly bias.} \citet{Borzyszkowski2017} showed with 7 zoom-in haloes that inter-filament (stalled) haloes have earlier formation epochs and higher concentrations than filament-fed haloes. The right panel of Fig.~\ref{fig:spin_flip} shows a concentration ordering consistent with this at $N \approx 3$--$5\times10^4$ per class (primary $51{,}329$, secondary $29{,}371$, isotropic $39{,}260$), but with Cliff's effect sizes of only $|\delta| \approx 0.04$--$0.11$, below the conventional $|\delta| = 0.147$ threshold for a small effect though coherent in direction; the ordering is what tidal stalling predicts, but it does not by itself establish that mechanism, and other formation-history differences between the environments could produce it. The most concentrated class, the isotropic groups, is also the one that shows no spin flip. The primary--secondary $\mflip$ offset is not resolved, so we do not compare the two orderings within the channelled classes. One interpretation is that the same environmental overdensity that shifts $\mflip$ upward also delays halo formation \citep[the `arrested development' mechanism of][]{Dalal2008}, so that $\mflip$ and concentration are joint manifestations of assembly bias \citep[which can modify clustering statistics at the ${\sim}15$\,per cent level on large scales;][]{Hearin2016} and tidal torque theory.

A more direct test uses the instantaneous mass accretion rate $\dot{M}$ from \rockstar (Fig.~\ref{fig:accretion}): splitting primary groups into stripped ($\dot{M} < 0$), passive ($0 \leq \dot{M} < \mathrm{median}$), and active ($\dot{M} \geq \mathrm{median}$) subpopulations, passive groups show the lowest $\mflip$ (${\approx}13.02$), with stripped groups intermediate (${\approx}13.09$) and showing randomised spin at high mass, consistent with tidal processing destroying the tidal torque imprint. The channelled spin flip (present for both primary and secondary, absent for isotropic) persists when restricting to active accretors only (Fig.~\ref{fig:accretion}b), indicating that the accretion rate is not what drives the topological signal.

\subsection{Orbital pericentre: location rather than delivery}
\label{sec:discussion_pericentre}

The predicted pericentres do not support a picture in which filaments deliver groups deeper into clusters. Pooled over all distances, primary groups have a ${\approx}16$ per cent smaller median predicted pericentre than isotropic groups ($0.733$ against $0.876\,R_{200}$). At matched distance, however, the isotropic groups are predicted to reach the smallest pericentres (Section~\ref{sec:pericentre}). The kinematics of Section~\ref{sec:kinematics_results} account for this ordering. Field groups fall in faster and more radially at a given distance, and in a static potential a faster, more radial orbit has a smaller pericentre. The near-radial pointing of the primary filaments (Section~\ref{sec:filament_geometry}) does not translate into more radial orbits for the groups moving along them. The pericentres recorded in the merger trees show no primary advantage either (Section~\ref{sec:pericentre}, Appendix~\ref{sec:appendix_pericentre}). To our knowledge no previous study has decomposed pericentre distributions by cosmic-web topology. Pericentres, like most orbital quantities, scale with cluster-centric distance, and groups near the cluster are more likely to lie close to a filament. Classes defined by the cosmic web must therefore be compared at matched distance. The same caution would apply to observational comparisons of filament and field galaxies around clusters.

The predicted pericentres are broadly consistent with previous studies of infalling satellite orbits. \citet{Tormen1997} and \citet{Gill2005} measured median first-infall pericentres of $r_\mathrm{peri}/r_\mathrm{vir} \approx 0.2$--$0.4$ from $N$-body cluster simulations; \citet{Iannuzzi2012} found similar distributions using semi-analytic models; and \citet{Wetzel2011} showed that orbital parameters are only weakly dependent on satellite mass but vary with host halo mass. Our pooled medians ($0.73$--$0.89\,R_{200}$) are larger than the subhalo-scale values reported in these earlier works, as expected for group-mass objects at predicted first infall rather than deeply embedded satellites on later passages.

\citet{Haggar2023} showed that galaxy groups do not survive cluster infall as coherent structures, losing mass rapidly after pericentric passage. The pooled median predicted pericentres of all three classes ($0.73$--$0.89\,R_{200}$) lie within the ram-pressure stripping zone identified by observations of jellyfish galaxies, which are concentrated at $0.5$--$1.0\,R_{200}$ in clusters \citep{Jaffe2018,Poggianti2017}, so groups arriving by every route are predicted to pass through it. Our results do not single out filament-channelled infall as a preferred pathway to that zone.

\subsection{Shape-filament alignment and the tidal stretching hierarchy}
\label{sec:discussion_shape}

The shape-filament alignment (Section~\ref{sec:shape_alignment}) provides a complementary test of the tidal origin of the spin flip. The monotonically increasing $S_\mathrm{shape}(\log_{10} M)$ with no zero-crossing is the expected counterpart of the spin flip in tidal torque theory: the same large-scale tidal field that stretches haloes so that their major axis lies along the filament ($\mathbf{A} \parallel \hat{f}$) generates angular momentum perpendicular to that stretching axis ($\mathbf{J} \perp \hat{f}$ at high mass). The anti-correlation between $S_\mathrm{shape}$ and $S_\mathrm{spin}$ shown in Fig.~\ref{fig:shape_alignment}(b) is consistent with this geometric relationship.

More massive haloes have undergone more tidal processing and are more elongated along the filament, increasing $S_\mathrm{shape}$. The vanishing signal for isotropic groups ($S_\mathrm{shape} \approx 0$) indicates that the alignment requires a coherent tidal field anchored to a filament, not merely proximity to a cluster. The primary-over-secondary ordering ($S_\mathrm{shape,pri} > S_\mathrm{shape,sec}$) mirrors the kinematic hierarchy and is consistent with a stronger tidal field in the cluster-feeding environment than around saddle-point tributaries, although we do not resolve the corresponding contrast in spin (Section~\ref{sec:discussion_spin}).

Our shape-filament alignment measurements extend the field-population results of \citet{GaneshaiahVeena2018}, who used the P-Millennium simulation with the NEXUS+ formalism to show that halo major axes are preferentially parallel to their host filament, with median $\langle\cos\theta_{s_a;e_3}\rangle$ increasing from ${\approx}0.52$ at $M_{200} \sim 10^{11}\,h^{-1}\,\msun$ to ${\approx}0.65$ at $10^{13}\,h^{-1}\,\msun$. A shape alignment along filaments that strengthens with halo mass was likewise found from the tidal- and velocity-shear fields by \citet{ForeroRomero2014}. Converting to our statistic, these correspond to $S_\mathrm{shape} \approx 0.01$--$0.10$ for the general filament population. Our primary channelled groups show $S_\mathrm{shape} = +0.08$ to $+0.18$ over a comparable mass range, systematically higher than the field values, as would be expected if the cluster-feeding environment imposes a stronger and more coherent tidal field along the filament spine than the average field filament. The secondary (saddle-point) groups, at $S_\mathrm{shape} = +0.05$ to $+0.12$, lie between the field and primary values, consistent with a gradient in tidal strength between the two channelled classes, although we do not resolve the corresponding contrast in spin. \citet{AragonCalvo2014} reported a similar environmental gradient for spin alignment, with the spin flip mass varying by cosmic web morphological type; our shape alignment hierarchy is the geometric complement of that finding. \citet{GaneshaiahVeena2018} also confirmed the absence of a shape flip in the P-Millennium; our results extend this to the cluster-outskirt environment.

Observationally, shape--filament alignment has been detected directly in galaxy surveys. \citet{Chen2019} measured a small but statistically significant alignment of galaxy major axes parallel to nearby filaments in the SDSS-III LOWZ sample, with the signal strongest for bright, early-forming galaxies. The Euclid Quick Data Release galaxy-alignment analysis of Laigle et al.\ \citep{EuclidQ1_2025} recently confirmed this at $0.5 < z < 0.9$: early-type galaxies show a significant preferential alignment of their major axes along projected filaments, with the signal increasing with stellar mass. These observational trends are qualitatively consistent with our simulation results, supporting the tidal stretching origin. The vanishing signal for isotropic groups ($S_\mathrm{shape} \approx 0$) further supports the tidal interpretation: groups that are not embedded in a coherent filament-scale tidal field show no systematic shape alignment, arguing against cluster proximity as the driver.

\subsection{Observational prospects}
\label{sec:discussion_obs}

The signals reported here differ in observational accessibility. The most readily testable is the kinematic alignment contrast. The 44-percentage-point difference between primary and isotropic velocity-filament alignment fractions at $2\,R_{200}$ is large enough to survive substantial dilution from projection effects, filament-finder disagreement, and interloper contamination. Spectroscopic cluster surveys extending to ${\sim}3\,R_{200}$ (such as the WEAVE cluster programme \citep{Cornwell2022} or DESI group catalogues around \textit{eROSITA}-selected clusters) could measure the amplitude of this contrast. A qualitatively simpler diagnostic is the sign of the radial velocity contrast: isotropic groups infalling $150$--$180\,\mathrm{km\,s^{-1}}$ faster than primary channelled groups (Section~\ref{sec:kinematics_results}), opposite to the naive expectation, is testable even in modest cluster samples.

The spin flip hierarchy is more demanding. The main falsifiable prediction is the rise of $\mflip$ with filament linear mass density within the channelled population, together with the absence of a spin flip in the isotropic class. Finding no density dependence, or a spin flip in the inter-filament population, would rule out the tidal-torque picture. HI 21-cm morphological position angles offer a spin-direction proxy accessible to wide-area surveys such as WALLABY \citep{Koribalski2020} and future SKA H\,{\sc i} programmes \citep{Blyth2015}. \citet{Tudorache2025} recently demonstrated the approach with MeerKAT, measuring the spin--filament alignment of galaxies along a $15$-Mpc filament. The sample sizes required for mass-binned detection nonetheless remain large. Integral-field spin measurements provide a more direct probe: \citet{Krolewski2019} searched for a spin--filament alignment in the MaNGA survey but found no significant signal, with only marginal ($2$--$3\sigma$) hints of the expected mass dependence, illustrating the large samples such a detection will require. All predictions are derived from the $z = 0$ snapshot; at $z \sim 0.3$--$0.5$, $\mflip$ is expected to shift by ${\sim}0.2$--$0.4$\,dex \citep{GaneshaiahVeena2021}, and the kinematic contrasts may weaken as the filament network is less evolved. These predictions should therefore be treated as qualitative guides rather than quantitative benchmarks until multi-snapshot calibrations are available.

\subsection{Caveats and limitations}
\label{sec:caveats}

\begin{figure*}
    \centering
    \includegraphics[width=\textwidth]{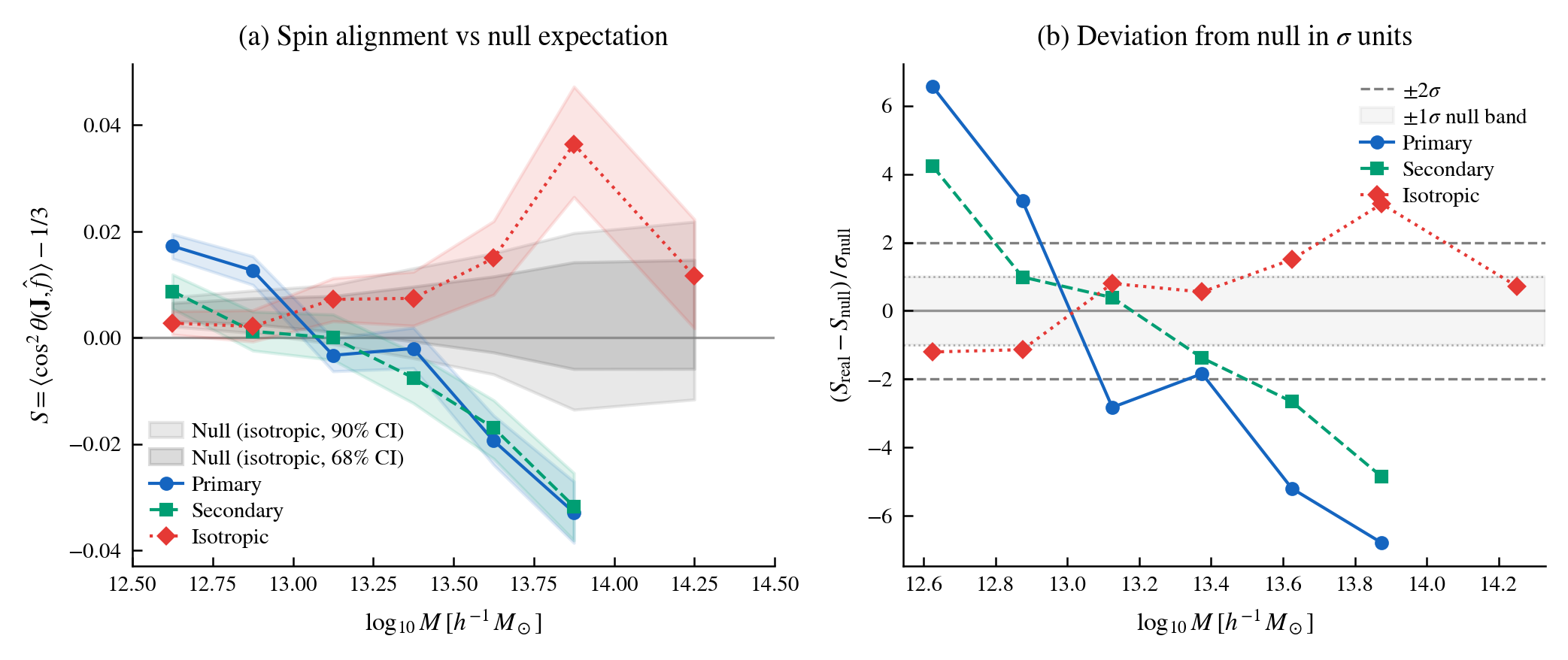}
    \caption{Null test for the spin-filament alignment statistic (500 permutation realisations). \textit{(a)} $S(\log_{10} M)$ for all three accretion classes (real tangents) with $\pm1\sigma$ bootstrap bands; grey bands: 68\,per cent and 90\,per cent CI of the null test for isotropic groups. \textit{(b)} Deviation from null in $\sigma$ units per class. Primary (dark blue) and secondary (green) deviate $>2\sigma$ at most masses, indicating a spin-filament signal. Isotropic (red) shows a marginal positive deviation at high mass ($+3.1\sigma$), consistent with weak residual large-scale tidal correlations and no spin flip.}
    \label{fig:null_test}
\end{figure*}

\textit{Isotropic spin control sample.} For isotropic groups, the filament tangent used to compute $\theta(\mathbf{J}, \hat{f})$ is the nearest cluster-connected filament at $d_\mathrm{conn} > 3\,h^{-1}$\,Mpc, which is not physically interacting with the group. Figure~\ref{fig:null_test} shows the full null test from 500 permutation realisations. Panel~(b) quantifies the deviation from null in units of $\sigma_\mathrm{null}$: primary groups deviate by $+6.6\sigma$ at $\log_{10} M \approx 12.6$ and by $-6.8\sigma$ at $\log_{10} M \approx 13.9$, confirming a real signal (parallel alignment at low mass, perpendicular at high mass) at both ends. Secondary groups show a significant high-mass deviation ($-4.9\sigma$ at $\log_{10} M \approx 13.9$). Isotropic groups show only a marginal positive deviation at high mass ($+3.1\sigma$ at $\log_{10} M \in [13.75, 14.0)$, positive $S$) and no sign reversal. The positive $S$ values at high mass ($N = 888$ at $\log_{10} M \in [13.75, 14.0)$) suggest a weak residual large-scale tidal alignment even for groups not proximate to any cluster-connected filament. This is physically plausible, since the large-scale density field that created the filament network also weakly imprints spin alignment on distant groups, and the consistently positive $S$ (parallel alignment) is consistent with the tidal torque prediction for groups that have not undergone the environmental processing that drives the spin flip.

\textit{Backsplash identification.} Backsplash groups are identified via merger-tree tracing as described in Section~\ref{sec:classification}. Excluding the self-pairs, confirmed backsplash groups make up only $1.2$\,per cent of the remaining dominant pairs (1,838 of 159,535 with the corrected flags; $1.0$\,per cent of all dominant pairs), $6$\,per cent of the pairs within $3\,R_{200}$ and $0.08$\,per cent of those beyond $2\,R_{200}$, but $93$\,per cent of the outward-moving pairs inside $1.5\,R_{200}$. They are concentrated just outside the splashback radius \citep[${\sim}1$--$1.5\,R_{200}$;][]{More2015}, with 16th--84th percentiles of $1.40$--$1.80\,R_{200}$ and 7\,per cent beyond $2\,R_{200}$. Two thirds are primary-outskirt groups and a quarter primary, with secondary and isotropic groups contributing 3 and 4\,per cent. This composition is largely geometric: backsplash groups lie at small cluster-centric radii, where the cluster's own filaments converge and the primary and primary-outskirt classes dominate.

Because the channelled and isotropic infall samples are populated almost entirely by first-infall groups, the accretion-channel contrasts reported here are not driven by returning backsplash orbits. Replacing the backsplash flags of the submitted version with the corrected ones (a net change of 4 pairs in the clean sample of $157{,}697$) shifts the quoted results by at most a few units in their last digit. For example, the intermediate tertile of the count-density test of Section~\ref{sec:spinflip_quality} moves from $13.12$ to $13.15$. The exception is the $v_r < 0$ sensitivity test of Section~\ref{sec:classification}, whose largest isotropic change, in the innermost bin ($1.5$--$2\,R_{200}$), falls from $27$ to $11\,\mathrm{km\,s^{-1}}$.

Table~\ref{tab:diagnostics} collects the main diagnostic checks discussed above and in the appendices.

\begin{table}
  \centering
  \caption{Summary of robustness diagnostics. Each row lists a test, the parameter varied or quantity checked, and the impact on the main results ($\mflip$ values, velocity contrast, alignment fraction).}
  \label{tab:diagnostics}
  \small
  \setlength{\tabcolsep}{4pt}
  \begin{tabular}{@{}>{\raggedright\arraybackslash}p{0.27\columnwidth}>{\raggedright\arraybackslash}p{0.28\columnwidth}>{\raggedright\arraybackslash}p{0.37\columnwidth}@{}}
    \hline
    Diagnostic & Variation & Impact \\
    \hline
    $d_\mathrm{conn}$ threshold & $0.5$--$2.0\,h^{-1}$\,Mpc & Primary $\mflip$ shift ${\sim}0.04$\,dex; secondary shift ${\sim}0.3$\,dex \\
    $d_\mathrm{iso}$ threshold & $2.0$--$5.0\,h^{-1}$\,Mpc & No stable iso.\ flip (absent at fiducial $3.0$) \\
    BFS $\epsilon$ tolerance & $0.3$--$1.0\,h^{-1}$\,Mpc & Near-equality preserved; sec.\ $\mflip$ $13.04$--$13.16$ \\
    $\kappa$ splits & $\kappa \leq 2$, $3$--$4$, $\geq 5$ & $S$ consistent within CI \\
    Low-mass floor & drop lowest bin ($\log_{10} M < 12.75$) & Both channelled $\mflip$ unchanged \\
    Backsplash removal & Merger-tree tracing & $1.0$\% of dominant pairs removed as confirmed backsplash (plus self-pairs); flags re-derived on the main branch and against $R_{200}$ (4-pair change to the clean sample) \\
    Pair independence & Unique-group and cluster-block bootstrap & $\sigma_S$ $+2$--$9$\% (groups), ${\lesssim}3$\% (clusters); $\mflip$ intervals unchanged within bootstrap noise; density split unchanged with filament blocks \\
    $\rho_\mathrm{lin}$ covariates & Matched in $M_\mathrm{gr}$, $M_\mathrm{cl}$, $d_\mathrm{cl}$ & $\Delta\log_{10}\mflip = 0.47$ ($0.07$--$0.58$) of $0.49$\,dex retained (median of 200 matching draws); sign favoured, size uncertain \\
    Pericentre estimator & Merger-tree pericentres (backsplash groups) & $\rho_s = +0.70$; median bias $0.89$ ($0.97$ in physical units); recorded medians equal for primary and isotropic groups ($N_\mathrm{iso} = 74$) \\
    Pericentre at fixed $d_\mathrm{cl}$ & Reweighting and matching in $d_\mathrm{cl}$, $M_\mathrm{cl}$, $M_\mathrm{gr}$ & No primary advantage in any match; isotropic up to $10$\% below primary \\
    Through-going filaments & Double crossings & excluded by BFS \\
    Strict kinematic cut & $v_r < 0$ \& $\theta_\mathrm{fil} < 60^\circ$ & Hierarchy unchanged \\
    Mass matching & $\log_{10} M \in [12.5, 13.5]$ & $\Delta v_r$ changes $\leq 8$\,km\,s$^{-1}$ \\
    Dominant-cluster assign. & Distance vs $v_r$ criterion & Iso-faster ordering preserved; contrast somewhat reduced \\
    \hline
  \end{tabular}
\end{table}

\textit{Snapshot limitation.} This analysis uses the $z = 0$ snapshot only. The spin flip mass evolves as $\mflip \propto M_*(z)$ \citep{GaneshaiahVeena2021}; the density-dependent modulation of $\mflip$ may also evolve if the filament density field changes with redshift. Multi-snapshot analyses would quantify this evolution. The single-snapshot limitation also propagates into the predicted pericentres (Section~\ref{sec:pericentre}). The turning-point calculation uses the instantaneous $z=0$ positions and velocities, so the pericentres are predictions under the frozen $z=0$ potential rather than orbits followed through cosmic time, and would shift if the clusters' mass growth between infall and pericentre were modelled. Appendix~\ref{sec:appendix_pericentre} quantifies the resulting bias against pericentres recorded in the merger trees.

\textit{Through-going filaments.} A filament that crosses the $1.5\,R_{200}$ shell can in principle pass through the cluster potential and out the far side (a double crossing). Such through-going arcs are excluded by construction. The BFS expansion traces connectivity only from the cluster maximum outward, so arcs whose endpoints lie elsewhere in the web are never classified as primary.

\textit{Dark-matter-only approximation.} MDPL2 is a gravity-only simulation; no baryonic processes (gas cooling, star formation, AGN/supernova feedback) are modelled. No hydrodynamical simulation currently exists at the 1\,Gpc scale with sufficient resolution for the statistical analysis performed here (27{,}029 clusters, 212{,}704 group--cluster pairs). Existing hydro comparisons suggest that baryons modify the magnitude but not the topology or qualitative ordering of the signals we measure. 
\citet{KraljicDave2020} found the same spin-flip transition in the SIMBA hydrodynamical simulation as in dark-matter-only runs, indicating that baryons modify the magnitude of the alignment but not the existence or sense of the transition. \citet{Laigle2015} confirmed consistent spin--filament alignment trends in hydrodynamical simulations. 
\citet{Sunseri2022} applied NEXUS to IllustrisTNG-300 (hydro vs.\ DM-only) and found that the large-scale filament topology is preserved, with volume fractions changing by $<1$\,per cent and filament density distributions boosted by ${\sim}10$\,per cent from expelled halo gas. 
The topological classification and tidal torque mechanisms identified here are gravitational and operate on scales $>1$\,Mpc where baryons are subdominant. The absolute values of $\mflip$, velocity contrasts, and channelled fractions should be treated as approximate, with baryons expected to shift the velocity contrasts at the ${\sim}10$--$20$\,per cent level and to modify $\mflip$ without altering the qualitative ordering, based on these comparisons.

\textit{Filament-finder dependence.} The entire accretion-channel classification rests on the \disperse skeleton at persistence threshold $n_\sigma = 6.5$. Different filament finders (the Multiscale Morphology Filter \citep[MMF;][]{AragonCalvo2007, AragonCalvo2010}, NEXUS+ \citep{Cautun2014, AragonCalvo2014}, the velocity-shear V-web \citep{Hoffman2012, Libeskind2012, Libeskind2013}, and the stochastic Bisous model \citep{Tempel2014}) produce different skeletons with different filament populations, and the BFS hierarchy is defined relative to the \disperse skeleton specifically. The bootstrap statistical uncertainties quoted throughout this paper ($\pm 0.05$\,dex on the primary $\mflip$, larger for the less-constrained secondary; $\pm 14\,\mathrm{km\,s^{-1}}$ on velocity contrasts) reflect sampling noise only. The systematic uncertainty from filament-finder choice almost certainly dominates.

Our internal sensitivity test (varying $n_\sigma$; Section~\ref{sec:caveats}) shows $\mflip$ shifts of $\lesssim 0.05$\,dex for the primary and up to $0.12$\,dex for the secondary, a lower bound on the external systematic, since it varies only one parameter within the same method. The full systematic from switching methods entirely is conservatively estimated at ${\sim}0.2$--$0.5$\,dex for $\mflip$ and ${\sim}30$--$50$\,per cent for the channelled fraction, based on the diversity of filament populations reported across methods in the literature \citep[see also][]{Wang2017}. The qualitative kinematic contrasts and the filament-density modulation of $\mflip$ are expected to be robust because they trace the underlying tidal field gradient regardless of how filaments are identified, but the absolute $\mflip$ values and sample fractions should be interpreted as indicative rather than as precise values.

The expected direction of the systematic is that methods which produce thicker or more numerous filaments (e.g.\ NEXUS+) would shift more groups into the channelled class, increasing the channelled fraction and potentially diluting the kinematic contrast by including groups with weaker filament coupling. Conversely, sparser skeletons (e.g.\ high-persistence \disperse or Bisous) would sharpen the contrast at the cost of sample size. Quantifying this dependence requires applying the BFS classification to multiple skeleton catalogues, which is beyond the scope of this paper.

\textit{BFS topology sensitivity.} The BFS classification depends on the persistence threshold $n_\sigma$ and the endpoint proximity tolerance $\epsilon$. Sensitivity tests varying $d_\mathrm{conn}$, the isotropic distance threshold $d_\mathrm{iso}$ (the $d_\mathrm{conn}$ value above which a group is labelled isotropic, $3\,h^{-1}$\,Mpc fiducially), and $\epsilon$ (Appendix~\ref{sec:appendix_sensitivity}) show that the channelled spin flip persists, with the primary $\mflip$ shifting by $\lesssim 0.04$\,dex and the less precisely determined secondary $\mflip$ by up to $0.3$\,dex (at $d_\mathrm{conn} = 0.5\,h^{-1}$\,Mpc) between parameter choices. The persistence threshold $n_\sigma$ itself controls which filaments exist in the skeleton and is therefore the most fundamental free parameter. Paper~I demonstrates that the filament network at $n_\sigma = 6.0$, $6.5$, and $7.0$ produces consistent connectivity statistics (slope and scatter of the $\kappa$--$M$ relation). We have verified that both channelled classes retain their spin flip at comparable mass at both $n_\sigma = 6.0$ (which adds ${\sim}15$\,per cent more cluster-connected filaments, increasing the channelled fraction) and $n_\sigma = 7.0$ (which removes ${\sim}20$\,per cent, reducing it), with the primary $\mflip$ shifting by $\lesssim 0.05$\,dex and the less precisely determined secondary $\mflip$ by up to $0.12$\,dex.

\textit{Dominant-cluster assignment.} Assigning groups by most-negative $v_r$ rather than nearest physical distance changes the cluster assignment for $16.9$\,per cent of pairs (reassigning the dominant cluster for $64$\,per cent of the multi-cluster groups). Because the most-negative-$v_r$ criterion by construction favours the fastest-infall pairing, nearest-distance assignment yields a somewhat smaller isotropic--primary velocity contrast (e.g.\ ${\sim}80$ versus ${\sim}130\,\mathrm{km\,s^{-1}}$ at $1.5$--$3\,R_{200}$), but leaves its sign and class ordering (isotropic groups falling faster than primary) unchanged, and shifts the primary population's own median $v_r$ by only ${\sim}13\,\mathrm{km\,s^{-1}}$. No conclusion depends on the choice of criterion.

\textit{Spin convergence at the low-mass limit.} Halo spin parameters from \rockstar are known to converge only above ${\sim}300$--$500$ particles \citep[e.g.][]{GaneshaiahVeena2018, Borzyszkowski2017}. Both channelled flip masses ($\log_{10}\mflip \approx 13.07$ and $13.12$, corresponding to ${\sim}8{,}000$ particles at MDPL2's particle mass of $1.51 \times 10^9\,h^{-1}\,\msun$) lie well above this threshold. The lowest sub-population probed (the thin-tendril density tertile at $\log_{10}\mflip = 12.96$, ${\sim}6{,}040$ particles) also remains above the convergence limit. We have verified that excluding the lowest mass bin ($\log_{10} M < 12.75$) leaves both the primary and secondary $\mflip$ unchanged (both zero-crossings lie above this floor) and preserves the finding that the two channelled classes flip at comparable mass. Particle-noise in the spin vector adds symmetric scatter to $\cos^2\theta$, biasing $S$ mildly toward zero at the lowest masses, but does not affect the class-averaged channelled flip masses or the $0.49$-dex density modulation.


\section{Conclusions}
\label{sec:conclusions}

We have used the \disperse\ (Discrete Persistent Structures Extractor) filament catalogue of the $(1\,h^{-1}\,\mathrm{Gpc})^3$ MDPL2 simulation to classify $212{,}704$ group-cluster infall pairs into primary channelled (51,444), secondary channelled (29,408), primary-outskirt (88,446), and isotropic (43,406) accretion classes, using a breadth-first-search (BFS) topological classification that distinguishes groups in cluster-feeding filaments (order-1), saddle-point tributary filaments (order-2), a primary-outskirt transition zone, and the inter-filament field. Three results are new measurements:

\begin{itemize}

\item Both channelled classes flip at comparable mass, well above the field. $\log_{10} \mflip\;[h^{-1}\,\msun] = 13.07$ (primary) and $13.12$ (secondary), to our knowledge the first measurement of the halo spin-flip mass relative to the filament spine within $7\,R_{200}$ of clusters with $M_\mathrm{vir} > 10^{14.0}\,h^{-1}\,\msun$, $1.3$-$1.4$\,dex above the general field population. Isotropic groups show no spin flip. The quoted uncertainties are not underestimated by groups that approach more than one cluster (Appendix~\ref{sec:appendix_independence}). The $\mflip(r)$ profile and the filament inclination $\varphi_\mathrm{seg}(r)$ are both approximately self-similar across the cluster influence zone (with only mild radial trends), consistent with a tidal environment set on large scales (Figs~\ref{fig:spin_flip}, \ref{fig:radial_spatial}).

\item Filament density, not just topology, tracks the spin-flip mass. Within the primary class alone, $\mflip$ rises by $0.49$\,dex from $12.96$ (thin tendrils) to $13.44$ (dense bridges), where filament density is the host-halo mass per unit length $\Sigma M_\mathrm{vir}/L$ (measured outside a fixed $2\,h^{-1}$\,Mpc cluster-core exclusion). When thin and dense filaments are matched in group mass, cluster mass and cluster-centric distance, the dense bridges still flip at the higher mass in $95$ per cent of bootstrap samples. The size of the difference is uncertain ($0.47$\,dex, $16$th--$84$th percentiles $0.07$--$0.58$; Appendix~\ref{sec:appendix_rholin}). A halo count per unit length in place of the mass density gives a weaker, non-monotonic trend. The primary $\mflip = 13.07$ is measured on all primary groups together, which include both cluster-feeding arms and thin tendrils that are topologically order-1 but flip at the lowest mass (Fig.~\ref{fig:filament_quality_spinflip}).

\item Isotropic groups are systematically under-aligned. At $2\,R_{200}$, only $41.4 \pm 2.0$ per cent of isotropic groups have velocities within $60^\circ$ of the nearest filament axis (Fig.~\ref{fig:alignment_radial}). A within-$60^\circ$ deficit implies an excess at larger angles, so they are mildly anti-aligned.

\end{itemize}

These new measurements are accompanied by results that extend known trends to the cluster outskirts at high statistics:

\begin{itemize}

\item Velocity-filament alignment. Primary groups stream coherently along their filaments, with the velocity--filament angle staying within $\approx 2$--$4^\circ$ of the velocity--radial angle at all radii and $85$ per cent aligned within $60^\circ$ at $2\,R_{200}$. Isotropic groups infall $150$--$180\,\mathrm{km\,s^{-1}}$ faster, mostly because they move faster rather than more radially (the speed term contributes $69$--$87$ per cent of the difference in mean $|v_r|$). Weaker tidal braking and void expansion in the field are plausible causes that we do not test. Primary filaments are strongly radially aligned ($\varphi_\mathrm{seg} \approx 12$--$19^\circ$ across $1.5$-$10\,R_{200}$): they point to within ${\approx}15^\circ$ of the direct line to the cluster centre, nearly four times more radial than the $57.3^\circ$ of random orientation, a fossil of the large-scale collapse geometry. The velocities of the groups moving along them are nonetheless less radially directed than those of isotropic groups at the same distance (Figs~\ref{fig:velocity_profiles}, \ref{fig:filament_inclination}). The pooled pericentre contrast therefore reflects distance, not channel. In a static NFW potential primary groups have the smallest pooled median ($0.733\,R_{200}$, against $0.892$ and $0.876$ for secondary and isotropic groups) only because they lie closer to the cluster. At matched distance isotropic groups are predicted to reach pericentres up to $10$ per cent smaller, so filament channelling does not deliver groups deeper into clusters. The pericentres recorded in the merger trees for backsplash groups are the same for primary and isotropic groups, although only $74$ isotropic groups enter (Fig.~\ref{fig:pericentre}, Section~\ref{sec:pericentre}, Appendix~\ref{sec:appendix_pericentre}).

\item Tidal origin of the spin flip. The spin flip is the mass $\mflip$ at which a group's spin switches from parallel to its filament (low-mass haloes, spun up within the plane of the collapsing wall) to perpendicular (high-mass haloes, torqued by material accreted along the filament). A denser, more anisotropic tidal field drives this crossover to higher mass. Both channelled classes show comparable perpendicular spin alignment at high mass. We do not resolve the node-versus-saddle contrast predicted by cTTT, but the rise of $\mflip$ with filament density within the primary class, in which the densest bridges flip at roughly three times the group mass of the thinnest tendrils (with a large uncertainty after covariate matching), is consistent with this tidal-torque dependence. The spin-orbital co-rotation transition occurs at the same $\mflip$ for each class. The NFW concentration ordering (isotropic~$>$~primary~$>$~secondary) is in the direction of the assembly-bias counterpart of the spin-flip result, in which inter-filament groups, with no filament to feed them fresh material, stall and form earlier and end up more concentrated. It is consistent with the mechanism of \citet{Borzyszkowski2017} at $N \approx 3$--$5\times10^4$ per class, although the effect sizes are weak ($|\delta| \leq 0.11$) and the ordering does not by itself establish the mechanism (Figs~\ref{fig:spin_flip}, \ref{fig:corotation}).

\item Shape-filament alignment. Halo major axes are preferentially parallel to the filament ($S_\mathrm{shape} = +0.08$ to $+0.18$), with alignment increasing with mass, opposite to the spin flip. The anti-correlation is consistent with the same tidal field stretching haloes along filaments while spinning them perpendicular to the stretching axis (Fig.~\ref{fig:shape_alignment}).

\end{itemize} 

These results are derived from the single $z=0$ snapshot of a gravity-only simulation using a single filament finder (\disperse). The qualitative ordering and physical mechanisms are expected to be robust. The absolute values of $\mflip$, velocity contrasts, and channelled fractions carry systematic uncertainties that likely exceed the quoted statistical errors. These come from the dark-matter-only approximation \citep[baryons modify magnitudes but not the qualitative ordering;][]{KraljicDave2020, Laigle2015}, the single-epoch measurement ($\mflip$ and the kinematic contrasts evolve with redshift; Section~\ref{sec:caveats}), and the choice of filament finder.

Together, these results show that the position of a group in the cluster-feeding filament network, and the linear density of its host filament, not just its proximity to a filament or its mass, are associated with systematic differences in its infall kinematics, spin alignment, and concentration. The new results (the rise of $\mflip$ with filament density and the isotropic under-alignment) are not predicted by any single existing framework, and complement the group-scale measurements of orbital angular momentum around filaments \citep{Rong2025, WangW2025} by locating the transition in the halo spin itself. They indicate that the cluster-connected filament network encodes physical information inaccessible from geometric filament distance alone. The kinematic alignment contrast is the most accessible observational test; the spin flip measurement will require larger samples from upcoming spectroscopic surveys.


\section*{Acknowledgements}

The MDPL2 simulation was carried out at LRZ Munich within the SuperMUC project. The CosmoSim database used in this paper is a service by the Leibniz-Institute for Astrophysics Potsdam (AIP). The MultiDark database was developed in cooperation with the Spanish MultiDark Consolider Project CSD2009-00064. AS acknowledges support from a University of Nottingham Vice-Chancellor's Scholarship  for Research Excellence. FRP acknowledges support from the UK Science and Technology Facilities Council (STFC) consolidated grant ST/T000244/1. MEG acknowledges support from the UK Science and Technology Facilities Council (STFC) grant ST/X000982/1.  UK acknowledges financial support from the UK Science and Technology Facilities Council (STFC; grant ref ST/T000171/1).  AK and GY acknowledge financial support from the Agencia Estatal de Investigación (AEI, Spain)  under project PID2024-156100NB-C21, funded by MCIN AEI/10.13039/501100011033.

This work made use of the University of Nottingham's Augusta HPC service.


\section*{Data Availability}

The MDPL2 simulation data are publicly available through the CosmoSim database (\url{https://www.cosmosim.org}). The filament catalogue described in Paper~I will be made publicly available upon acceptance. The accretion channel classification catalogue, including group-cluster pair assignments and topological classes, will be made available upon acceptance of this paper.


\bibliographystyle{mnras}
\bibliography{paper_refs_v2}


\appendix

\section{Sensitivity to classification thresholds}
\label{sec:appendix_sensitivity}

We test the robustness of the spin-flip hierarchy and kinematic results under three types of threshold variation. Figure~\ref{fig:sensitivity} summarises the results.

\subsubsection{Connectivity distance threshold $d_\mathrm{conn}$}

The primary/secondary channelled classification requires the group to lie within $d_\mathrm{conn} < 1\,h^{-1}$\,Mpc of a cluster-connected filament spine (Section~\ref{sec:classification}). We vary this threshold over $\{0.5,\, 1.0,\, 2.0\}\,h^{-1}$\,Mpc and compute bootstrap $\pm1\sigma$ confidence intervals on $\mflip$ at each threshold (Fig.~\ref{fig:sensitivity}a). The stricter threshold $d_\mathrm{conn} = 0.5\,h^{-1}$\,Mpc reduces sample sizes by ${\sim}48$\,per cent and shifts $\mflip$ downward: primary shifts by ${\sim}0.04$\,dex (within the $\pm0.05$\,dex bootstrap interval), while secondary shifts by ${\sim}0.3$\,dex. This secondary shift is physically consistent: groups within $0.5\,h^{-1}$\,Mpc of the filament spine are in the densest core of the filament where the tidal environment differs from the full $<1\,h^{-1}$\,Mpc sample. At $d_\mathrm{conn} = 2.0\,h^{-1}$\,Mpc the results are identical to $d_\mathrm{conn} = 1.0\,h^{-1}$\,Mpc by construction, since the primary class already requires $d_\mathrm{conn} < 1\,h^{-1}$\,Mpc and raising the ceiling admits no new groups. Only the tightening to $0.5\,h^{-1}$\,Mpc is an informative test. The near-equality of the primary and secondary flip masses (offset consistent with zero) is preserved at all thresholds.

\subsubsection{Isotropic distance threshold $d_\mathrm{iso}$}

\begin{figure*}
    \centering
    \includegraphics[width=\textwidth]{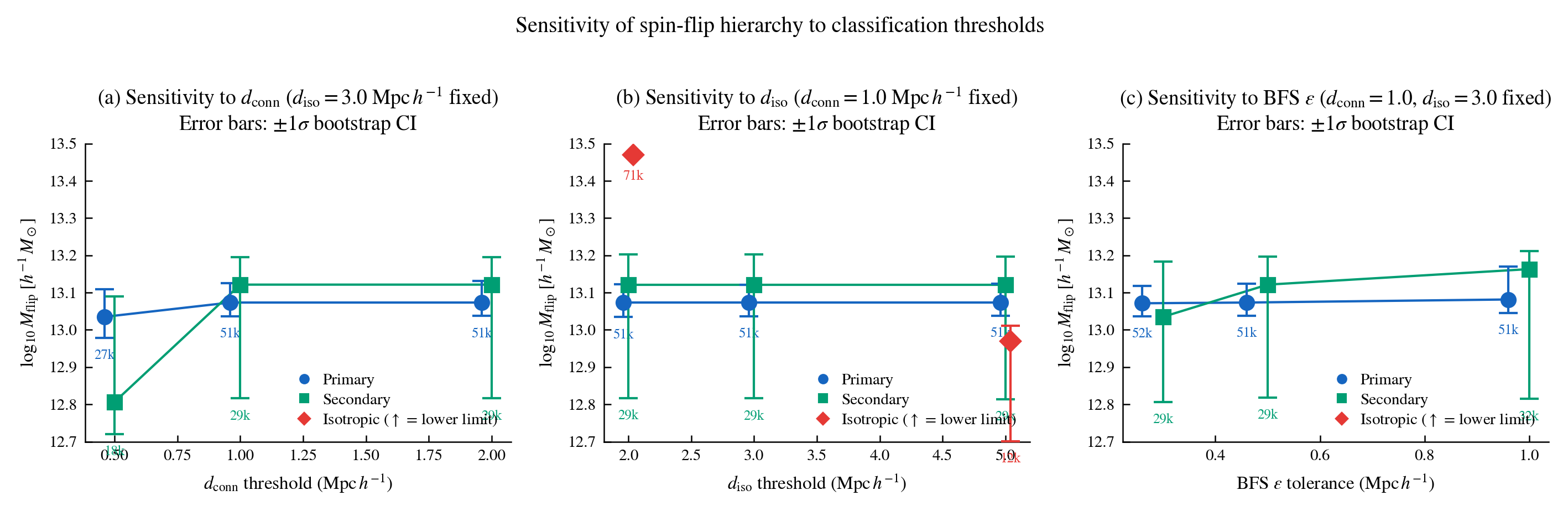}
    \caption{Sensitivity of the spin-flip hierarchy to classification thresholds, with $\pm1\sigma$ bootstrap confidence intervals on $\mflip$. \textit{(a)} $\mflip$ vs $d_\mathrm{conn}$ threshold; primary is robust (shift ${\sim}0.04$\,dex, within the $\pm0.05$\,dex bootstrap interval); secondary shifts ${\sim}0.3$\,dex at the strictest threshold. \textit{(b)} $\mflip$ vs $d_\mathrm{iso}$ threshold; the isotropic class shows no stable flip at any threshold (the interpolated crossing lies above the reliable mass range at $d_\mathrm{iso} = 2\,h^{-1}$\,Mpc, is absent at the fiducial $d_\mathrm{iso} = 3\,h^{-1}$\,Mpc, and becomes noise-dominated in a much-reduced sample at $d_\mathrm{iso} = 5\,h^{-1}$\,Mpc), supporting the null-control interpretation. \textit{(c)} $\mflip$ vs BFS endpoint tolerance $\epsilon$; primary is $\epsilon$-independent (defined by shell-crossing) and secondary varies by ${\lesssim}0.13$\,dex across $\epsilon = 0.3$--$1.0\,h^{-1}$\,Mpc. The near-equality of the primary and secondary flip masses (offset consistent with zero) is preserved at all threshold choices.}
    \label{fig:sensitivity}
\end{figure*}

Groups are classified as isotropic when their distance to any filament exceeds $d_\mathrm{iso} = 3\,h^{-1}$\,Mpc. We vary this over $\{2.0,\, 3.0,\, 5.0\}\,h^{-1}$\,Mpc (Fig.~\ref{fig:sensitivity}b). The isotropic class shows no stable spin flip at any threshold. At $d_\mathrm{iso} = 2.0\,h^{-1}$\,Mpc the interpolated zero-crossing lies at $\log_{10}\mflip \approx 14.1$, above the reliable mass range ($\log_{10} M < 13.75$, beyond which the high-mass bins are sparse), with a bootstrap confidence interval spanning the entire high-mass end. At the fiducial $d_\mathrm{iso} = 3.0\,h^{-1}$\,Mpc there is no zero-crossing at all (consistent with the $S > 0$ isotropic behaviour of Section~\ref{sec:spin_flip}). At $d_\mathrm{iso} = 5.0\,h^{-1}$\,Mpc a marginal low-mass crossing reappears (point estimate $\log_{10}\mflip \approx 13.0$) in a much-reduced, noise-dominated sample ($N \approx 12{,}000$). No threshold yields a well-defined isotropic flip within the regime probed, consistent with the $S > 0$ behaviour throughout the reliable mass range reported in Section~\ref{sec:spin_flip}. This supports treating the isotropic class as a null control sample. Primary and secondary are unchanged by construction.

\subsubsection{BFS endpoint proximity tolerance $\epsilon$}

The BFS expansion of Section~\ref{sec:connectivity_network} links filaments whose endpoints lie within $\epsilon = 0.5\,h^{-1}$\,Mpc. This tolerance controls the secondary/isotropic boundary: a filament whose outer endpoint lies $0.4\,h^{-1}$\,Mpc from a primary endpoint is classified as secondary, while one at $0.7\,h^{-1}$\,Mpc is unconnected and its groups become isotropic. We vary $\epsilon$ over $\{0.3,\, 0.5,\, 1.0\}\,h^{-1}$\,Mpc and recompute the full classification and spin-flip hierarchy.

At $\epsilon = 0.3\,h^{-1}$\,Mpc, the secondary sample shrinks by ${\sim}2$\,per cent (to $28{,}894$) as fewer tributary filaments are linked to the primary network; $\mflip^\mathrm{sec}$ shifts to $13.04$ (a $0.09$-dex downward shift, within the bootstrap CI). At $\epsilon = 1.0\,h^{-1}$\,Mpc, the secondary sample grows by ${\sim}9$\,per cent (to $32{,}180$) as more distant filaments are connected; $\mflip^\mathrm{sec}$ shifts to $13.16$. Primary $\mflip$ values are essentially unchanged ($13.07$--$13.08$; $N \approx 51{,}150$--$51{,}550$) since primary filaments are defined by shell-crossing, not BFS linking. The near-equality of the primary and secondary flip masses is preserved at all three $\epsilon$ values, the offset remaining below ${\approx}0.1$\,dex throughout. The most consequential effect of $\epsilon$ is on the secondary sample size, not on the spin-flip mass. The BFS topology is robust to the precise endpoint matching tolerance.

\subsubsection{Strict kinematic criterion}

We also test the strict kinematic criterion (Step~4 of Section~\ref{sec:classification}), requiring $v_r < 0$ and $\theta_\mathrm{fil} < 60^\circ$ in addition to spatial proximity. Of the $51{,}444$ primary groups, $67$\,per cent satisfy the strict criterion; of the $29{,}408$ secondary groups, $60$\,per cent satisfy it. The strict-criterion $v_r$ profiles and $S(\log_{10} M)$ curves are consistent with the topological results within bootstrap uncertainties, and the near-equality of the primary and secondary flip masses is unchanged.

\section{Spin alignment by accretion state}
\label{sec:appendix_accretion}

\begin{figure*}
    \centering
    \includegraphics[width=\textwidth]{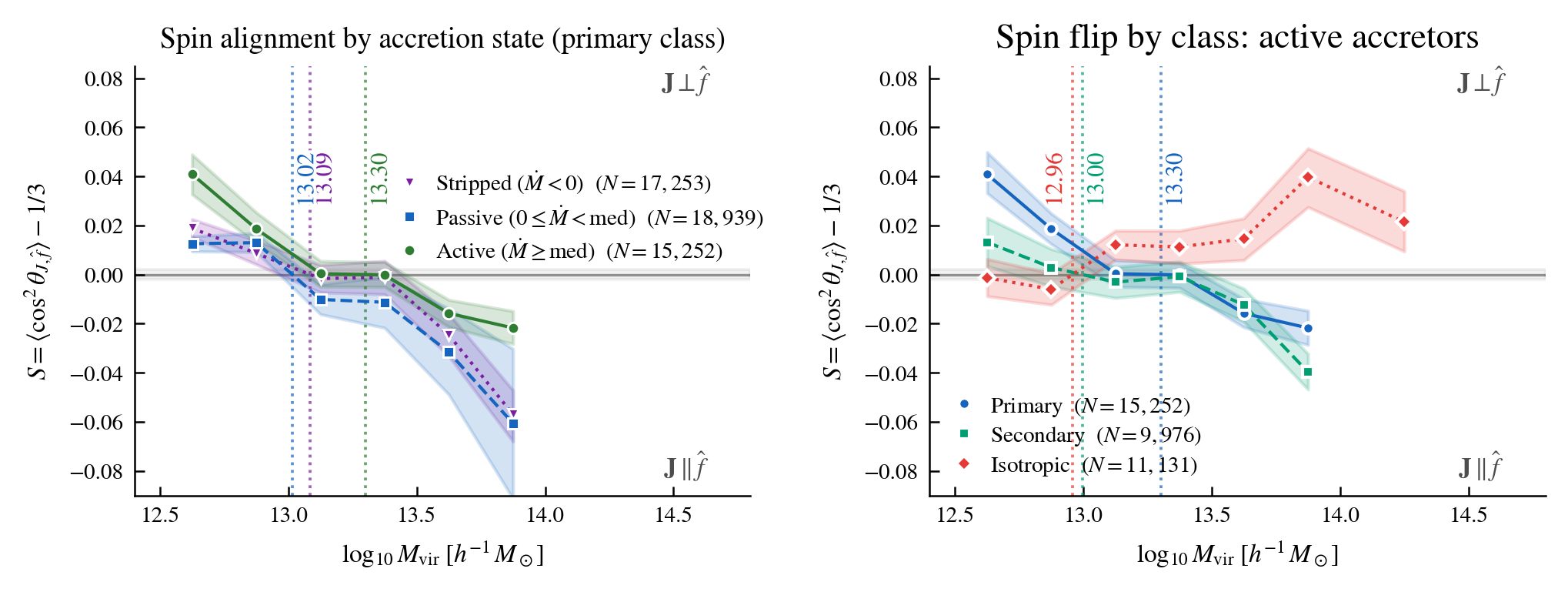}
    \caption{\textit{(a)} Spin-filament alignment $S(\log_{10} M)$ for primary groups split by instantaneous accretion rate: stripped ($\dot{M} < 0$, purple triangles), passive ($0 \leq \dot{M} < \mathrm{median}$, blue squares), active ($\dot{M} \geq \mathrm{median}$, green circles). Vertical dotted lines: $\mflip$ per subpopulation. \textit{(b)} $S(\log_{10} M)$ restricted to active accretors only for primary (dark blue), secondary (green), and isotropic (red). The channelled-versus-isotropic spin flip persists, indicating that the accretion rate is not what drives the topological signal.}
    \label{fig:accretion}
\end{figure*}

Figure~\ref{fig:accretion} tests whether the spin flip is driven by the instantaneous mass accretion rate $\dot{M}$ from \rockstar rather than by the topological classification. Panel~(a) splits primary groups into three accretion states: stripped ($\dot{M} < 0$; $N = 17{,}253$), passive ($0 \leq \dot{M} < \mathrm{median}$; $N = 18{,}939$), and active ($\dot{M} \geq \mathrm{median}$; $N = 15{,}252$). Passive groups have the lowest $\mflip$ (${\approx}13.02$), with stripped groups intermediate (${\approx}13.09$) and showing randomised spin alignment at high mass, consistent with tidal processing partially erasing the primordial tidal torque imprint. Active groups have the highest $\mflip$ (${\approx}13.30$), possibly because ongoing mass assembly is fastest in the strongest tidal environments.

Panel~(b) restricts the sample to active accretors only ($\dot{M} \geq \mathrm{median}$) and retains a spin flip in both channelled classes, with $\mflip$ values of $13.30$ (primary) and $13.00$ (secondary), whose ordering we do not interpret. The isotropic class shows only a marginal low-mass crossing (${\approx}12.96$), comparable to the secondary crossing rather than well below it, and consistent with its near-null full-sample signal. This indicates that the topological signal is not a selection effect of differential accretion rates across classes: the channelled-versus-isotropic contrast persists even among groups that are all actively growing.

\section{Connectivity and concentration dependence of the spin flip}
\label{sec:appendix_environment}

\begin{figure*}
    \centering
    \includegraphics[width=\textwidth]{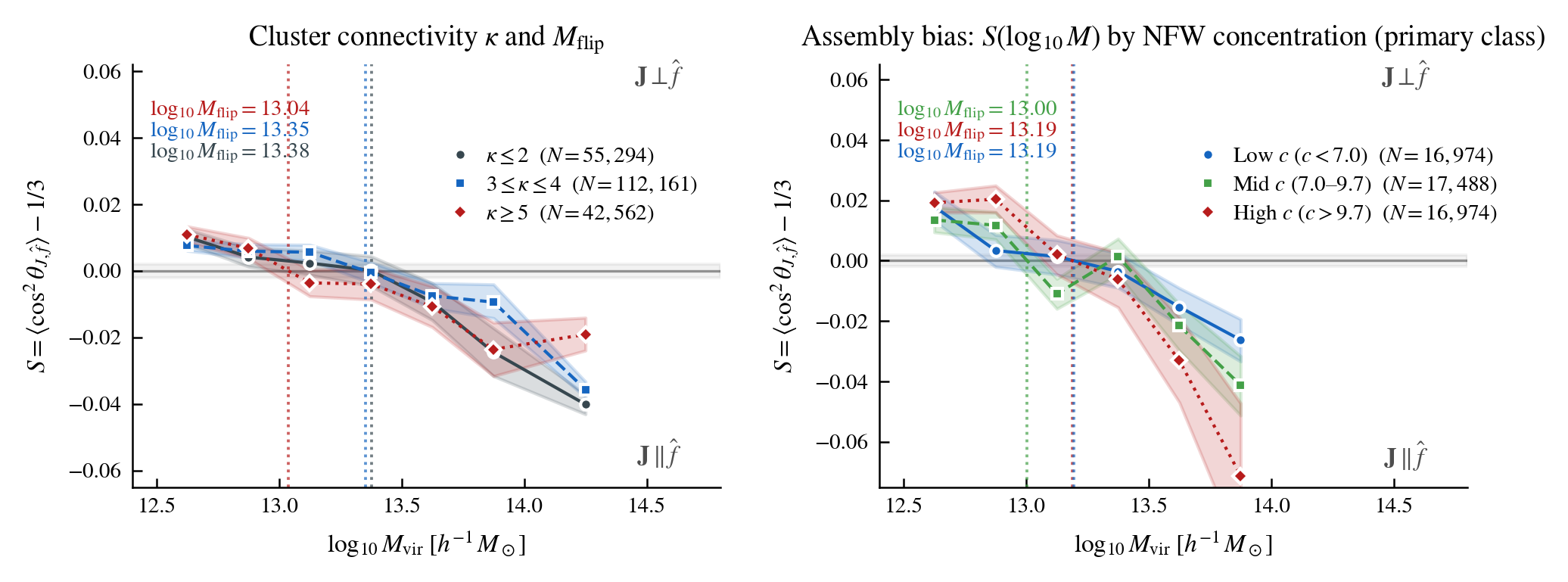}
    \caption{\textit{(a)} Spin-filament alignment $S(\log_{10} M)$ split by cluster connectivity into three bins: low $\kappa$ ($\kappa \leq 2$, black circles), intermediate $\kappa$ ($3 \leq \kappa \leq 4$, blue squares), and high $\kappa$ ($\kappa \geq 5$, red diamonds); curves overlap within uncertainties, consistent with no $\kappa$ dependence. \textit{(b)} $S(\log_{10} M)$ for primary groups split by NFW concentration tertiles: low $c$ (blue circles), medium $c$ (green squares), high $c$ (red diamonds). Vertical dotted lines: $\mflip$ per tertile (differences within the bootstrap uncertainty). Higher-concentration (earlier-forming) groups show stronger high-mass perpendicular alignment, the direction expected if $\mflip$ tracks formation history.}
    \label{fig:environment}
\end{figure*}

Figure~\ref{fig:environment} tests two environmental variables that could modulate $\mflip$. Panel~(a) splits the full sample by cluster connectivity $\kappa$ into three bins: low ($\kappa \leq 2$), intermediate ($3 \leq \kappa \leq 4$), and high ($\kappa \geq 5$). The $S(\log_{10} M)$ curves overlap within bootstrap uncertainties at all masses, showing no significant $\kappa$ dependence (Section~\ref{sec:spin_flip}). This is consistent with the expectation that $\kappa$ modulates the number of feeding channels but not the tidal environment within each channel.

Panel~(b) splits primary groups by NFW concentration tertiles ($c < 7.0$, $7.0$--$9.7$, $> 9.7$). Higher-concentration (earlier-forming) primaries show stronger perpendicular spin alignment at high mass, the direction predicted by ZOMG/TTT, in which haloes that formed earlier in denser environments have higher concentrations and a lower spin-flip mass. The flip-mass zero-crossings of the three tertiles differ by less than the bootstrap uncertainty, however: because concentration anticorrelates with halo mass, the highest-concentration tertile contains few high-mass groups and its crossing is poorly constrained. We therefore read this panel as a directional rather than a quantitative confirmation that $\mflip$ tracks formation history; the clearer modulation of $\mflip$ comes instead from the filament linear density (Section~\ref{sec:spinflip_quality}).

\section{Robustness of the main results}
\label{sec:appendix_robustness}

\subsection{Statistical independence of the spin-alignment pairs}
\label{sec:appendix_independence}

The spin-alignment statistic is measured over all classified pairs with a defined filament tangent (Section~\ref{sec:spin_method}), so a group in the infall zones of several clusters enters it more than once with the same $\mathbf{J}$. Multiplicity is small. The $210{,}017$ pairs belong to $179{,}065$ unique groups. These pairs include the clusters' $26{,}601$ self-pairs, which are almost all primary-outskirt and all at $\log_{10} M_\mathrm{vir} \geq 14.0$, above the flip-mass range (Section~\ref{sec:catalogues}). Within a single class a group appears on average $1.04$ (primary), $1.06$ (secondary) and $1.10$ (isotropic) times, because a group near two clusters is usually assigned a different class with respect to each. To test whether the pair-level bootstrap underestimates the uncertainties, we implement three bootstrap schemes in a common block form. Each of 2000 resamples draws $N_\mathrm{blocks}$ blocks with replacement and recomputes $S$ in every mass bin, and hence $\mflip$. The block is a pair (the published scheme), a unique group (all pairs of a group together) or a cluster (all pairs of a cluster together).

Resampling unique groups inflates the uncertainty on $S$ by ${\approx}5$ per cent. The median inflation over mass bins is $2$--$9$ per cent across the three classes and eight random seeds, and the inflation is at most $12$ per cent in any single bin. The cluster-block scheme changes the uncertainty by at most a few per cent (medians $-2$ to $+3$ per cent, within the Monte Carlo noise). Cluster blocks are not the more conservative test here. Alignments of different groups around the same cluster are nearly uncorrelated (each cluster contributes $2.2$--$2.6$ pairs on average). Because a repeated group's pairs belong to different clusters by construction, cluster blocks cannot capture the repetition of $\mathbf{J}$, and the unique-group scheme is the relevant test for it.

The $16$th--$84$th percentile range of the primary $\mflip$ is $[13.04, 13.12]$ for pairs, $[13.04, 13.13]$ for unique groups and $[13.04, 13.12]$ for clusters, and that of the secondary $\mflip$ is $[12.82, 13.20]$, $[12.81, 13.20]$ and $[12.82, 13.21]$. The pair-level scheme reproduces the intervals quoted in Section~\ref{sec:spin_flip} to within $0.01$\,dex, and across random seeds the three schemes' intervals differ by at most $0.02$\,dex, which is bootstrap noise. For the isotropic class, the observed $S$ is positive in every mass bin, and $47$--$53$ per cent of resamples produce a zero crossing somewhere from noise alone.

The density split of Section~\ref{sec:spinflip_quality} uses the dominant-cluster sample and so contains each group at most once, but groups on a common filament share its density and environment. The thin, intermediate and dense tertiles hold $14{,}915$, $15{,}366$ and $14{,}915$ pairs on $11{,}821$, $10{,}617$ and $9{,}490$ filaments, and $38$, $54$ and $62$ per cent of their pairs share a filament with another pair. We also resample whole filaments, or whole clusters, as blocks. The $16$th--$84$th percentile ranges of the tertile flip masses, and of the dense--thin difference before and after covariate matching (Appendix~\ref{sec:appendix_rholin}) change by at most $0.01$\,dex from the single-pair values. With either block, the matched difference stays positive in $95$ per cent of the pooled bootstrap samples. Groups that share a filament share its density but have nearly independent spin alignments. We therefore retain the pair-level uncertainties throughout.

\subsection{Covariates of the filament linear density}
\label{sec:appendix_rholin}

\begin{figure*}
    \centering
    \includegraphics[width=\textwidth]{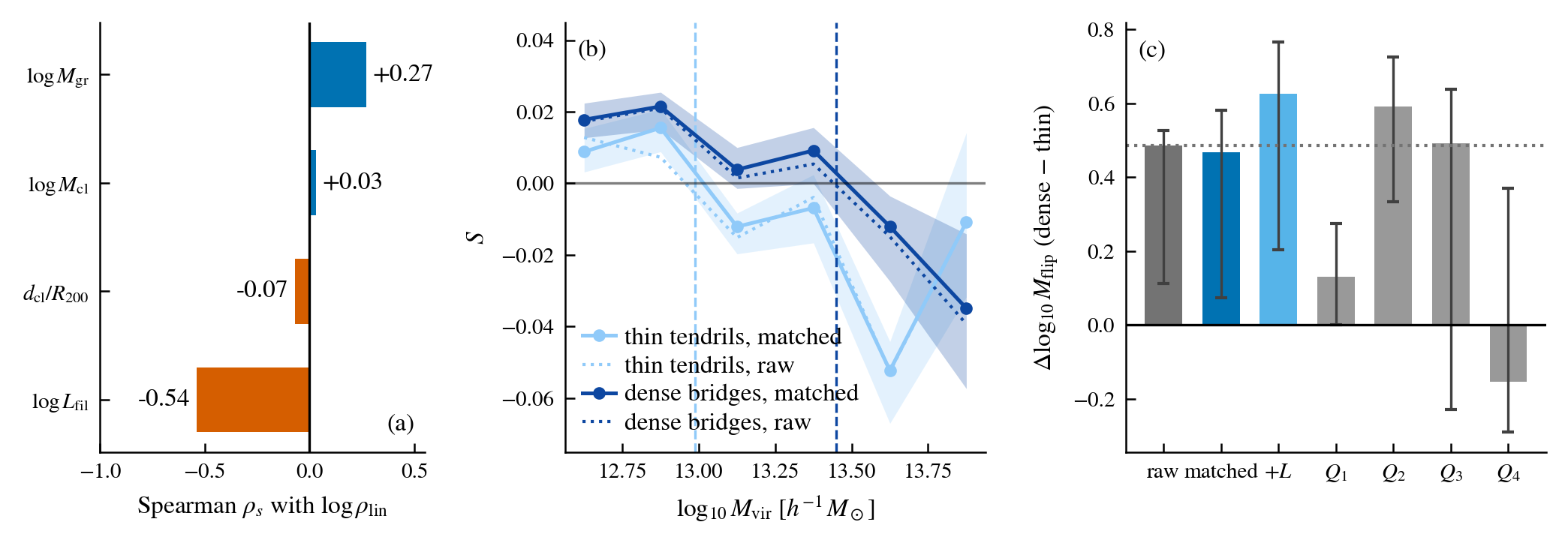}
    \caption{Covariates of the filament linear density. \textit{(a)} Spearman rank correlation of $\log\rho_\mathrm{lin}$ with filament length, cluster-centric distance, cluster mass and group mass, for $45{,}196$ primary pairs. \textit{(b)} $S(\log_{10} M)$ for the thin-tendril (light blue) and dense-bridge (dark blue) tertiles, unmatched (dotted) and matched in group mass, cluster mass and cluster-centric distance (solid, with $16$th--$84$th percentile bands), for the matching draw whose difference is closest to the median. Vertical dashed lines mark the median matched $\mflip$ over 200 draws. \textit{(c)} Dense-minus-thin difference in $\log_{10}\mflip$, unmatched, matched, additionally matched in filament length ($+L$), and within quartiles $Q_1$--$Q_4$ of filament length (grey). Error bars are $16$th--$84$th percentiles; for the matched bars they are percentiles of the bootstrap pooled over 200 matching draws, around their median. The dotted line marks the unmatched value.}
    \label{fig:robust_rholin}
\end{figure*}

The flip-mass sequence of Section~\ref{sec:spinflip_quality} could reflect a covariate of $\rho_\mathrm{lin}$ rather than $\rho_\mathrm{lin}$ itself. Among the $45{,}196$ primary pairs with a measured density, the Spearman rank correlations of $\log\rho_\mathrm{lin}$ are $-0.54$ with filament length $L_\mathrm{fil}$, $+0.27$ with group mass, $-0.07$ with cluster-centric distance and $+0.03$ with cluster mass (Fig.~\ref{fig:robust_rholin}a).

To control for them we match the thin-tendril and dense-bridge tertiles cell by cell. The covariate space is binned. In each cell the more populous tertile is randomly reduced to the size of the less populous one, so that the matched samples share the same joint covariate distribution to the resolution of the binning. Because the alignment statistic of the dense bridges is shallow near zero, a single random matched sample can move their flip mass by up to ${\approx}0.35$\,dex. The draw-to-draw r.m.s.\ is $0.16$\,dex, against $0.02$\,dex for the thin tendrils. We therefore repeat the matching 200 times with independent random draws, bootstrap each matched sample 100 times, and quote the median over the draws with the $16$th--$84$th percentiles of the pooled bootstrap, which include the draw-to-draw scatter.

Matching in group mass, cluster mass and cluster-centric distance uses $825$ occupied cells and ${\approx}2 \times 10{,}200$ pairs per draw. This matching gives median flip masses of $12.99$ (thin) and $13.45$ (dense), and a median per-draw difference of $0.47$\,dex ($0.07$--$0.58$) against $0.49$\,dex ($0.11$--$0.53$) for the unmatched tertiles (Fig.~\ref{fig:robust_rholin}b,c). The difference is positive in all 200 draws (which are not independent, since the less populous tertile of each cell enters every draw whole) and in $95$ per cent of the pooled bootstrap samples. Its sign is therefore favoured, but its size is less certain. The unmatched dense-bridge $S$ at $\log_{10} M = 13.125$ is $+0.0015$ ($16$th--$84$th percentiles $-0.005$ to $+0.008$). Small changes to the sample therefore move the first zero crossing of the dense-bridge statistic between this mass bin and the next, and the 200 matched draws give $13.09$--$13.77$. Additionally matching filament length (${\approx}2 \times 5{,}470$ pairs per draw) increases the difference to $0.63$\,dex ($0.20$--$0.77$). This increase is expected, since $L_\mathrm{fil}$ is the denominator of $\rho_\mathrm{lin}$. At fixed length, the density contrast is purely a contrast in mass.

As a diagnostic we also split the primary pairs into quartiles of filament length (Fig.~\ref{fig:robust_rholin}c). The difference is $+0.13$\,dex ($0.00$--$0.28$) in the shortest quartile, $+0.59$ ($0.33$--$0.72$) and $+0.49$ ($-0.23$--$0.64$) in the middle two, and $-0.15$ ($-0.29$--$0.37$) in the longest. These subsamples are strongly unbalanced, because long filaments are rarely dense. The longest quartile contains $6{,}959$ thin-tendril pairs but only $675$ dense-bridge pairs, and the shortest the reverse ($1{,}389$ against $6{,}973$). The matched test, which compares the tertiles on a common covariate distribution, is the appropriate control. In the quartile split the contrast is largest in the second quartile, similar in size but unconstrained in the third, and small ($Q_1$) or consistent with zero ($Q_4$) in the extreme quartiles, where the tertiles are most unbalanced.

\subsection{Validation and distance dependence of the predicted pericentre}
\label{sec:appendix_pericentre}

The predicted pericentres of Section~\ref{sec:pericentre} are turning points of a static, spherical NFW potential calibrated at $z = 0$. The subsequent pericentres of infalling groups lie in the future of the $z = 0$ snapshot, but confirmed backsplash groups have already passed pericentre and their previous closest approach is recorded in the merger trees. In a static spherical potential the next turning point equals the previous one. Applying the same estimator to backsplash groups at $z = 0$ and comparing with the recorded closest approach therefore measures the combined effect of the approximations: the cluster's growth since the passage, its asphericity, dynamical friction and tidal mass loss.

The validation sample consists of the $1{,}926$ dominant pairs (self-pairs excluded) whose group's main progenitor passed within the virial radius $R_\mathrm{vir}$ of the cluster's main progenitor. This is slightly looser than the $R_{200}$ criterion of the backsplash cut (Section~\ref{sec:classification}), and contains all $1{,}838$ groups that criterion flags. For the purpose of testing the estimator, any group that has passed pericentre will do. For each of them we follow the main progenitor branches of the group and of the cluster through the consistent-trees catalogues and record the minimum of their separation in units of the cluster progenitor's virial radius. The minimum is bracketed by snapshots on both sides for $99.9$ per cent of systems, so it is an interior turning point, and a parabola through the three bracketing snapshots refines it (lowering it by a median $0.9$ per cent). We convert it to units of the progenitor's own $R_{200c}$ by solving its NFW profile. The conversion is needed because at the epoch of closest approach (median scale factor $a = 0.84$), $R_\mathrm{vir}/R_{200c}$ is $1.27$ (16th--84th percentiles $1.23$--$1.29$), which differs from its $z = 0$ value. The predicted pericentre is in units of the $z = 0$ $R_{200}$, as in Section~\ref{sec:pericentre}.

\begin{figure*}
    \centering
    \includegraphics[width=\textwidth]{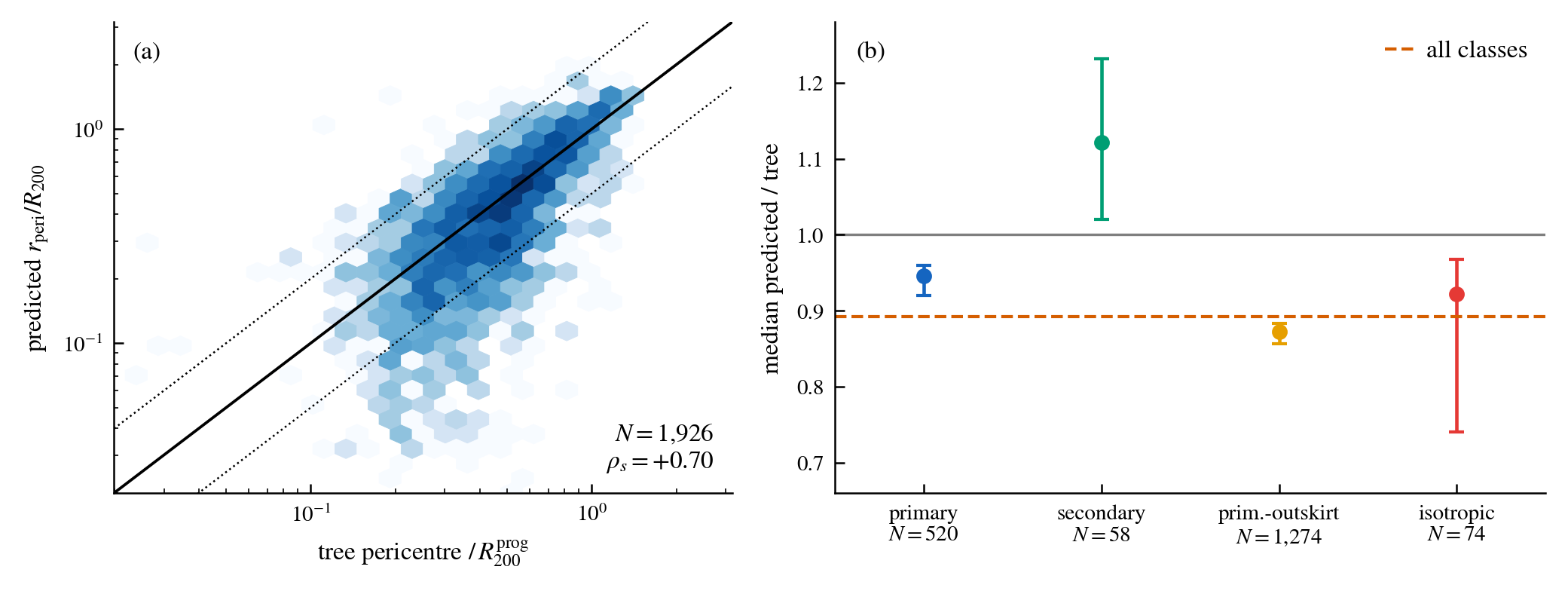}
    \caption{Merger-tree validation of the predicted pericentre for the $1{,}926$ groups whose progenitor has passed within $R_\mathrm{vir}$ of the cluster's main progenitor. \textit{(a)} Predicted pericentre (static NFW potential of mass $M_{200c}$ at $z = 0$, in units of the $z = 0$ $R_{200}$) against the closest approach recorded in the merger trees (in units of the cluster progenitor's $R_{200}$ at that epoch). The solid line is equality, the dotted lines a factor of two, and the shading the number of systems per hexagonal cell. \textit{(b)} Median ratio of predicted to recorded pericentre by accretion class, with $16$th--$84$th percentile bootstrap intervals; the dashed line is the median over all classes and the solid line a ratio of one.}
    \label{fig:robust_pericentre}
    \label{lastpage}
\end{figure*}

The predicted and recorded pericentres are well correlated (Fig.~\ref{fig:robust_pericentre}a; Spearman $\rho_s = +0.70$). The prediction is smaller than the recorded value by a median factor of $0.89$ ($0.88$--$0.91$), with $78$ per cent of systems within a factor of two and a scatter of $0.28$\,dex. Most of this offset is the growth of $R_{200}$ since the passage (a median factor $1.07$). In physical units the median ratio is $0.97$ ($0.96$--$0.98$), so the predicted turning point itself is only slightly too small. Groups retain a median $0.69$ of the virial mass they had at closest approach, but the estimator bias is essentially uncorrelated with the fraction of mass retained (Spearman $+0.07$), so tidal stripping is not what drives it. Replacing $M_{200c}$ by $M_\mathrm{vir}$ as the mass within $R_{200}$ deepens the potential and lowers the median ratio to $0.77$ on the same pairs. We therefore adopt $M_{200c}$.

The quantity that matters for this paper is the comparison between classes, which a class-dependent bias could in principle distort. The median ratio is $0.95$ ($0.92$--$0.96$; $N = 520$) for primary and $0.92$ ($0.74$--$0.97$; $N = 74$) for isotropic groups (Fig.~\ref{fig:robust_pericentre}b). Their difference is $+0.02$ ($16$th--$84$th percentiles $-0.03$ to $+0.21$). Its 95 per cent interval ($-0.07$ to $+0.28$) includes zero, so the two biases are consistent within wide errors. The recorded pericentres themselves, which involve no estimator, are the same for the two classes. The medians are $0.453\,[0.427,\,0.476]$ for primary and $0.453\,[0.359,\,0.516]\,R_{200}$ for isotropic groups (95 per cent intervals), at the same median current distance ($1.59$ and $1.60\,R_{200}$). The ratio is lower for primary-outskirt groups ($0.87$; $N = 1{,}274$) and higher for the small secondary sample ($1.12$, $1.02$--$1.23$; $N = 58$), so if anything the secondary predicted pericentres are overestimated relative to the isotropic ones. Backsplash groups lie closer to the cluster than the first-infall sample of Section~\ref{sec:pericentre} (median $1.5\,R_{200}$, two thirds moving outward, against $4$--$6\,R_{200}$ and $96$--$97$ per cent infalling) and are on a later passage. This test therefore checks the estimator in a different regime; it does not calibrate it for first-infall groups.

The estimator also uses peculiar velocities only (omitting the Hubble flow, a median $H_0 r = 437\,\mathrm{km\,s^{-1}}$) and an untruncated NFW profile. Adding the Hubble flow gives medians of $0.80$, $1.02$ and $1.01\,R_{200}$ (primary, secondary, isotropic; $97.6$ per cent bound), and truncating the profile at $R_{200}$ gives $0.71$, $0.88$ and $0.87\,R_{200}$ ($46.5$ per cent bound). The primary--isotropic contrasts in these two variants are $21$ and $19$ per cent. Including the $9{,}476$ bound pairs whose energies lie beyond the $r = 20$\,Mpc circular-orbit grid used for $\eta$ gives $0.75$, $0.93$ and $0.92\,R_{200}$ and $19$ per cent. The absolute pericentres are thus model-dependent at the ${\lesssim}16$ per cent level, but the pooled ordering of the classes is not.

\textit{Comparison at fixed distance.} Because predicted pericentres grow with cluster-centric distance and the classes have different distance distributions (Fig.~\ref{fig:pericentre}b), we compare them at fixed distance by reweighting one class to the distance distribution of the other in $0.1\,R_{200}$ cells. Reweighted to the primary distances, the isotropic median is $0.660\,R_{200}$, $9.9$ per cent ($95$ per cent interval $7.8$--$11.8$) below the primary $0.733$. Reweighting the primary groups to the isotropic distances instead gives $0.960$, against the isotropic $0.876$. Matching jointly on distance and cluster mass gives $0.684$ ($6.7$ per cent below primary; $4.0$--$9.1$). Matching on distance, cluster mass and group mass gives $0.677$ ($7.7$; $4.5$--$9.5$). Matching on physical distance in $h^{-1}$\,Mpc gives $0.706$ ($3.7$; $0.9$--$6.4$), and reweighting the primary groups to the isotropic physical distances instead gives $1.5$ per cent ($-0.7$ to $+3.5$). Matching on distance for groups within $7\,R_{200}$ of a single cluster gives $10.3$ per cent ($8.0$--$12.8$).

Comparing the two classes within each cluster, in $0.5\,R_{200}$ shells that contain both, the primary median is the larger by $0.077\,R_{200}$ ($0.044$--$0.108$; $3{,}586$ shells). The estimator variants give the same reversal. Reweighted to the primary distances, the isotropic and primary medians are $0.683$ and $0.752$ (all bound pairs), $0.744$ and $0.803$ (Hubble flow added), $0.629$ and $0.706$ (NFW truncated at $R_{200}$), and $0.597$ and $0.657$ ($M_\mathrm{vir}$ normalisation). Secondary groups reweighted to the primary distances have a median of $0.761$, $0.10\,R_{200}$ ($0.08$--$0.12$) above the reweighted isotropic one. Restricting the pooled comparison to small distances gives primary and isotropic medians of $0.525$ and $0.478$ within $3\,R_{200}$, $0.583$ and $0.537$ within $4\,R_{200}$, and $0.626$ and $0.613$ within the turnaround radius (95 per cent interval of the difference $-0.004$ to $+0.030$).

\end{document}